\documentclass[useAMS,usenatbib]{mn2e}

\usepackage{rotating}
\usepackage{longtable}
\usepackage{graphics}

\usepackage{aas_macros}

\newcommand{\teff}{$T_{\rm eff}$}

\def\lgg{log\,${g}$}
\def\vt{$\xi_{\rm t}$}

\newcommand{\kms}{km\,s$^{-1}$}

\def\ione{\,{\sc i}}
\def\ii{\,{\sc ii}}
\def\iii{\,{\sc iii}}
\def\iv{\,{\sc iv}}
\newcommand{\eps}{\log\varepsilon}

\newcommand{\kH}{$S_{\rm H}$}    

\title[Chemical abundances of metal-poor stars in Seg 1, ComaBer, and Tri\ii\ UFDs]
{The formation of the Milky Way halo and its dwarf satellites: a NLTE-1D abundance analysis. \\ IV. Segue 1, Triangulum\ii, and Coma Berenices UFDs. 
\thanks{Based on observations obtained at the W. M. Keck Observatory, which is operated jointly by the California Institute of Technology and the University of California, and the National Aeronautics and Space Administration.}
}

\author[Sitnova et al.]{
\parbox{\textwidth}{
T. M. Sitnova$^{1}$\thanks{E-mail:sitnova@inasan.ru},
L. I. Mashonkina$^{1}$, A. M. Tatarnikov$^{2}$, O. V. Voziakova$^{2}$,  
M.~A.~Burlak$^{2}$, Yu. V. Pakhomov$^{1}$, 
P. Jablonka$^{3,4}$,
M. D. Neretina$^{1,2}$,
A. Frebel$^{5,6}$
}\\
\\
$^{1}$Institute of Astronomy, Russian Academy of Sciences, Pyatnitskaya 48, 119017, Moscow, Russia\\
$^{2}$Sternberg Astronomical Institute, Moscow M. V. Lomonosov State University, Universitetskiy 13, Moscow 119992, Russia\\
$^{3}$Laboratoire d' Astrophysique, Ecole Polytechnique F\'ed\'erale de Lausanne (EPFL), Observatoire de Sauverny, CH-1290 Versoix, \\
Switzerland \\
$^{4}$Observatoire de Paris, GEPI, CNRS, Sorbonn\'e University, PSL Research Universty, 75014 Paris, France \\
$^{5}$Department of Physics \& Kavli Institute for Astrophysics and Space Research, Massachusetts Institute of Technology,\\ Cambridge, MA 02139, USA \\
$^{6}$Joint Institute for Nuclear Astrophysics--Center for Evolution of the Elements, East Lansing, MI 48824, USA
}

\begin{document}

\date{}

\pagerange{\pageref{firstpage}--\pageref{lastpage}} \pubyear{2019}

\maketitle

\label{firstpage}

\begin{abstract}

%We present atmospheric parameters and the NLTE abundances for important chemical elements from carbon to barium in  metal-poor stars in Coma Berenices and Segue~1 systems. Effective temperatures for the sample stars are based on new photometric observations in the visible and infra-red bands obtained with the 2.5-m telescope of the SAI MSU Caucasian observatory. We found element abundance patterns in ComaBer S1, S2, Seg~1 SA, and SB to be peculiar and show low [Na/Mg] from $-1.08$ dex to $-1.67$, which is usually attributed in the literature to nucleosynthesis in explosions of massive metal-free stars. Our results for Coma Berenices stars confirm lower [Sr/Ba] abundance ratio in UFD compared to those in classical dwarfs and the Milky Way halo stars found in the literature. We suppose that the additional Sr production at the earliest epoch of galactic formation could be attributed to evolution of massive metal-free binaries.
%lima
%We revised the effective temperatures of six metal-poor stars in the ultra-faint dwarf galaxy (UFD) Coma Berenices and the Segue 1  system using new photometric observations in the visible and infra-red bands obtained with the 2.5-m telescope of the SAI MSU Caucasian observatory.

We present atmospheric parameters and abundances for chemical elements from carbon to barium in metal-poor stars in Segue~1 (seven stars), Coma Berenices (three stars), and Triangulum\ii\ (one star) ultra-faint dwarf galaxies (UFDs).
The effective temperatures rely on new photometric observations in the visible and infra-red bands, obtained with the 2.5~m telescope of the SAI MSU Caucasian observatory.  
Abundances of  up to fourteen chemical elements were derived under the non-local thermodynamic equilibrium (NLTE) line formation, and LTE abundances were obtained for up to five more elements. For the first time we present abundance of  oxygen in Seg~1 S1 and S4, silicon in ComaBer S2 and Tri\ii\ S40, potassium in Seg~1 S1$-$S6 and ComaBer S1$-$S3, and barium in Seg~1 S7.
Three stars in Segue~1, two stars in Coma Berenices, and Triangulum\ii\ star have very low [Na/Mg] of $-1.08$ to $-1.67$~dex, which is usually attributed in the literature to an odd-even effect produced by nucleosynthesis in massive metal-free stars. 
We interpret this chemical property as a footprint of first stars, which is not blurred due to a small number of nucleosynthesis events that contributed to chemical abundance patterns of the sample stars.
Our NLTE abundances of Sr and Ba in Coma Berenices, Segue~1, and Triangulum\ii\ report on lower [Sr/Ba] abundance ratio in the UFDs compared to that in  classical dwarf spheroidal galaxies and the Milky Way halo.   However, in UFDs, just as in massive galaxies, [Sr/Ba] is not constant and it can be higher than the pure r-process ratio.
We suggest a hypothesis of Sr production in metal-poor binaries at the earliest epoch of galactic evolution.
%We suppose that an additional production of Sr, occurred in massive galaxies at the earliest epoch of their evolution, could be attributed to massive metal-free binaries. 
%The stars in Segue 1 do not reveal a  [Mg/Na] anti-correlation excluding thus a hypothesis that it is a typical Galactic globular cluster. 

\end{abstract}


%Skuladottir 2017
%Thechemical signature of these zero-metallicity PISN might be pre-served in long-lived, relatively metal-rich stars at [Fe/H]&−2(Salvadori et al. 2007; de Bennassuti et al. 2017), which shouldshow low [Zn/Fe] values accompanied with a high abundanceratios  between  even  and  odd  elements  (e.g.  Heger & Woosley2002).

\section{Introduction}

Ultra-faint dwarf galaxies (UFDs) are  the least luminous, oldest, most dark matter-dominated, most metal-poor, and least chemically evolved stellar systems known \citep[for a review, see][]{2019ARA&A..57..375S,2009ARA&A..47..371T}.
From analyses of colour-magnitude diagrams and spectra, it is known that some faint dwarf galaxies formed their stars within a short period, less than 1~Gyr \citep{2008AA...487..103O,2012ApJ...744...96O,2012ApJ...759..115F,2014ApJ...786...74F,2017MNRAS.466.3741V}.
Short star formation timescale together with a low stellar mass 
result in a small number of nucleosynthesis events that contributed to chemical composition of long-lived stars observed today in faint dwarf galaxies \citep[see, for example,][]{2014ApJ...786...74F}.

%The observed abundance patterns in individual stars contain information about those nucleosynthesis events, their properties and progenitors, and the efficiency of gas mixing at the early epoch of galactic formation.
The observed abundance patterns in individual stars contain information about properties of nucleosynthesis events and their progenitors and the efficiency of gas mixing at the early epoch of galactic formation.
The more enrichment events a system encounters, the more blurred the signature of an individual event becomes. 
Unusual chemical element abundance ratios found in some metal-poor stars are interpreted as evidence of individual nucleosynthesis episodes.
Nearby dwarf galaxies are the best places to find such stars. 
%Such footprints are visible in outlying chemical element abundance ratios found in metal-poor stars.
%The more enrichment events a system encounters, the more washed-out the signature of any  individual event becomes.  UFDs are the best candidates for search footprints of individual nucleosynthesis events.
%Number of studies report on peculiar chemical element abundance patterns found in metal-poor stars in nearby dwarf galaxies.
%, which are explained by inhomogeneous mixing and with small number of SN\ii\ explosions of stars with different progenitor masses, contributed to abundances of the observed stars. 
For example, \citet{2008ApJ...688L..13K} reported the chemical composition of two stars in the Hercules Dwarf Spheroidal Galaxy (dSph) with high ratio of hydrostatic to explosive $\alpha$-element abundance ratios, [Mg/Ca]\footnote{We use a standard designation, [X/Y] = log(N$_{\rm X}$/N$_{\rm Y}$)$_{*}$ - log(N$_{\rm X}$/N$_{\rm Y}$)$_{\odot}$, where N$_{\rm X}$ and N$_{\rm Y}$ are total number densities of element X and Y, respectively.} = 0.94 and 0.58 dex and [O/Ca] = 1.25 and $<$ 0.70 dex, respectively. A similar chemical feature, [Mg/Ca] = 0.61, was found in  S119  in Draco dSph \citep{2004ApJ...612..447F}. However, for this star, \citet{1998AJ....115.1888S}  found a normal  [Ca/Mg] ratio of 0.12. 
\citet{2020ApJ...897..183H} found [Mg/Ca] = 0.3 in a star in Grus\ii\ UFD. We note the large errors of 0.20~dex and 0.26~dex for Mg and Ca abundances, respectively, found by \citet{2020ApJ...897..183H}. 
% Aoki_he1327,
In the Milky Way (MW) halo, there are stars with high [Mg/Ca] $> 0.5$  \citep[for example,][]{2008ApJ...684..588F,2013ApJ...762...28N,2014Natur.506..463K,Depagne2002,2008ApJ...672..320C}. 
Thus, similar chemical abundance peculiarities could be found in stars in different galaxies.
%We note, that the listed MW stars with have [Fe/H] $< -4$,  while in dwarf galaxies one can find stars with peculiar abundance pattern at higher [Fe/H] of $-2$.

Some dwarf galaxies may have individual endemic chemical peculiarities.
For example,  in Hor\ione\ UFD stars, \citet{2018ApJ...852...99N} found surprisingly low [$\alpha$/Fe] $\simeq$ 0 in contrast to typical [$\alpha$/Fe] $\sim$ 0.3 found in the literature in the majority of metal-poor (MP) stars in the MW and dwarf galaxies. 
Ret\ii, Tuc\iii, and Gru\ii\ UFDs are known for their stars enhanced in r-process elements \citep{2016ApJ...830...93J,2017ApJ...838...44H,2020ApJ...897..183H} in contrast to the extremely low neutron-capture (n-capture) element abundances found in stars in other UFDs \citep[see, for example,][]{2017AA...608A..89M,2019ApJ...870...83J}.

There are stars with normal chemical composition in dwarf galaxies as well.
For example, in Boo\ii\ UFD,   \citet{2014ApJ...794...89K} and \citet{2016ApJ...817...41J} found no chemical peculiarities and abundance patterns to be similar to the normal MW halo stars for all elements except low neutron-capture element abundances.

Chemical properties of stars in different galaxies depend on the formation history of their host galaxies.
To understand how different or similar the first stars in different UFDs were and how well the gas in a given UFD  was mixed, it is necessary to perform a careful homogeneous abundance determination in stars in different UFDs. 
%For our analysis, we selected stars in the  smallest UFDs known to date, where high-resolution spectra of individual stars are available in the archives.
In the present analysis, we selected stars in the smallest UFDs known to date with individual high-resolution spectra available in the archives.
The selected UFDs, namely, Segue~1, Triangulum\ii, and Coma Berenices meet the criteria of \citet{2012ApJ...759..115F} and could be recognized as surviving first galaxies.

%Segue~1 (Seg~1) with M$_{\rm V}$ = $-3.0$ and r$_{\rm 1/2}$ = 30~pc \citep{2007ApJ...654..897B,2009ApJ...692.1464G,2011ApJ...733...46S}
Segue~1 (Seg~1) with M$_{\rm V}$ = $-1.5$ and half-light radius r$_{\rm 1/2}$ = 30~pc
\citep{2008ApJ...684.1075M,2009ApJ...692.1464G,2011ApJ...733...46S} is one of the smallest  and faintest known UFDs, that first was thought to be a globular cluster \citep{2007ApJ...654..897B,2009MNRAS.398.1771N}.
Using high-resolution spectra, \citet[][hereafter F14]{2014ApJ...786...74F} and \citet[][hereafter N10]{2010ApJ...722L.104N} determined the chemical composition  of Segue~1 stars.
Stars in Seg~1 span a wide metallicity range of $-3.78 <$ [Fe/H] $< -1.42$ and show high [$\alpha$/Fe], indicating enrichment only from massive stars.
This result is in line with medium resolution observations of 
%\citet{2013ApJ...767..134V}.
\citet[][hereafter V13]{2013ApJ...767..134V}. 
%S2 (P=0.00142), S3 (P=0.3), S6 (P=0.37), S7 (P=0.00024)

Triangulum\ii\ (Tri\ii) has M$_{\rm V}$ = $-1.8$ and r$_{\rm 1/2}$ = 34~pc  \citep{2015ApJ...802L..18L,2016ApJ...818...40M,2017ApJ...838...83K}, which is very similar to Seg~1. 
%From low-resolution observations of 13 members, \citet[][hereafter K17]{2017ApJ...838...83K} found a metallicity range of $-2.86 < $ [Fe/H] $< -1.40$. 
From low-resolution observations of 13 Tri\ii\ members, \citet[][hereafter K17]{2017ApJ...838...83K} found that their metallicities range from $-2.86$ to $-1.40$.
The two brightest stars in Tri\ii\ were studied with high-resolution spectra \citep{2017ApJ...838...83K,2017MNRAS.466.3741V,2019ApJ...870...83J}. Tri\ii\ S40 with [Fe/H] = $-2.9$ shows normal [$\alpha$/Fe] of 0.3, while low [Na/Fe] from $-1.1$ \citep{2017ApJ...838...83K} to $-0.8$ \citep{2019ApJ...870...83J}, high [K/Fe] of 0.8, and high [Ni/Fe] of 0.5.
Another star, S46, with [Fe/H] $\simeq -2$ shows low [$\alpha$/Fe] $< 0$. 
%The quality of the observed spectrum of S46 does not enable measurements of abundances of iron-peak elements  to reconcile whether low [$\alpha$/Fe] is caused by a prompt   SN  Ia enrichment or the stochastic sampling of few massive SNe\ii\ \citep{2017MNRAS.466.3741V,2019ApJ...870...83J}.
The quality of S46 observed spectrum does not enable iron-peak element abundances measurements  to reconcile whether low [$\alpha$/Fe] is caused by a prompt type \ione a  supernova (SN) enrichment or the stochastic sampling of a few massive SNe\ii\ \citep{2017MNRAS.466.3741V,2019ApJ...870...83J}.
%It is worth mentioning that \citet{2020arXiv200705011M} found low membership probability (p = 0.04) for S46 in Tri\ii.
It is worth noting that S46 is likely not a member of Tri\ii, since its membership probability p = 0.04 found by \citet{2020arXiv200705011M} is low.
%It is worth noting that it is likely the S46 is not a member of Tri\ii\ with low membership probability p = 0.04 found by \citet{2020arXiv200705011M}.

%Coma Berenices (ComaBer) UFD is another small  galaxy, with M$_{\rm V}$ = $-3.7$, half-light radius r$_{\rm 1/2}$ = 70 pc \citep{2007ApJ...654..897B}, and stellar mass of 3700~M$_{\odot}$ \citep{2012AJ....144....4M}. 
Coma Berenices (ComaBer) has M$_{\rm V}$ = $-4.1$,  r$_{\rm 1/2}$ = 70~pc \citep{2007ApJ...654..897B,2008ApJ...684.1075M}, and  is thus larger than Seg~1 and Tri\ii. 
Chemical composition of the three brightest stars with V $\simeq$ 18 mag  was determined by \citet[][hereafter F10]{2010ApJ...708..560F} from analyses of high-resolution spectra. 
F10 found that the metallicity spans $-2.88$ $<$ [Fe/H] $<$ $-2.31$, and that the most metal-rich star shows slightly lower [$\alpha$/Fe] abundance ratio compared to more metal-poor stars. F10 concluded that more observations of stars with [Fe/H]$>-2$  are required to understand whether the higher  iron abundance of this star could be due to an onset of iron production in SN\ione a or  stochastic mixing of few massive SNe\ii.
V13 increased the statistics and determined [Fe/H] and [$\alpha$/Fe] of ten stars using medium resolution spectra. 
%Stars in the V13  stellar sample show metallicities with $-3.38 <$ [Fe/H] $< -2.12$ 
Stars in the V13  stellar sample  have metallicities in the range  $-3.38 <$ [Fe/H] $< -2.12$
and various [$\alpha$/Fe] ratios: low [$\alpha$/Fe] $\simeq$ 0 in two stars with [Fe/H] $> -2.4$ and high [$\alpha$/Fe] $\simeq$ 0.3 to 0.9 dex in eight stars with [Fe/H] $< -2.4$. V13 concluded that low [$\alpha$/Fe] ratios in two stars with the highest [Fe/H] is caused by the production of iron  in SN\ione a.

In this study, we revisit existing data on the stars in Segue~1, Triangulum\ii, and Coma Berenices UFDs  and determine  accurate stellar parameters and chemical abundances, using new multiband photometric observations and accurate line formation calculations, taking into account deviations from the local thermodynamic equilibrium (LTE), i. e. using the NLTE approach.

%}
%Frebel and Bromm 2012
% https://iopscience.iop.org/article/10.1088/0004-637X/759/2/115/pdf

This study is performed within a project on a homogeneous NLTE abundance analysis of stars in the Galactic dwarf satellites. The studies by \citet[][hereafter M17a]{2017AA...604A.129M} and \citet[][hereafter M17b]{2017AA...608A..89M}, where accurate stellar parameters and the NLTE abundances were presented for 59 very metal-poor (VMP, $-4 < $[Fe/H]$ < -2$) stars in seven dwarf spheroidal galaxies (dSphs) and in the MW halo, gave rise to the project. Further, the stellar sample was complemented by \citet[][hereafter P19]{2019AstL...45..259P}. This paper employs the same method of stellar parameter and the NLTE abundance determinations as in M17a, M17b, and P19. 
%An exception is iron abundance, which is determined using the updated NLTE method presented by \citet{2019AA...631A..43M}. Here, we adopt iron abundance, calculated according to \citet{2019AA...631A..43M} for our total sample of stars (M17b, P19, and this study).
%	Here, we recalculated iron abundances in all the sample stars in order to avoid a systematic difference between M17b and this study.

Photometric observations and data reduction are described in Section~\ref{photometry}. 
We outline the method of stellar parameter determination in Section~\ref{param}.
The derived stellar abundances and the methods are presented in  Section~\ref{abundances}. The discussion of the results  and our conclusions are given in Sections \ref{discussion} and \ref{con}, respectively.

\section{Photometric observations and data reduction}\label{photometry}

%Photometric observations were carried out with 2.5-m telescope of Caucasian Mountain Observatory (CMO) of Sternberg Astronomical Institute of Moscow State University (SAI MSU) on February-April 2017, on nights with stable extinction, which was monitored by the astroclimate site monitor of CMO \citep{2014PASP..126..482K}. 
Photometric observations were carried out with Moscow State University Sternberg Astronomical Institute (SAI MSU) Caucasian Mountain Observatory (CMO) 2.5~m telescope in February to April 2017, on nights with stable extinction, which was monitored by the astroclimate site monitor of CMO \citep{2014PASP..126..482K}.

%Observations in the BVRcIc bands were carried out using an NBI 4k x 4k camera produced by N. Bohr Institute, Copenhagen. 
For observations in the BVRcIc bands, we used an NBI 4k x 4k camera produced by Niels Bohr Institute, Copenhagen. 
The preliminary reduction procedure comprises bias subtraction, nonlinearity and flat field correction. Magnitudes were derived with the source extractor package \citep{1996A&AS..117..393B} taking into account  aperture corrections. Coordinates of the objects were derived with the astrometry package \citep{2010AJ....139.1782L}. Extinction measurements and photometric calibrations were made using the Landolt standards \citep{2013noao.prop....2L}, observed on the same nights. The equations of the transformation to the standard photometric system were derived earlier using Landolt standards \citep{2013AJ....146...88C}. 

The $JHK$ photometric data were obtained 
%on 18 February and 27 April, 2017 
with the ASTRONIRCAM camera \citep{2017AstBu..72..349N} in dithering mode. 
%During each night, 30 frames for each filter were observed with a total exposure time of 900 seconds.
Depending on the seeing and exposure time, 30 to 60 frames, 30 seconds each,  were observed with each filter.
Nonlinearity, dark current, and flat field corrections were taken into account. Photometry was carried out in the MKO system, with a further transformation to 2MASS according to equations from \citet{2006MNRAS.373..781L}. 
%As the main standard, we adopted the star GSPC P264-F from the list of standards of the MKO photometric system \citep{2006MNRAS.373..781L}, to which bright enough stars in the field were referred and served as reference stars for every investigated object. 
GSPC P264-F star from the list of MKO photometric system standards  \citep{2006MNRAS.373..781L} was adopted as the main standard to which bright enough stars in the field were referred and served as reference stars for each investigated object.
The derived magnitudes are listed in Table~\ref{magnitudes}. 
An error of each magnitude is within several hundredths.

%{\bf
%The observations in the visible (BVRcIc) bands were performed only for three stars in Coma Berenices. For the other stars, we carried out observations  only in the infrared JHK bands. 

In addition to Coma Berenices and Segue~1 stars, we observed four stars investigated by 
%\citet{2017AA...604A.129M}
\citet{2017AA...604A.129M}, namely three stars in Ursa~Major\ii\ (UMa\ii) UFD and one star in Leo\iv\ UFD,   for which available literature  infrared (IR) magnitudes are either of low accuracy or missing. 
%Looking ahead, we note that, for the Ursa~Major\ii\ S1 star, the new photometric magnitudes result in  a significant difference in \teff\ of 150~K compared to those adopted by \citet{2017AA...604A.129M}. Using the revised stellar parameters, we recalculated abundances for this star  as described in Section~\ref{abundances}. For the other stars in Ursa~Major\ii\ and Leo\iv, we found the effective temperatures to be consistent with those adopted in  \citet{2017AA...604A.129M}.  
%The observations in the infrared JHK bands were performed for all sample stars, while the observations in the visible ($BVRcIc$) bands were performed only for three stars in Coma Berenices. 
In total, the observations in the infrared JHK bands were performed for ComBer S1--S3, Seg~1 S1--S3, UMa\ii\ S1--S3, and Leo\iv\ S1 stars, while the observations in the visible ($BVRcIc$) bands were performed only for three stars in Coma Berenices. For the remaining sample stars, the literature photometric data are accurate enough and were used for  \teff\ determination. For example, the errors in {\it gri} magnitudes  for Seg~1 S1, S2, and S3 are within 0.011~dex, which translates in errors in \teff\ smaller than 20~K for any of these stars, colour, and calibration.

\begin{table*}
	\caption{
Photometric magnitudes obtained with the 2.5~­m telescope of CMO of SAI MSU together with  literature data.}
%Photometric magnitudes of stars in different systems.}
		\label{magnitudes}
				\setlength{\tabcolsep}{2.5mm}
		\begin{tabular}{|l|l|l|c|c|c|c|c|c|c|}
			\hline
Star &  RA & DEC &  $B$   &  $V$   &    $R_c$   &   $I_c$    &   $J$   &   $H$    & $K$  \\
			\hline
			\multicolumn{10}{l}{Coma Berenices:} \\
 S1 & 12 26 43.44 & +23 57 02.6 & 18.920 &  18.164 &  17.678 &  17.197 &  16.514 $\pm$ 0.010  &  15.996 $\pm$ 0.020  &  15.955 $\pm$ 0.020 \\
 S2 & 12 26 55.45 & +23 56 09.5 & 18.854 &  18.099 &  17.608 &  17.115 &  16.407 $\pm$ 0.020  &  15.885 $\pm$ 0.030  &  15.834 $\pm$ 0.030  \\
 S3 & 12 26 56.66 & +23 56 11.8 & 18.375 &  17.572 &  17.058 &  16.552 &  15.818 $\pm$ 0.020  &  15.272 $\pm$ 0.020  &  15.213 $\pm$ 0.020  \\
 			\multicolumn{10}{l}{Segue 1:} \\	
S1 & 10 07 10.07 & +16 06 23.9 &       &  18.96$^1$   &       &      & 17.415 $\pm$ 0.009 & 16.931 $\pm$ 0.010 & 16.813 $\pm$ 0.013 \\
S2 & 10 07 02.46 & +15 50 55.2 &       &  18.22$^1$   &       &      & 16.691 $\pm$ 0.010 & 16.247 $\pm$ 0.013 & 16.136 $\pm$ 0.019 \\
S3 & 10 07 42.71 & +16 01 06.8 &       &  18.39$^1$   &       &      & 16.913 $\pm$ 0.017 & 16.451 $\pm$ 0.038 & 16.397 $\pm$ 0.017 \\
S4 & 10 07 14.6  & +16 01 54.5 &       &  18.58$^1$   &       &      & & & \\
S5 & 10 06 52.3  & +16 02 35.8 &       &  18.64$^1$   &       &      & & & \\
S6 & 10 06 39.3  & +16 00 08.9 &       &  19.26$^1$   &       &      & & & \\
S7 & 10 08 14.4  & +16 05 01.2 &       &  17.73$^1$   &       &      & 16.056$^2$ & 15.616$^2$ & 15.542$^2$ \\
 			\multicolumn{10}{l}{Ursa Major\ii:} \\	
S1 & 08 49 53.41 & +63 08 21.6 &       &  18.20$^1$   &       &      & 16.383 $\pm$ 0.017 & 15.912 $\pm$ 0.013 & 15.860 $\pm$ 0.029 \\
S2 & 08 52 33.50 & +63 05 00.9 &       &  17.74$^1$   &       &      & 15.805 $\pm$ 0.011 & 15.303 $\pm$ 0.007 & 15.186 $\pm$ 0.016 \\
S3 & 08 52 59.02 & +63 05 54.4 &       &  16.85$^1$   &       &      & 14.744 $\pm$ 0.010 & 14.139 $\pm$ 0.009 & 14.024 $\pm$ 0.014 \\ 			
 			\multicolumn{10}{l}{Leo\iv:} \\ 	
S1 & 11 32 55.99 & $-$00 30 27.8 &       &  19.22$^1$   &       &      & 17.193 $\pm$ 0.010 & 16.616 $\pm$ 0.009 & 16.494 $\pm$ 0.009 \\ 
 			\multicolumn{10}{l}{Tri\ii:} \\ 
S40 & 02 13 16.55 & +36 10 45.8 &      &  17.25$^3$	  &     &    & 15.44$^2$  & 14.99$^2$  & 14.77$^2$  \\
			\hline
\multicolumn{10}{p{.98\textwidth}}{$^1 - V$ magnitudes  calculated from the $gri$ magnitudes \citep[SDSS Photometric Catalog, Release 7,][]{2009yCat.2294....0A}   and transformation of \citet{jordi2006}. $^2 -$ Data from \citet[][2MASS]{2003yCat.2246....0C}. $^3 -$ Data from \citet[][GSC-II]{2008AJ....136..735L}. 
	
%	We do not provide  errors for individual magnitudes in the visible ($BVRcIc$) bands, we only note that they never  exceed several hundredths.
The  errors for individual magnitudes in the visible ($BVRcIc$) bands are within several hundredths.  
	}\\			
			\hline
		\end{tabular}
\end{table*}  

\section{Stellar parameters}\label{param}
%\section{Atmospheric parameters}\label{param}

%To determine stellar atmosphere parameters, we employed the same method as in \citet{2017AA...604A.129M}. 
\subsection{Effective temperature}
Effective temperatures (\teff) were calculated from the V$-$K and V$-$J colours using colour-\teff\ relation from \citet{1999A&AS..140..261A} and \citet[][hereafter RM05]{ramirez_melendez}.
Employing the \citet{1998ApJ...500..525S} maps, we adopted an extinction of E$_{\rm B-V}$ = 0.018, 0.030 and 0.080 for ComaBer, Seg~1, and Tri\ii, respectively. The effective temperatures derived from different colours and calibrations are presented in Table~\ref{tcol}.
The new photometric observations significantly improve the accuracy of  \teff\ determination.
For example, for ComaBer S1, the data from SDSS \citep{2009yCat.2294....0A} and 2MASS 
\citep{2MASS2006AJ1311163S} 
surveys result in  T(V$-$K) = 4954~K and T(V$-$J) =  5222~K, while the magnitudes obtained in our study provide consistent T(V$-$K) = 4915~K and T(V$-$J) = 4904~K. For each of the Coma Berenices stars, the differences in \teff\ between (V$-$K) and (V$-$J) colours are within 12~K.

For the other stars, $V$ magnitudes were calculated using the {\it gri} magnitudes from the SDSS Photometric Catalog, Release 7 \citep{2009yCat.2294....0A}  and transformation from \citet{jordi2006}. 	
%{\bf The accuracy of the adopted literature photometry is high.  For example, the errors in {\it gri} magnitudes  for Seg~1 S1, S2, and S3 are within 0.011~dex, which translates in errors in \teff\ smaller than 20~K for any star, colour, and calibration. }
For S1, S2, and S3 in Seg~1 UFD, perfect agreement was found between our \teff\	and those determined by \citet[][hereafter SGM11]{2011ApJ...733...46S}. For each star,  the difference in \teff\ between this study and  SGM11 is within 25~K. 
In addition to the Seg~1 S1, S2, and S3 stars, for which accurate photometry was obtained in this study, we also study several carbon rich stars, namely, Seg~1 S4, S5, S6 stars from F14 and SGM11, and the S7 star from \citet{2010ApJ...722L.104N}. 
Taking into account that this study and SGM11 provide similar photometric temperatures for Segue~1 S1, S2, and S3, we adopted effective temperatures
from SGM11 for the S4, S5, S6 stars.
%For the S4, S5, S6 stars, we adopted effective temperatures from SGM11. 
For S7, we adopted  \teff\ = 5000~K, which is between  T(V$-$K) =  5014~K and T(V$-$J) =  4983~K and  consistent with  \teff\ = 4960~K from \citet{2010ApJ...722L.104N}. 	
%{\it Stars in Ursa~Major~II and Leo~IV UFDs} 

For Tri\ii~S40, our \teff\ determinations are based on photometry from the literature, see Table~\ref{magnitudes} for references. Using the RM05 calibration, we calculated  T(V$-$K) = 4839~K and T(V$-$J) = 4930~K and finally adopted \teff\ = 4900~K.

%In addition to stars in UFDs investigated in this study, we perform photometry and calculate effective temperatures for three stars in Ursa~Major\ii\ and one star in Leo\iv, investigated by \citet{2017AA...604A.129M}.
%For each of UMa\ii\ and Leo\iv\ stars, we found consistent within 50~K effective temperatures from (V$-$K) and (V$-$J) colours with the RM05 calibration, and (V$-$K) colour with A99 and RM05 calibrations. 
%For each of UMa\ii\ and Leo\iv\ stars, different colours and calibrations provide consistent within 50~K effective temperatures. 

For each of UMa\ii\ and Leo\iv\ stars, (V$-$K) and (V$-$J) colours with the RM05 calibration and (V$-$K) colour with A99 and RM05 calibrations provide consistent within 50~K effective temperatures.
%The effective temperatures, adopted by \citet{2017AA...604A.129M} for stars in Ursa~Major\ii\ and Leo\iv\ UFDs are consistent with those derived in this study. 
The difference between the mean photometric temperature and those of M17a does not exceed 50~K for any star. The exception is UMa\ii~S1, where our average  \teff\ is 150~K higher compared to that of M17a. For this star, we revised the atmospheric parameters  by adopting \teff\ = 5000~K and surface gravity of log~g = 2.10 and recalculated abundances  as described in Section~\ref{abundances}. 
%Based on new photometric observations, we adopt \teff\ = 5000~K and log~g = 2.10. The recalculated LTE and NLTE abundances from individual lines are presented in Table~\ref{atomic}.
%}

%seg 1-7 distance from Bailer-Jones 2018
%d=[3362.396690950210086,2699.493825709740122,4223.807522946170138]
%print,4.44+alog10(0.8)+4.0*alog10(4960/5780.0)-2.0*alog10(d/10.0)+0.4*(17.73-0.421-4.75),format='(7f6.2)' 
%4.05  4.24  3.85

\begin{table}
	\caption{Effective temperatures from different colours and calibrations of \citet[][A99]{1999A&AS..140..261A} and \citet[][RM05]{ramirez_melendez}. }
		%Based on photometric magnitudes derived in this study.}
	\label{tcol}
	\setlength{\tabcolsep}{1.8mm}
	\begin{tabular}{|l|c|c|c|c|l|}
		\hline
		star & T(V$-$K) & T(V$-$K) & T(V$-$J) & final & ref.$^1$ \\  
	       	& A99    &  RM05  &  RM05  &       &  \\  
		\hline
		ComaBer S1 & 4910 & 4915 & 4904 & 4900 & TS \\
		ComaBer S2 & 4862 & 4887 & 4898 & 4875 & TS \\
		ComaBer S3 & 4755 & 4783 & 4796 & 4785 & TS \\
		Seg1 S1 & 5029 & 4990 & 5082 & 5030 & TS \\
		Seg1 S2 & 5123 & 5112 & 5168 & 5130 & TS \\
		Seg1 S3 & 5244 & 5228 & 5265 & 5250 & TS \\
		Seg1 S7$^2$ & 5014 & 4997 & 4983 & 5000 & TS \\
		Tri\ii\ S40$^2$ & 4807 & 4839 & 4930 & 4900 & TS \\
%		UMaII S1 & 5041 & 5019 & 4968 & 4850 & M17a \\
		UMa\ii\ S1 & 5041 & 5019 & 4968 & 5000 & TS \\
		UMa\ii\ S2 & 4800 & 4794 & 4814 & 4780 & M17a \\
		UMa\ii\ S3 & 4528 & 4569 & 4603 & 4560 & M17a \\
		Leo\iv\ S1 & 4442 & 4490 & 4520 & 4530 & M17a \\	
		\hline
%\multicolumn{6}{p{.45\textwidth}}{$^1$ - indicates a study, where the adopted \teff\ was employed.  }\\
\multicolumn{6}{p{.45\textwidth}}{$^1 -$ Indicates the reference for the adopted value of \teff,  }\\
%\multicolumn{6}{p{.45\textwidth}}{$^2$ - this study, $^3$ - \citet{2017AA...604A.129M}.  }\\
\multicolumn{6}{p{.45\textwidth}}{this study (TS) or  \citet[][M17a]{2017AA...604A.129M}.  }\\
\multicolumn{6}{p{.45\textwidth}}{$^2 -$ Based on photometry from the literature.  }\\
%\multicolumn{6}{p{.45\textwidth}}{For Coma Berenices, Segue~1, Ursa Major~II, and Leo~IV, we adopted an extinction, E(B-V) = 0.018, 0.030, 0.100, and 0.025, respectively \citep{1998ApJ...500..525S}. }\\
			\hline		
	\end{tabular}
\end{table}

\subsection{Surface gravity}
Surface gravities (log~g) were calculated using the well-known relation between log~g, apparent magnitude, bolometric correction, distance,
%\citet[][]{2007ApJ...654..897B}
 \teff, and the star's mass, adopted as 0.8~M$_{\odot}$. 
%(d$_{\rm ComaBer}$ = 44 kpc and d$_{\rm Seg1}$ = 23 kpc as determined by Belokurov et al., 2007)
For ComaBer and Seg~1, we adopted distances of d = 44 kpc and 23 kpc from \citet{2007ApJ...654..897B},  and d = 30 kpc from \citet{2015ApJ...802L..18L} for Tri\ii.
Bolometric corrections were taken  from \citet{2014MNRAS.444..392C}.

\subsection{Microturbulent velocity}
Microturbulent velocities (\vt) were determined from lines of iron and titanium, 
by reducing the slope of the relation between element NLTE abundance from different lines and their equivalent widths.

%assuming that the lines of different strength, from 15 to 120~m\AA, provide the same element abundance.
%{\bf Stellar atmosphere parameters for the sample stars are presented in Table~\ref{ionazation}.}

\subsection{Ionization equilibrium}
Using photometric \teff\ and distance-based log~g, we checked Fe\ione\--Fe\ii, Ti\ione\--Ti\ii, and Ca\ione\--Ca\ii\ ionization equilibria in NLTE and LTE. 
See Section~\ref{abundances} for the method of abundance determination.
The determined \teff,  log~g, \vt, [Fe/H], and NLTE and LTE abundance differences from lines of neutral and ionized species ($\Delta_{\rm I-II} = \eps_{\rm I} - \eps_{\rm II}$) are presented in Table~\ref{ionization}. 
In NLTE the abundance difference between the two ionization stages is within the abundance uncertainties. 
%For each star, the NLTE the abundance difference between the two ionization stages is within the abundance uncertainties. 
The abundance error is calculated as the dispersion in the single line measurements around the mean
 $\sigma = \sqrt{ \Sigma (x - x_i )^2 /(N - 1)}$, where N is a total number of lines.
For $\Delta_{\rm I-II}$, the error is calculated as $\sqrt{\sigma_I^2 +\sigma_{II}^2}$, where $\sigma_I^2$ and $\sigma_{II}^2$ are the dispersions in abundance from neutral and ionized species, respectively.
In Seg~1 S5 and S6 lines of Fe\ii\ and Ti\ione\ are not detected, which does not allow to check Fe\ione\--Fe\ii\ and Ti\ione\--Ti\ii\ ionization balance. The quality of spectra near the Ca\ii\ IR triplet lines is poor due to fringes and does not allow to carry out an accurate abundance determination. However, for the Ca\ii\ IR triplet lines in S5 and S6,   we found a reasonable agreement between observed spectra and theoretical spectra calculated with calcium NLTE abundances from Ca\ione\ lines. 
Thus, our spectroscopic NLTE analysis supports atmospheric parameters as determined from photometry and distances. 

%The method of abundance determination is described in Section~\ref{abundances}.

\begin{table*}
	\caption{Atmospheric parameters and NLTE and LTE abundance difference between neutral and ionized species.}
		\label{ionization}
				\setlength{\tabcolsep}{1.3mm}
		
		\begin{tabular}{|l|c|c|c|c|r|r|r|r|r|r|}
			\hline
Star  & \teff & log~g & [Fe/H] & \vt & \multicolumn{2}{c}{Fe\ione\--Fe\ii} & \multicolumn{2}{c}{Ti\ione\--Ti\ii}  & \multicolumn{2}{c}{Ca\ione\--Ca\ii} \\
 &  K & SGC &  & km/s  & LTE & NLTE & LTE  & NLTE & LTE & NLTE \\
			\hline
				\multicolumn{11}{l}{Coma Berenices:} \\
%Ti SY2019, Fe B18kau
S1 &  4900 & 2.00 & --2.12 & 2.1 & --0.20$\pm$0.23 & 0.03$\pm$0.23 & --0.14$\pm$0.23 & --0.01$\pm$0.26 & --0.10$\pm$0.16 & 0.00$\pm$0.20 \\
S2 &  4875 & 1.95 & --2.52 & 1.8 & --0.26$\pm$0.19 & --0.07$\pm$0.19 & --0.02$\pm$0.30 &  0.17$\pm$0.30 & --0.10$\pm$0.20 &  0.00$\pm$0.20 \\
S3 &  4785 & 1.70 & --2.39 & 1.7 & --0.14$\pm$0.18 & 0.05$\pm$0.18 & --0.30$\pm$0.18 & --0.10$\pm$0.19 & --0.03$\pm$0.10 &  0.01$\pm$0.10 \\
%Ti SY2019
%S1 & 4900 & 2.00 & -2.14 & 2.1 & -0.08$\pm$0.25 & 0.00$\pm$0.25 & -0.14$\pm$0.23 & -0.01$\pm$0.26 & -0.10$\pm$0.16 & 0.00$\pm$0.20 \\
%S2 & 4875 & 1.95 & -2.50 & 1.7 & -0.04$\pm$0.25 & 0.03$\pm$0.25 & -0.25$\pm$0.26 & -0.05$\pm$0.26 & -0.09$\pm$0.13 & -0.01$\pm$0.13 \\
%S3 & 4785 & 1.70 & -2.56 & 1.8 & -0.07$\pm$0.18 & 0.03$\pm$0.18 & -0.06$\pm$0.32 &  0.12$\pm$0.27 & -0.03$\pm$0.14 & -0.08$\pm$0.14 \\
%S1 & 4900 & 2.00 & -2.14 & 2.1 & -0.08$\pm$0.25 & 0.00$\pm$0.25 & -0.03$\pm$0.31 &  0.11$\pm$0.31 & -0.10$\pm$0.16 & 0.00$\pm$0.20 \\
%S2 & 4875 & 1.95 & -2.50 & 1.7 & -0.04$\pm$0.25 & 0.03$\pm$0.25 & -0.22$\pm$0.25 & -0.04$\pm$0.25 & -0.09$\pm$0.13 & -0.01$\pm$0.13 \\
%S3 & 4785 & 1.70 & -2.56 & 1.8 & -0.07$\pm$0.18 & 0.03$\pm$0.18 & -0.14$\pm$0.18 &  0.02$\pm$0.18 & -0.03$\pm$0.14 & -0.08$\pm$0.14 \\
	\multicolumn{11}{l}{Segue 1:} \\
% Fe B18 kau, Ti SY2019
%S1 & 5030 & 2.91 & --1.47 & 1.4 & --0.19$\pm$0.23 & --0.05$\pm$0.25 & --0.13$\pm$0.29 & --0.09$\pm$0.29 &  & \\
%S2 & 5130 & 2.64 & --2.25 & 1.6 & --0.06$\pm$0.24 & 0.20$\pm$0.24 & --0.18$\pm$0.26 & --0.01$\pm$0.26 &  &  \\
%S3 & 5250 & 2.75 & --2.26 & 1.6 & --0.13$\pm$0.19 & 0.16$\pm$0.20 & --0.12$\pm$0.16 &  0.08$\pm$0.19 &  &  \\
% H=0.5 gf Fe 2 RU +0.11 (Paper I)
%SA & 5030 & 2.91 & -1.47 & 1.4 & -0.08$\pm$0.23 & -0.06$\pm$0.25 & -0.06$\pm$0.32 & -0.02$\pm$0.31 &  & \\
%SB & 5130 & 2.64 & -2.25 & 1.6 &  0.05$\pm$0.24 & 0.07$\pm$0.25 & -0.19$\pm$0.26 & -0.03$\pm$0.26 &  &  \\
%SC & 5250 & 2.75 & -2.26 & 1.6 & -0.02$\pm$0.19 & 0.06$\pm$0.20 & -0.05$\pm$0.17 &  0.13$\pm$0.18 &  &  \\	
%  H=0.5 gf Fe 2 RU +0.11 (Paper I), Ti SY2019
S1&   5030 & 2.91 & --1.71 & 1.6 &  0.02$\pm$0.31 & 0.04$\pm$0.31 & --0.35$\pm$0.33 & --0.30$\pm$0.33 &  & \\
S2&   5130 & 2.64 & --2.38 & 1.8 &  0.06$\pm$0.27 & 0.13$\pm$0.27 & --0.15$\pm$0.18 &   0.02$\pm$0.21 &  &  \\
S3&   5250 & 2.75 & --2.29 & 1.6 & --0.09$\pm$0.21 & --0.01$\pm$0.21 & 0.00$\pm$0.25 &  0.20$\pm$0.29 &  &  \\
%			\hline
S4 &  5100 & 2.77 & --1.69 & 1.5 & 0.13$\pm$0.32 & 0.15$\pm$0.32 & --0.12$\pm$0.18 & --0.01$\pm$0.21 & & \\
%S5 &  5270 & 2.85 & --3.63 & 1.6 &               &               &                 &  & --0.29$\pm$0.04     & 0.01$\pm$0.02  \\
%S6 &  5640 & 3.23 & --3.13 & 1.7 &               &               &                 &  & --0.27$\pm$0.15 & 0.05$\pm$0.13    \\
S5 &  5270 & 2.85 & --3.56 & 1.6 &               &               &                 &  &  &  \\
S6 &  5640 & 3.23 & --3.19 & 1.7 &               &               &                 &  &  &  \\
S7 &  5000 & 2.40 & --3.48 & 1.3 & --0.18$\pm$0.22 & --0.03$\pm$0.22	&  &  &  &   \\
	\multicolumn{11}{l}{Triangulum\ii:} \\
S40 & 	4900 & 1.87 & --2.77 & 1.9 & --0.13$\pm$0.17 & 0.00$\pm$0.17 & --0.05$\pm$0.10 & 0.23$\pm$0.12 &  &  \\
	\hline
		\end{tabular}
\end{table*}
    
\subsection{Isochrones}       
\begin{figure}
	\includegraphics[trim=1.0cm 0.0cm 0.0cm 0.0cm,clip,width=90mm]{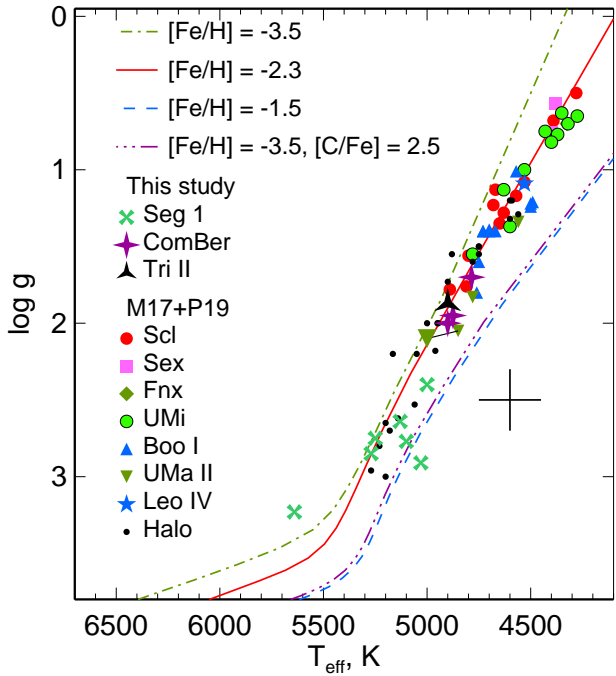}
%	iso_seg_com_tri.ps
	\caption{Investigated stars with 12 Gyr isochrones from \citet[][]{yi2004}. %and \citet[][D]{2007AJ....134..376D}. 
		The revised stellar parameters for the UMa\ii~S1 star are linked with the previous ones with a dash. Data from M17a and P19 are shown for comparison. The large cross indicates log~g and \teff\ error bars of 0.2 dex and 150~K, respectively.}
	\label{isochrones}
\end{figure}

As a sanity check, we compared the positions of the stars on \teff\--log~g diagram with  12~Gyr isochrones with [Fe/H] = $-3.5$,  $-2.3$, and $-1.5$   from the grid of \citet{yi2004}.
% and [Fe/H] = $-1.5$ and [$\alpha$/Fe] = 0 from \citet{2007AJ....134..376D}\footnote{We adopted isochrone interpolation tool available at http://stellar.dartmouth.edu/models/isolf\_new.html} to fit the most metal rich Seg~1 stars S1 and S4. 
To calculate metallicities of the isochrones,  we adopted solar metal-to-hydrogen ratio Z$_{\odot} = 0.0224$ \citep{1996ASPC...99..117G}, and [$\alpha$/Fe] = 0.6. 
% O = 8.87
The stars sit reasonably well on the isochrones in line with their age and stellar parameters (Fig.~\ref{isochrones}). 
%An exception is Seg~1 S7 with \teff/log~g/[Fe/H] = 4960~K/2.39/-3.6, which requires 0.4~dex lower log~g or 200~K higher \teff to sit on the corresponding isochrone. See Sect.~\ref{s7_status} for a discussion of the status of Seg~1 S7.
%Fig.~\ref{isochrones} shows the position of the sample stars  on the isochrones. 
For comparison, we plotted in Figure~\ref{isochrones} data from M17a and P19. 
	
%It is worth noted, that there is a systematic shift in position of MP giants compared to the corresponding isochrones, such that stars need to have up to 150~K hotter \teff\ or  0.5~dex higher log~g to fit perfectly the isochrones. This problem is most likely attributed to stellar evolution modeling, but not stellar parameter determination. For example, the MW halo benchmark star HD~122563	with well-fixed \teff/log~g/[Fe/H] of 4600~K $\pm$ 40~K/1.40 $\pm$ 0.03/$-2.6 \pm 0.1$
%\citep{2012A&A...545A..17C,2018MNRAS.475L..81K,2019A&A...625A..33C,2019AA...631A..43M} does not sit on the corresponding isochrone \citep{2020arXiv200513883C}.
 %\citep{2012A&A...545A..17C,2018MNRAS.475L..81K,2019A&A...625A..33C,2019AA...631A..43M} does not sit on the corresponding isochrone.

\subsection{Comparison with the literature}\label{litparam}

\begin{figure}
	\includegraphics[trim=0.5cm 1.40cm 0.0cm 0.0cm,clip,width=90mm]{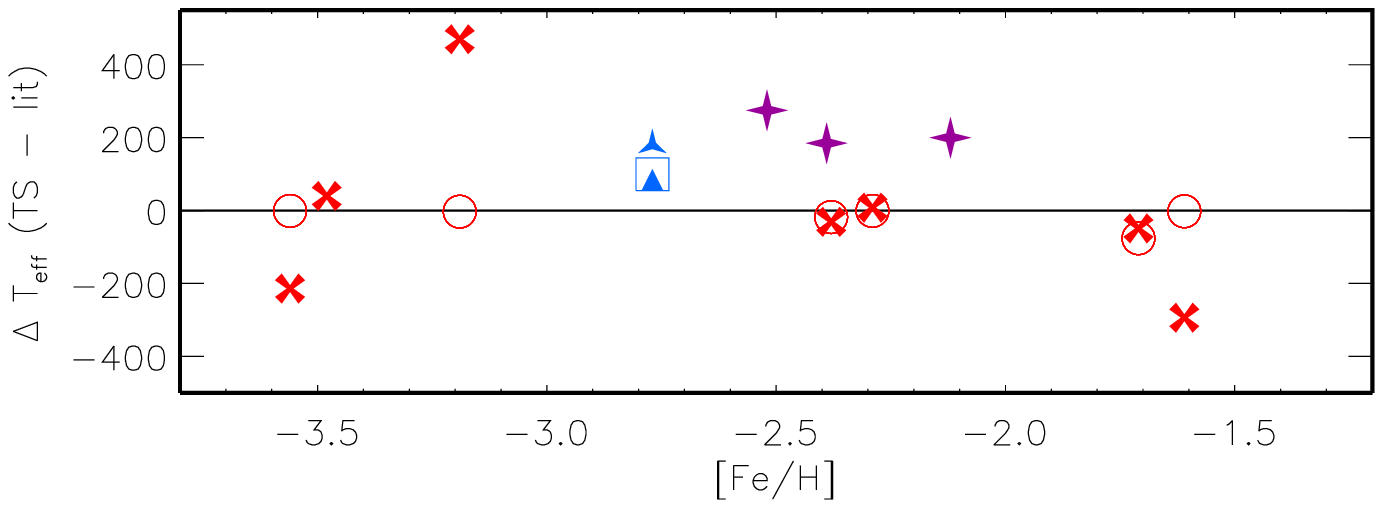}
	\includegraphics[trim=0.5cm 0cm 0.0cm 0.72cm,clip,width=90mm]{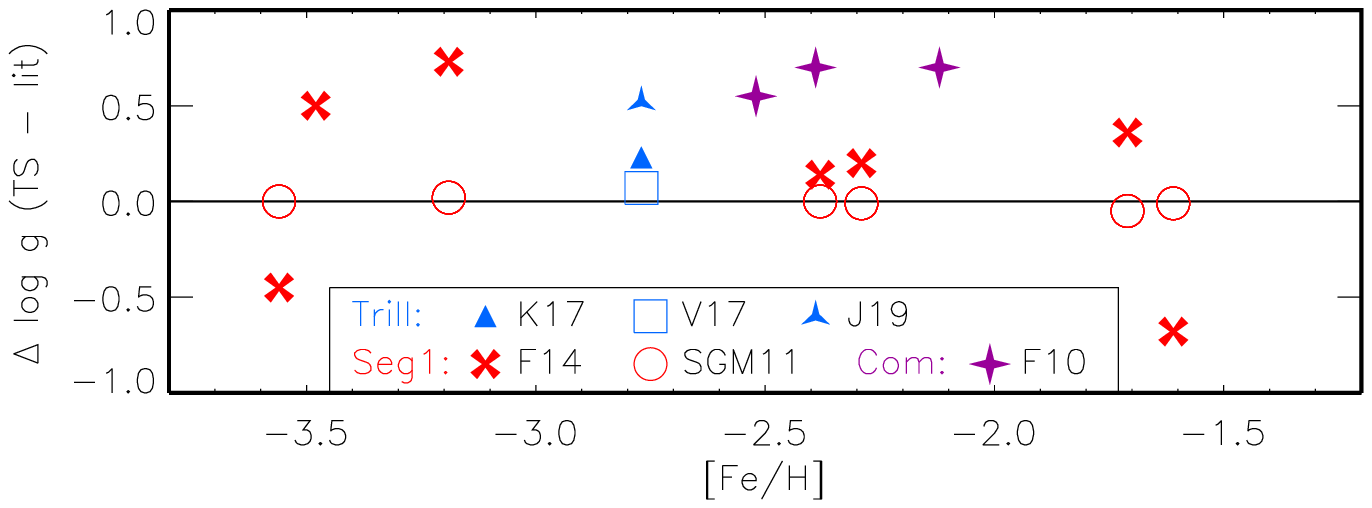}	
	\caption{Comparison of the derived \teff\ and log~g with the literature data. See Sect.~\ref{litparam} for details.}
	\label{paramcomp}	
\end{figure}

%(Table~\ref{ionazation})
\underline{Coma Berenices.}
Using spectroscopic method based on the LTE analyses of the Fe\ione\ excitation and Fe\ione\--Fe\ii\ ionization balance, F10 derived \teff\ /log~g  = 4700/1.3, 4600/1.4, and 4600/1.0  for S1, S2, and S3  stars, respectively. Our \teff\ and log~g values  are higher by up to 200~K and 0.7~dex, respectively (Fig.~\ref{paramcomp}), and this is understandable. Spectroscopic temperatures determined from the Fe\ione\ lines appear to be systematically lower compared with photometric \teff, by 70 to 270~K depending on \teff, as shown by \citet{2013ApJ...769...57F}, and LTE leads to systematically lower surface gravity determined from lines of Fe\ione\ and Fe\ii\  \citep[see, for example,][]{mash_fe}.

%{\bf 
\underline{Segue~1.}
For S1, S2, and S3 stars, the adopted effective temperatures are consistent with those of   \citet{2014ApJ...786...74F} and the difference in \teff\ between the two studies is within 50~K. \citet{2014ApJ...786...74F} corrected their spectroscopic \teff\ with an  empirical relation between photometric and spectroscopic temperatures, as determined by \citet{2013ApJ...769...57F}. 
For Seg~1 S1, S2, and  S3, the differences in surface gravities log~g~(TS) -- log~g~(F14) = 0.36, 0.14, and 0.20, respectively. Lower log~g in F14 can be explained by application of LTE analysis of Fe\ione\ and Fe\ii\ lines for log~g determination.
%Different cases are Segue 1 S4, S5, and S6 stars.
Seg~1 S4, S5, and S6 stars are a different case.
%For the Segue~1 S4, S5, and S6 stars, our temperatures are
Photometric temperatures of SGM11 adopted in this study are
 293~K and 214~K lower and 470~K higher, respectively,  compared to those of F14.  
For S4, S5, and S6, the difference log~g~(TS) -- log~g~(F14) = $-0.68$, $-0.45$, and 0.73, respectively, and it is  mainly caused by differences in \teff.   
%The surface gravities derived in this study differ from those of F14, and the differences vary from 0.14~dex for S2 up to 0.73~dex for S6.

%We found a perfect agreement between \teff\ and log~g adopted in this study and S11, since similar methods, based on photometry and distances, were adopted. 

For Seg~1 S7, its \teff\ is well fixed. We adopted \teff\ = 5000~K, which is consistent with photometric T(V$-$K) = 4997~K and T(V$-$J) = 4983~K values derived with the  RM05 calibration and \teff\ = 4960~K from \citet{2010ApJ...722L.104N}. We adopted the distance based log~g = 2.40, which is 0.5~dex higher compared to that of   \citet{2010ApJ...722L.104N}, derived from isochrones.

\underline{Triangulum\ii.}
For Tri\ii\ S40, our \teff/log~g = 4900/1.87 are in agreement with 4800/1.80 derived by \citet{2017MNRAS.466.3741V}, using similar methods.
Parameter determinations based on Fe\ione\ and Fe\ii\ line analysis in LTE results in lower \teff\ and log~g of 4720/1.35 and 4816/1.64, found by \citet{2019ApJ...870...83J} and \citet{2017ApJ...838...83K}, respectively.

% }

\section{Chemical abundances}\label{abundances}

\subsection{Spectral observations}\label{obs}

For ComaBer S1, S2, S3, Tri\ii\ S40, and Seg~1 S3,  observed spectra (R = 34~000) are taken from the archive of the Keck telescope\footnote{https://koa.ipac.caltech.edu/cgi-bin/KOA/nph-KOAlogin} (Program ID C168Hr, C222Hr, and C332Hr). \citet{2010ApJ...722L.104N} observed  Seg~1 S7 with UVES/VLT, and we took its spectrum from the ESO archive (Programme ID 383.B-0093(A)). For the remaining Seg~1 stars, we use spectra taken with MIKE/Magellan spectrograph/telescope and reduced in the original study of \citet{2014ApJ...786...74F}.

%In this study, we exclude from our abundance analyses saturated lines, with EW $>$120~m\AA. The exceptions are Na\ione\ 5889 and 5895 \AA, Al\ione\ 3961 \AA, Si\ione\ 4102 \AA, and the Ca\ii\ infrared triplet lines. The latter are used for checking the Ca\ione\--Ca\ii\ ionization balance.

\subsection{Method of calculation}

For our sample stars, element abundances are determined using the synthetic spectrum method by fitting line profiles to high-resolution observed spectra.
Theoretical spectra were calculated using
the \textsc {synthV\_NLTE} code \citep{Tsymbal2018} integrated to the \textsc {idl binmag} code by O. Kochukhov\footnote{http://www.astro.uu.se/$\sim$oleg/download.html}. This  technique allows to obtain the best fit to the observed spectrum with the NLTE effects taken into account via  pre-calculated departure coefficients for a given model atmosphere.

The coupled radiative transfer and statistical equilibrium equations were solved with a revised version of the  \textsc {detail} code  \citep{detail}. The update was presented by  \citet{mash_fe}. 

%{\bf
%When calculating abundances from equivalent widths, we first compute the LTE abundances with the \textsc {width} code \citep[][updated by Vadim Tsymbal by including the recent data for opacity and the equation of state calculation packages]{2005MSAIS...8...14K}. NLTE abundances are then derived by applying the NLTE abundance corrections. For each line and set of stellar atmospheric parameters, these corrections were obtained with the code \textsc {linec} \citep{Sakhibullin1983}  and the pre-calculated 
%with the \textsc {detail} code 
%departure coefficients. 
%}

We used classical plane-parallel model atmospheres  interpolated for a given \teff, \lgg, and [Fe/H] from the \textsc {marcs} model grid \citep{MARCS}. 

The full list of the investigated lines is presented in Table~\ref{atomic_data} along with gf-values, excitation energies of the lower levels,
and the derived NLTE and LTE abundances for the sample stars.
%damping constants (log~$\gamma_{rad}$, log~$\gamma_4$, log~$\gamma_6$)
The line list for synthetic spectrum calculation was extracted from the Vienna Atomic Line Database  \citep[VALD,][]{vald2015}.  
The isotopic splitting is taken into account for the Ca\ii\ resonance and IR triplet lines and Ba\ii\ lines. 
For the lines of Sc\ii, V\ione, Mn\ione, Co\ione, and Ba\ii\ we take into account their hyperfine splitting (HFS) structure, using the HFS data implemented to the VALD database by \citet{2019ARep...63.1010P}.
To determine barium abundances, we assume the r-process isotopic ratio $\rm {Ba^{\rm 135}:Ba^{\rm 137}:Ba^{\rm 138}}$ = 0.26:0.20:0.54 from \citet{1999ApJ...525..886A}.

\begin{table}
	\caption{NLTE and LTE abundances from individual lines of the sample stars.}
	\label{atomic_data}
	\setlength{\tabcolsep}{1.9mm}
	
	\begin{tabular}{|l|l|c|r|r|c|c|}
		\hline
		%\multicolumn{4}{c}{}  &  \multicolumn{2}{c}{LTE} & \multicolumn{4}{c}{NLTE} \\
		%Z & Species &  $\eps$ & N$_l$ &  $\eps$ & $\sigma$ & $\eps$ & $\sigma$ & [El/H] & [El/Fe] \\
		%\textbf{$\lambda$} & \multicolumn{1}{c}{\textbf{ E$_{exc}$}} & \multicolumn{1}{c}{\textbf{log gf}} & \multicolumn{1}{c}{\textbf{log  $\gamma_6 $}} & $\eps$LTE &  $\eps$NLTE & EW \\
		%\textbf{ \AA} & \multicolumn{1}{c}{\textbf{  eV}} &  & \multicolumn{1}{c}{\textbf{\tiny{rad/s cm$^{3}$}}} \\  & m\AA \\
		species & $\lambda$ &  E$_{exc}$ & log gf & EW & $\eps_{\rm NLTE}$ &  $\eps_{\rm LTE}$  \\
		& \AA & eV &    & m\AA &  & \\
		\hline
		\multicolumn{7}{l}{Seg1 S1: } \\
		Na I & 5889.95 &0.00 & 0.11 &190.8 & 3.88 & 4.19 \\
		Na I & 5895.92 &0.00 &--0.19 &171.7 & 3.90 & 4.30 \\
%		Mg I & 5528.40 &4.35 &--0.50 &136.8 & 6.22 & 6.34 \\
%		Al I & 3961.52 &0.01 &--0.32 &174.3 & 3.99 & 3.77 \\
%		Si I & 4102.94 &1.91 &--3.14 &113.9 & 6.40 & 6.40 \\
		\hline	
		%             \multicolumn{7}{l}{p{0.5\textwidth}}{This table is available in its entirety  in a machine-readable form in the online journal. A portion is shown here for guidance  regarding its form and content.} \\
		\multicolumn{7}{p{0.48\textwidth}}{This table is available in its entirety  in a machine-readable form in the online journal. A portion is shown here for guidance  regarding its form and content.}\\	         
		\hline
	\end{tabular}
\end{table}

\subsection{NLTE effects}\label{nlte}

%Abundances of O, Na, Mg, Al, Si, Ca, Ti, Fe, Sr, and Ba  based on NLTE line formation calculations with comprehensive model atoms developed and tested in our earlier studies have been determined for our sample stars. 
Abundances of O, Na, Mg, Al, Si, Ca, Ti, Fe, Sr, and Ba were determined using the NLTE line formation calculations with comprehensive model atoms developed and tested in our earlier studies.
For mechanisms of the departures from LTE and the details of the NLTE calculations for the above chemical elements,  see references in Table~\ref{model_atoms}. Here we briefly describe the impact of NLTE on abundance determinations. 

In atmospheres of cool metal-poor giants, NLTE may lead to significant changes in abundances and abundance ratios compared to LTE. 
For Mg\ione, Al\ione,  Ca\ione, Ti\ione, and Fe\ione, NLTE leads to higher abundances, up to 0.2~dex, depending on species, atmospheric parameters, and adopted spectral lines. For O\ione, Na\ione, and K\ione, NLTE leads to lower abundances, up to 0.5~dex in our sample stars.
For Sr\ii\ and Ba\ii\ in the sample stars, NLTE leads to nearly 0.1~dex higher average abundance.
For Si\ione, Fe\ii, and Ti\ii, NLTE effects are negligible.
%abundance corrections (differences between NLTE and LTE abundances from individual lines, %$\Delta_{\rm NLTE}$ = $\eps_{\rm NLTE} - \eps_{\rm LTE}$) are small.

Our [Fe/H] is based on Fe\ii\ lines, thus shifts in [El/Fe] abundance ratios due to NLTE are dominated by NLTE effects for a given element. The largest shift is found for [Na/Mg] abundance ratios, since, for Na\ione\ and Mg\ione, NLTE leads to changes of different signs.  
For more details on the impact of NLTE on abundances of the  above chemical elements in the atmospheres of VMP giants see Fig.~1 in \citet{2017AA...608A..89M}.

\begin{table}
	\caption{Model atoms for the NLTE calculations performed in this study.}
	\label{model_atoms}
	\begin{tabular}{|l|l|l|}
		\hline
		species & reference &  H\ione\ collisions   \\
		\hline
%		\multicolumn{3}{l}{ } \\
%        O  I & \citet{Przybilla2000,2018AstL...44..411S} & BVM19 \\
      O  I & \citet{2018AstL...44..411S} & BVM19 \\
		Na I & \citet{2014AstL...40..406A} & BBD10 \\ 
		Mg I & \citet{mash_mg13} & BBS12 \\ 
		Al I & \citet{Baumueller_al1} & B13 \\ 
%		Si I & \citet{Shi_si_sun} & BYB14 \\
		Si I & \citet{2020MNRAS.493.6095M} & BYB14 \\
		K  I & \citet{k1_neretina} & YVB19 \\
		Ca I-II & \citet{mash_ca} & BVY17-BVG18\\
		Ti I-II  & \citet{Sitnova_ti} & SYB20\\
%		Fe I-II  & \citet{mash_fe}  & YBK18-YBK19\\
		Fe I-II  & \citet{mash_fe}  & \kH\ = 0.5 \\
		Sr II & \citet{1997ARep...41..530B} & \kH\ = 0.01 \\
		Ba II & \citet{Mashonkina1999} & BY18 \\
		\hline	  
\multicolumn{3}{p{0.48\textwidth}}{
	BVM19 = \citet{2019MNRAS.487.5097B},
	BBD10 = \citet{2010AA...519A..20B},
	BBS12 = \citet{2012AA...541A..80B},
	B13 = \citet{2013AA...560A..60B},
	BYB14 = \citet{2014AA...572A.103B},
	YBV19 = \citet{2019OptSp.127..207Y}, BVY17 = \citet{Belyaev2017_ca1},
	BVG18 = \citet{ca2_hydrogen}, SYB20 = \citet{sitnova_tihyd}, BY18 = \citet{ba_col},
	\kH\ is a scaling coefficient to the \citet{drawin1968} formula.
%	\kH\ is a scaling factor to the \citet{} formula.
	} \\       
		\hline
	\end{tabular}
\end{table}

For C, Sc, V, Cr, Mn, Ni, Co, and Zn, we determined LTE abundances and applied the NLTE abundance corrections (differences between NLTE and LTE abundances from individual lines, $\Delta_{\rm NLTE}$ = $\eps_{\rm NLTE} - \eps_{\rm LTE}$) from  the literature, where available.

Carbon abundances were derived from molecular CH bands in the blue spectral region at 4300-4400 \AA.
%\ under LTE assumption. 
NLTE calculations for lines of CH molecules, as well as for  other molecules in stellar atmospheres are not available. 
%\citet{2015MNRAS.453.1619A} showed that the LTE abundances from CH lines agree with the NLTE abundance from C\ione\  lines in a sample of cool metal-poor stars, when using classical model atmospheres. 
\citet{2015MNRAS.453.1619A} showed that the LTE abundances from CH lines are consistent with the NLTE abundance from C\ione\  lines in a sample of cool metal-poor stars, when using classical model atmospheres.
We thus adopted the LTE abundances from CH lines.

We determined scandium abundances from lines of Sc\ii. In atmospheres of cool stars, Sc\ii\
is a majority species, and minor NLTE abundance corrections  could be expected.
NLTE calculations for Sc\ii\ in the atmospheres of cool dwarf stars were presented by \citet{2008A&A...481..489Z} and \citet{zhao2016}. At metallicity below [Fe/H] $\simeq -1$,  the NLTE abundance corrections for  Sc\ii\ lines are slightly positive and are less than 0.05~dex. 
The NLTE effects for Sc\ii\ were thus ignored.

Zinc abundances were determined from  only one line, Zn\ione\ 4810 \AA. To account for deviations from LTE, we applied the NLTE abundance corrections from \citet{Takeda2005zn}. 
For very metal-poor (VMP) giants, $\Delta_{\rm NLTE}$ for  Zn\ione\ 4810 \AA\ ranges from $-0.02$ to 0.12~dex depending on \teff, log~g, and EW. 
We applied $\Delta_{\rm NLTE}$ = 0.05~dex, which corresponds to  lines with EW $\simeq$ 20~m\AA, as observed in our sample stars. 

NLTE calculations for neutral species with ionization energy  below 8~eV, such as Ti\ione\ \citep{2011MNRAS.413.2184B}, Cr\ione\ \citep{2010A&A...522A...9B}, Mn\ione\ \citep{2008A&A...492..823B}, Fe\ione\ \citep{mash_fe},  and Co\ione\ \citep{2010MNRAS.401.1334B} report an  overionization, which leads to weakened lines and positive NLTE abundance corrections. 

%\citet{2010A&A...522A...9B} presented their Cr\ione-Cr\ii\ NLTE calculations for dwarfs, but not giants. We note that the differences between the NLTE and LTE average abundances from the Cr\ione\ lines, $\Delta_{\rm Cr I}$ = 0.33, 0.43, and 0.28~dex, obtained by \citet{2010A&A...522A...9B} for the very metal-poor (VMP) dwarfs HD~84937, HD~140283, and BD~$-4^\circ$3208, respectively, are similar to the LTE abundance difference Cr\ione\ - Cr\ii\ = -0.39~dex in a giant HE~2252-4225  \citep{HE2252}, which has atmospheric parameters (4710/1.65/-2.63) close to those of our Coma Ber stars. If the NLTE effects for Cr\ii\ are small, the LTE abundance difference between Cr\ione\ and Cr\ii\ gives an idea of the NLTE abundance corrections for lines of Cr\ione. \citet{2010A&A...522A...9B} predict $\Delta_{\rm NLTE}$ = 0.08~dex for lines of Cr\ii\ in the [Fe/H] = -2.5 dwarfs. Assuming similarly small value for the [Fe/H] = -2.5 giants, we expect $\Delta_{\rm NLTE} \simeq$ 0.3~dex for lines of Cr\ione\ in S1, S2, and S3. 
%For Cr\ione-Cr\ii, NLTE calculations were performed by \citet{2010A&A...522A...9B}.
%To determine the NLTE abundance from lines of Cr\ione, we adopted the NLTE abundance corrections provided by on-line service http://nlte.mpia.de/gui-siuAC\_secE.php. 
To determine the NLTE abundance from lines of Cr\ione, we adopted abundance corrections
calculated with the model atom of \citet{2010A&A...522A...9B} and available at the MPIA website\footnote{http://nlte.mpia.de/gui-siuAC\_secE.php}. 
For the sample stars, NLTE abundance corrections for Cr\ione\ lines vary from 0.20 to 0.45~dex depending on the line and stellar parameters. 

Our manganese abundances or upper limits rely on the subordinate Mn\ione\ 4783~\AA\ and 4823~\AA\ lines. For all sample stars, we applied the same NLTE abundance corrections of 0.3~dex as calculated by \citet{2019A&A...631A..80B} for model atmosphere with \teff/log~g/[Fe/H] = 4500/1.5/$-2$. 
%We note, that the NLTE abundance corrections for Mn\ione, available at MPIA website, are calculated with the model atom of \citet{2008A&A...492..823B} and provide overestimated NLTE abundances from the subordinate lines.

%The only line, Mn\ione\ 4823~\AA, can be measured in the observed spectra of S1, S2, nd S3 to derive the Mn abundances. \citet{2008A&A...492..823B} report the NLTE and LTE average abundances from seven lines of Mn\ione\ in HD~122563, which has atmospheric parameters (4600/1.6/-2.5) close to those of our Coma Ber stars. They obtained $\Delta_{\rm Mn I}$ = 0.44~dex. This is higher than the LTE abundance difference Mn\ii\ - Mn\ione\ = 0.21~dex in HE~2252-4225 \citep{HE2252}, with close atmospheric parameters. As discussed by \citet{HE2327}, the NLTE effects for the Mn\ione\ subordinate lines are probably overestimated by \citet{2008A&A...492..823B}. In this study, we adopt $\Delta_{\rm NLTE}$ = 0.2~dex for Mn\ione\ 4823~\AA\ in the Coma Ber stars. 

%Our vanadium, nickel, and cobalt abundances were determined from  V\ione, Ni\ione\ and Co\ione, since no lines of V\ii, Ni\ii\ and Co\ii\ are available in the observations of the sample stars.
For V\ione, Co\ione\ and Ni\ione, the literature data on NLTE effects are either missing or poor and cannot be applied for abundance correction.
NLTE calculations for V\ione\ have never been performed. 
The statistical equilibrium  calculations for Ni\ione\ in the solar atmosphere were presented by \citet{Bruls93} and \citet{2013ApJ...769..103V}, however, they did not study NLTE effects on atmospheres with different parameters and the impact of NLTE on abundance determination.
NLTE calculations for  Co\ione-\ii\ were performed by \citet{2010MNRAS.401.1334B}. In cool MP stars, they found large positive NLTE abundance corrections of up to 0.8~dex. 
For Co\ione\ 4121~\AA\ in our sample stars, $\Delta_{\rm NLTE}$ ranges from 0.41 to  0.75~dex according to the data from the MPIA website. 
Observations of MP stars argue for   overestimated  NLTE effects in \citet{2010MNRAS.401.1334B}.
For example, in a halo giant HD~122563
with well-determined stellar parameters, we found LTE abundance difference between lines of neutral and ionized species of 
$\eps$(Co\ione\ 4110,4121~\AA) -- $\eps$(Co\ii\ 3501~\AA) = $-0.08$~dex (Table~\ref{hd122563}). 
The NLTE calculations of \citet{2010MNRAS.401.1334B} predict $\Delta_{\rm NLTE} = 0.72$~dex and 0.69~dex for Co\ione\ 4110~\AA\ and 4121~\AA, respectively, while minor $\Delta_{\rm NLTE}$ for Co\ii\ 3501~\AA, and thus do not allow to fulfil an ionization balance. 
To make these estimates for HD~122563, we adopted \teff\ = 4600~K, log~g = 1.4, [Fe/H] = $-2.55$, and \vt = 1.6 \kms\ as in  \citet{2019AA...631A..43M}.

%{\bf 
For V\ione, Co\ione\ and Ni\ione, we applied an empirical strategy to account for deviations from LTE.
For a sample of VMP giants, \citet{2017AA...608A..89M} found  [Ni\ione/Fe\ione] $\sim$ 0, when abundances from Ni\ione\ and Fe\ione\ are taken in LTE.
This argues for similar values for the NLTE abundance corrections for both Ni\ione\ and Fe\ione. 
To account for NLTE effects for Ni\ione, V\ione, and Co\ione, we provide the final abundance ratios as [El/Fe\ione]$_{\rm LTE}$.
We note that for Ni\ione\ and Co\ione, this strategy provides realistic results, while it underestimates NLTE effects for V\ione. 
For example, in HD~122563, V\ione\ -- V\ii\ = $-0.52$ dex in LTE (Table~\ref{hd122563}), while Fe\ione\ -- Fe\ii\ = $-0.25$ dex as determined in \citet{2019AA...631A..43M}. 
%We assume, that for V\ione\ an average difference between NLTE and LTE abundance is two times larger than for Fe\ione. 
%Vanadium abundances of the Seg~1 stars were determined from V\ii\ lines, for which the departures from LTE are expected to be minor.

% 0.3 dex for Cr I is a good estimation
%       CrI(NLTE-LTE) Fe I(NLTE-LTE)
% 84937     0.33  0.07
% 140283    0.43  0.05
% BD-4 3208 0.28  0.09

% V I 6.75; Cr I 6.75; Mn I 7.43; Ni I 7.64

%In atmospheres of VMP giants The only line of Zn\ione\ 4810 \AA

%EW Zn
%S3 18.9 S1 19.4(upper limit of F10)
% 4950. 2.00 -3.30 4810.53     4.8 1.50  1.692 1.645  0.047  CS30325-094     cayrel04 
%  4900. 1.70 -3.33 4810.53     6.0 1.80  1.755 1.688  0.067  CS22956-050     cayrel04  
%  4700. 1.20 -2.88 4810.53    17.6 1.90  2.139 2.063  0.076  BS17569-049     cayrel04    
%  4700. 1.20 -3.04 4810.53    10.0 2.20  1.839 1.767  0.072  CS22169-035     cayrel04 

\subsection{Stellar abundances}\label{results}

\begin{table}
	\caption{LTE abundances of selected chemical species in HD~122563.}
	\label{hd122563}
	\setlength{\tabcolsep}{1.4mm}	
	%		\begin{tabular}{|l|l|c|r|r|r|r|r|r|}
	\begin{tabular}{|l|l|c|r|r|r|r|r|}
		\hline
		Species           & Sc\ii\  & V\ione\ & V\ii\   & Co\ione\ & Co\ii\ & Ni\ione\ & Zn\ione\ \\
		${\rm [El/Fe]}$ &  0.05   & --0.52   & 0.00      & --0.09   & 0.02   & --0.18    & --0.07 \\
		$\sigma$          & 0.03    &         & 0.07       &  0.06    &        &  0.10    & \\
		N$_{\rm lines}$   & 3       & 1       & 3          & 3        & 1      & 9        & 1 \\
		gf from           & L19     & L14     & W14        & L15      & R98    & WL14, & W68 \\
		&         &         &            &          &        & F88   &   \\
		%
		%Species           & Sc\ii\  & V\ione\ & V\ii\ & Cr\ione\  & Co\ione\ & Co\ii\ & Ni\ione\ & Zn\ione\ \\
		%${\rm [El/Fe]}$ &  0.05   & --0.52   & 0.00  & --0.56     & --0.09    & 0.02   & --0.18    & --0.07 \\
		%$\sigma$          & 0.03    &         & 0.07  &  0.03     &  0.06    &        &  0.10    & \\
		%N$_{\rm lines}$   & 3       & 1       & 3     & 4         & 3        & 1      & 9        & 1 \\
		%gf from           & L19     & L14     & W14   & S07       & L15      & R98    & WL14, & W68 \\
		%&         &         &       &           &          &        & F88   &   \\                  
		\hline
		\multicolumn{8}{p{0.45\textwidth}}{We adopted [Fe/H] = $-2.55$ from NLTE analyses of \cite{2019AA...631A..43M}.}\\
		\multicolumn{8}{p{0.45\textwidth}}{L19 - \citet{2019ApJS..241...21L}, L14 - \citet{2014ApJS..215...20L}, W14 - \citet{2014ApJS..214...18W}, 
			%S07 - \citet{2007ApJ...667.1267S}, 
			L15 - \citet{2015ApJS..220...13L}, R98 - \citet{1998AAS..130..541R}, WL14 - \citet{2014ApJS..211...20W}, F88 - \citet{1988JPCRD..17S....F}, W68 - \citet{1968MNRAS.140...53W}.}\\
		\hline
	\end{tabular}
\end{table}

The LTE and NLTE (where available) abundances of up to 18 chemical elements for the sample stars are presented in Table~\ref{atomic_data}. In the Appendix, we show standard [El/Fe] versus [Fe/H] diagrams, where our results for Coma~Berenices, Segue~1, and Triangulum\ii\ 
are plotted together with those from M17b and P19 for the MW halo, classical  dSphs Sculptor, Ursa~Minor, Fornax, and Sextans, and the UFDs Bo\"otes~I, UMa\ii, and Leo\iv. 

\begin{figure}
\includegraphics[trim=0.5cm 1.42cm 0.0cm 0.0cm,clip,width=90mm]{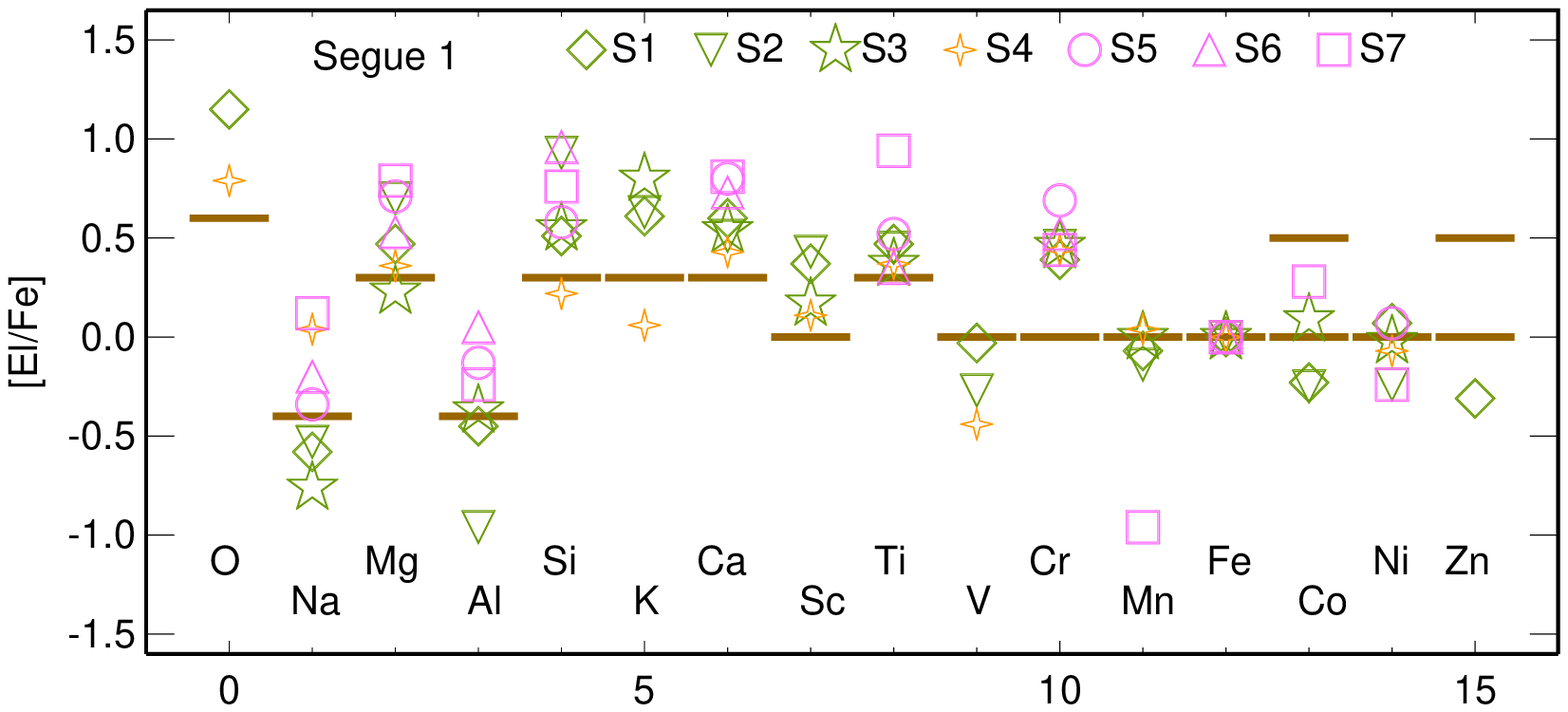}	
\includegraphics[trim=0.5cm 1.40cm 0.0cm 0.85cm,clip,width=90mm]{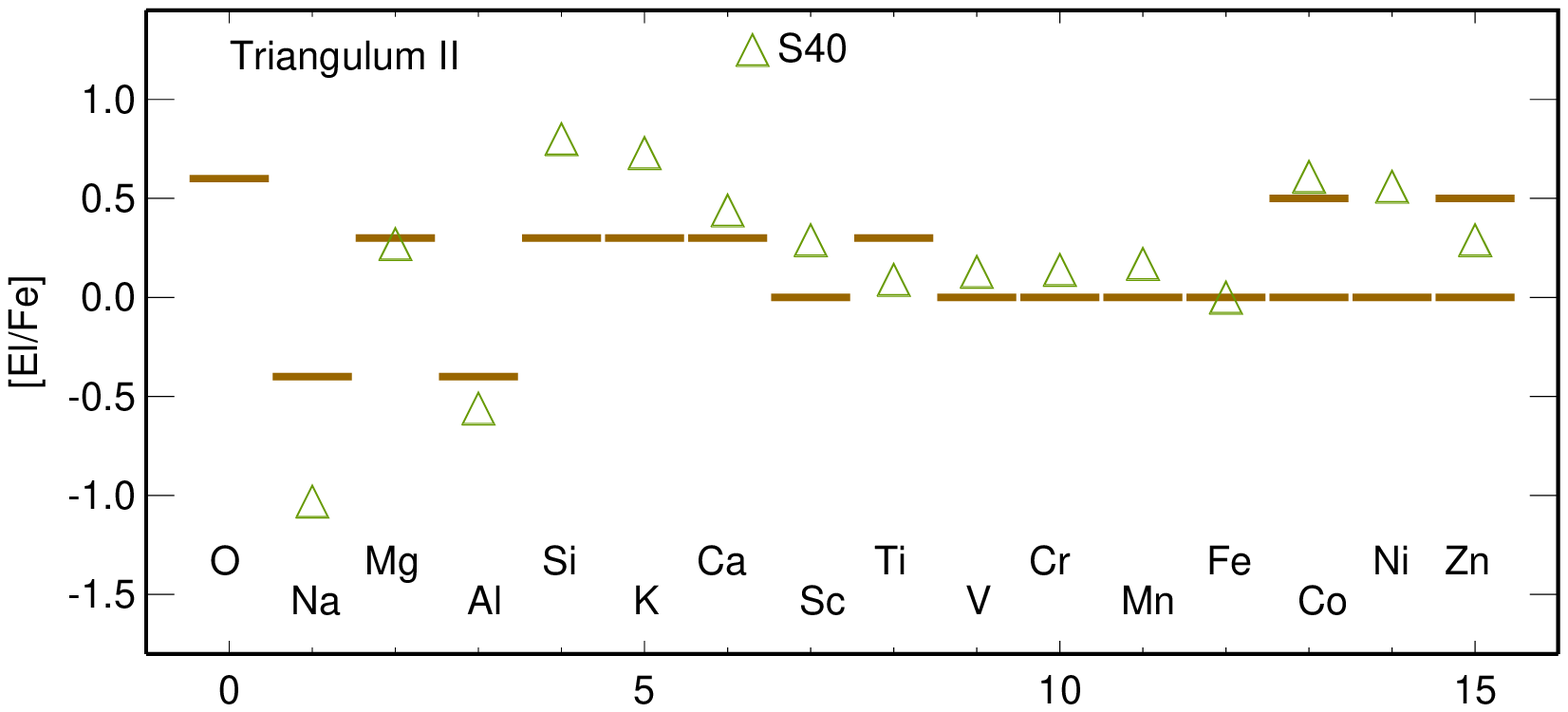}
\includegraphics[trim=0.5cm 1.300cm 0.0cm 0.85cm,clip,width=90mm]{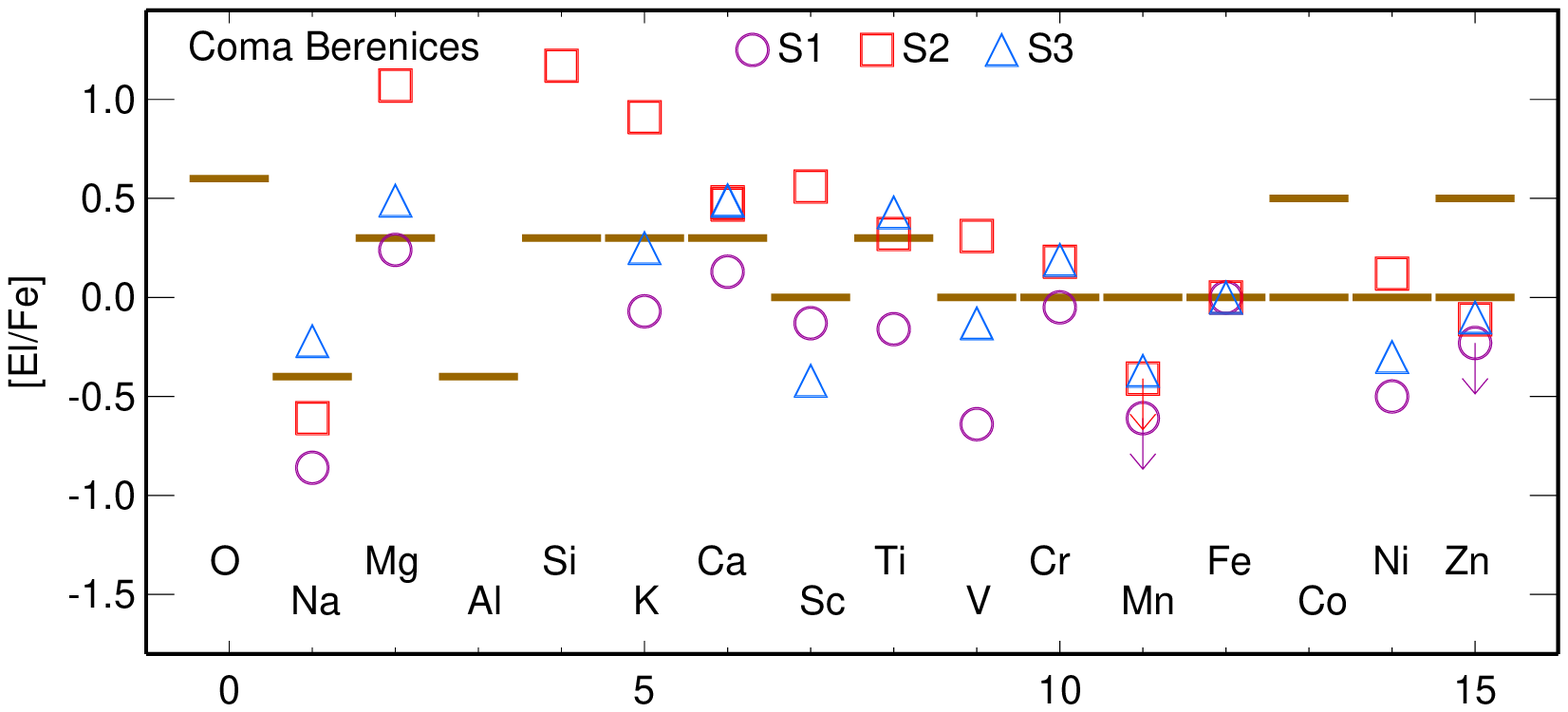}
\caption{The NLTE element  abundance patterns of stars in Seg~1 (top panel), Tri\ii\ (middle panel), and ComaBer  (bottom panel) UFDs. Mean abundance ratios for  metal-poor MW halo stars are shown with dashes. }
\label{elfe_patterns}	
\end{figure}

Element abundance patterns in the range from  O to Zn are displayed in Fig.~\ref{elfe_patterns} for individual stars in Seg~1, Tri\ii, and ComaBer UFDs. For comparison, we indicate the [El/Fe] elemental ratios, which are typical for the MW halo in the $-4 <$ [Fe/H] $< -2$ regime, as derived in the literature. They are based on the NLTE abundances from M17b (Na, Mg, Al, Si, Ca, Ti), \citet[][K, Sc]{zhao2016}, \citet[][Cr]{2010A&A...522A...9B}, and \citet[][Mn]{2008A&A...492..823B}. Based on our LTE analysis of V\ii\ lines in HD~122563 (Table~\ref{hd122563}), we adopted [V/Fe] = 0 for the MW halo. This is in line with a comprehensive LTE study of vanadium abundance in 255 halo stars  performed by \citet{2020ApJ...900..106O}. For [Ni/Fe], we took [Ni\ione /Fe\ione] = 0, as obtained by M17b. Abundances of Co and Zn grow toward lower metallicity in the MW halo \citep[see, for example][] {cayrel2004,Takeda2005zn,Yong2013}. Therefore, for each of [Co/Fe] and [Zn/Fe], we indicate in Fig.~\ref{elfe_patterns} two values of 0 and 0.5, which correspond to [Fe/H] $\sim -2$ and $\sim -4$, respectively.
%, as reported by \citet{cayrel2004}.

%Abundances of the neutron-capture elements will be discussed in Section~\ref{ncapture}.
Abundance ratios of the neutron-capture elements, represented by Sr and Ba,  are discussed in more detail in Section~\ref{ncapture}.

%We present the derived chemical abundances in Seg~1, ComaBer, and Tri\ii\ stars with references to the earlier  studies, 
%of F10, F14, N09, K17, J17, V17, 
%where  the sample stars were originally investigated.

\subsubsection{Segue 1 UFD}

\underline{Carbon.}
Seg~1  S1, S2, and S3 stars have normal nearly solar  [C/Fe], ranging from $-0.11$ to 0.08. 
Seg~1 S4 has [C/Fe] = 0.74 and gained this carbon enhancement from its companion. 
S5 has carbon enhancement [C/Fe] of 1~dex. 
%The strength of CH lines decreases steeply with increasing \teff\ and, in S6 with \teff\ = 5640~K, CH lines are weak and we estimated [C/Fe]~$< 1$dex for this star. 
The strength of CH lines decreases steeply with increasing \teff\ and they become weak in S6 with \teff\ = 5640~K. We estimated [C/Fe]~$< 1$~dex for this star.
A very high [C/Fe] = 2.30 was found in S7. 
Our results are in agreement with earlier determinations of F14 and N10.
%We note that in S5, S6, and S7 with luminosity of L/L$_{\odot}$ = 1.33, 1.07, and 1.69, respectively, the carbon enhancement  is caused by high carbon abundance in the gas at the epoch of formation of these stars, but not carbon dredge-up from the stellar interiors due to stellar evolution.
We note that in S5, S6, and S7 with luminosity of L/L$_{\odot}$ = 1.33, 1.07, and 1.69, respectively, the carbon enhancement  is caused by high carbon abundance in the gas at the epoch of their formation, but not carbon dredge-up from the stellar interiors due to stellar evolution.

%For other stars in Seg~1, we cannot determine carbon abundance due to the absence of observed spectra, thus we rely on the literature data on carbon abundances from F14 and \citet{2010ApJ...722L.104N}, who found normal carbon abundance for S1 and S2, while S4, S5, S6, and S7 are carbon enhanced. 

\underline{$\alpha$-elements}.
In the most metal-rich sample stars S1 and S4, we detected the O\ione\ 7771 \AA\ triplet lines and found high [O/Fe] = 1.15 and 0.79, respectively. This is the first measurement of oxygen abundance in Segue~1.

For each of Mg, Si, Ca, and Ti, the stars in
Segue~1 reveal enhancements relative to Fe in line with the earlier
conclusions of F14.
%However, in contrast to similar [$\alpha$/Fe] of 0.5~dex found by F14 for different elements in all sample stars, our [$\alpha$/Fe] ranges from 0.2 to 0.9~dex depending on a star and chemical element.
However, in contrast to similar [$\alpha$/Fe] of 0.5~dex found by F14 for different elements and stars, our [$\alpha$/Fe] ranges from 0.2 to 0.9~dex depending on a star and chemical element.
%In S1, S2, S3, S4, and S6, a similar [Mg/Fe] of 0.3 was found, while Si, Ca, and Ti show higher [$\alpha$/Fe] of 0.6.
%S5 and S7 have similar [Mg/Fe] and [Ca/Fe] of 0.8, while [Ti/Fe] = 0.5 and 0.9, respectively.

\underline{Sodium and Aluminium}.
%are underabundant with respect to iron in all sample stars. An exception is [Na/Fe] = 0.12 found in Seg~1 S7.
%We found increasing [Na/Fe] toward lower [Fe/H], with [Na/Fe] from $-0.76$ to 0.12~dex in S3 and S7 with [Fe/H] = $-2.29$ and $-3.48$, respectively.
We found low [Na/Fe] from $-0.76$ to $-0.53$ in S1, S2, and S3 stars, which have  [Fe/H] $> -2.4$, while more metal-poor S5, S6, and S7 have higher [Na/Fe] from $-0.34$ to 0.12~dex. S4 does not follow this trend, since it gained high sodium from its binary companion.
This result is qualitatively in line with F14.
%However, due to using NLTE, our [Na/Fe] are significantly lower, compared to LTE ratios of F14.
However, due to application of NLTE,  [Na/Fe] in this study are significantly lower, compared to LTE ratios of F14.

A similar behaviour is found for aluminium, with [Al/Fe] from $-0.24$ to 0.05 in stars with [Fe/H]$ < -3.1$, and lower [Al/Fe]  from $-0.96$ to $-0.37$ in stars with [Fe/H] $> -2.4$.
%Aluminium, being another light odd-Z element follows sodium within 0.1~dex in Seg~1 S1, S2, S5 and S6, while in S3 it is higher  with [Al/Na] = 0.52 and lower in S7 with [Al/Fe] = $-0.36$.

\underline{Potassium.}
Inspecting the spectra of Seg~1 stars, we detected the resonance line K\ione\ 7698~\AA\ in S1, S2, S3, and S4. 
In S1, S2, and S3, we found high [K/Fe] of 0.61, 0.63, and 0.80, respectively, while S4 has lower [K/Fe] = 0.06.  
This is the first measurement of potassium in Segue 1.

\underline{Scandium and Vanadium}.
We found nearly solar [V/Fe] in S1, while lower [V/Fe] = $-0.26$ and $-0.44$ in S2 and S4, respectively. 
Seg~1 stars are enhanced in scandium with [Sc/Fe] from 0.11 in S4 to 0.43 in S2. 
Scandium abundance does not follow vanadium, and [Sc/V] ratio is about 0.5~dex, on average.
%Seg~1 S1, S2, and S3 stars have very similar and high [V/Fe] from 0.66 to 0.86~dex. F14 do not present a discussion on vanadium, while online table contains EW  and high abundances of vanadium, resulting in [V/Fe] from 0.6 to 1.0~dex in different stars.  

%The majority stars in the MW and dwarf galaxies have nearly solar [V/Fe]. However, there are some outliers with high [V/Fe] $> 0.5$~dex in the MW \citep[]{2013ApJ...778...56C,2014ApJ...781...40P,2020ApJ...900..106O}  and dwarf galaxies Horologium\ione\ \citep{2018ApJ...852...99N}, Tucana\ii\ \citep{2018ApJ...857...74C}, and Carina \citep{2012ApJ...751..102V,2017ApJS..230...28N}. It is worth mentioning that, in the above studies, for stars with high [V/Fe], vanadium abundance was determined from the one line. We adopted from 1 to 3 lines of V\ii\ for abundance determination in Seg~1 stars.

%High [V/Fe] in Seg~1 stars argues for their common origin from gas enriched by an explosion with unusual nucleosynthesis.
%\citet{2020arXiv200705011M} found membership probabilities P = 0.71, 0.0042, and 0.30 for Seg~1 S1, S2, and S3, respectively. Our abundance analysis, namely common abundance peculiarities as high [V/Fe], does not support the idea that S1, S3, and S3 belong to different galactic subsystems.

%Scandium abundance in Seg~1 S3 is normal, with [Sc/Fe] = 0.14. For the remaining stars, we cannot determine scandium abundance, since observed spectra are not available. Our [Sc/Fe] in S3 is consistent with results of F14, who found slightly positive [Sc/Fe] in Seg~1 stars. 

\underline{Iron-peak elements.}
In all Seg~1 stars, we found \underline{chromium} abundance to be enhanced with respect to iron, with [Cr/Fe]  from 0.14 to 0.69~dex.
Our [Cr/Fe] ratios are higher compared to those of F14, derived in LTE.

In the majority of Seg~1 stars, \underline{manganese} follows iron with [Mn/Fe] from $-0.14$ to 0.04~dex. An exception is S7, with low [Mn/Fe] = $-0.96$~dex.

\underline{Cobalt} abundance  follows iron within the error bars, however, [Co/Fe] slightly decreases with [Fe/H] from 0.28~dex in S5 to $-0.24$ in S1 and S2.
This result is in line with F14 and [Co/Fe] of metal-poor MW halo stars \citep{cayrel2004,Yong2013}.

In the majority of Seg~1 stars, \underline{nickel} abundance  follows iron within 0.1~dex, while lower [Ni/Fe] = $-0.23$ and $-0.24$ are found in S2 and S7, respectively.

An abundance of Zn is determined only in S1 and it yields [Zn/Fe] = $-0.31$.

\begin{figure}
	\includegraphics[trim=0.5cm 0.0cm 0.0cm 0.0cm,clip,width=90mm]{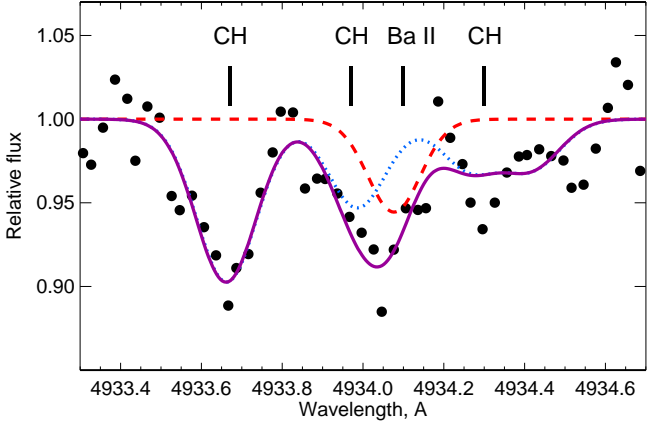}	
	\caption{Ba\ii\ 4934 line blended with CH molecular lines in the observed spectrum of Seg~1 S7 (circles). The best fit synthetic spectrum (solid line) is plotted together with spectrum calculated without carbon (dashed line) and without barium (dotted line). }
	\label{ba_s7}	
\end{figure}

\underline{Neutron capture elements.}
In  five Segue~1 stars, lines of Sr and Ba  stars are too weak to be detected even in the S1 star with [Fe/H] = $-1.6$, which argues for missing sources of neutron-capture element production at the early stage of Segue~1 formation. 
The lines of strontium and barium are detected in Seg~1 S7, and this is the first barium abundance determination  in this star. Due to high carbon overabundance, the only Ba\ii\ 4934 line available in the spectrum is blended with the CH lines, as shown in Fig.~\ref{ba_s7}. 
The nearby  lines are fitted well, and we believe that their  contribution to the  Ba\ii\ 4934 line is evaluated correctly.
Another star, where strontium and barium lines can be measured is S4, a carbon rich star, enhanced with neutron-capture elements as found by F14. 
The abundance pattern of this star follows that of the s-process.
Its initial chemical composition  was contaminated by its binary companion. 
We excluded neutron-capture element abundances of S4 from our analysis.

\subsubsection{Coma Berenices UFD}

\underline{Carbon.} 
Coma Berenices  S1 and  S2  with luminosities  log(L/L$_{\odot}$) $<$ 2.3 show [C/Fe] = $-0.21$ and $-0.11$, while more luminous  S3 with log(L/L$_{\odot}$) = 2.4 is depleted in carbon with [C/Fe] = $-0.78$. 
This corresponds to the carbon-normal stellar populations of dwarf spheroidal  galaxies 
\citep[see, for example, Fig. 7 in][]{2015A&A...583A..67J}. 
These ratios are slightly lower compared to those, found in F10. However, this difference does not impact on classification of ComaBer stars as carbon normal.

\underline{$\alpha$-elements}. %  in Coma Berenices are represented by Mg, Ca, and Ti.
ComaBer stars are enhanced in $\alpha$-elements relative to iron. 
However, different elements in a given star and a given element in different stars show different [El/Fe] abundance ratios. 
S3 shows [$\alpha$/Fe] of 0.5 for Mg and Ca and 0.4 for Ti.
S2 is extremely enhanced in magnesium and silicon with [Mg/Fe] = 1.07 and [Si/Fe] = 1.17, while calcium and titanium enhancements are lower with [Ca/Fe] = 0.48 and [Ti/Fe] = 0.32. 
S1 has the highest [Fe/H] = $-2.1$ and shows lower  compared to S2 and S3 [$\alpha$/Fe] = 0.24 and 0.13 for Mg and Ca and [Ti/Fe] = $-0.28$. Our results are in line with F10.  The possible causes of low [$\alpha$/Fe] in ComaBer S1 are discussed in Section~\ref{mixing}.

\underline{Sodium} is underabundant with respect to iron in all ComaBer stars.
Different stars show different [Na/Fe] = $-0.86$, $-0.61$, and $-0.22$ in S1, S2, and S3, respectively. 
We confirm a very low sodium abundance in ComaBer S1  compared to the MW halo stars and stars in other dwarf galaxies, as found by F10.
However, our NLTE [Na/Fe] are much lower than those of F10 determined in LTE.

\underline{Potassium.}
Inspecting the spectra of  ComaBer S1, S2, and S3, we detected the resonance K\ione\ 7698~\AA. 
We found high  overabundance [K/Fe] = 0.91  in ComaBer S2.  ComaBer S3 shows moderate [K/Fe] = 0.25, while lower [K/Fe] = $-0.07$ was found in ComaBer S1.
This is the first measurement of potassium in ComaBer.

\underline{Scandium and Vanadium}.
ComaBer  stars show a scatter of 1~dex in [V/Fe] and [Sc/Fe]. S2 is enhanced in Sc and V with [Sc/Fe] = 0.56 and [V/Fe] = 0.34, while S1 and S3 are depleted and show negative [Sc/Fe] and [V/Fe].
For scandium and vanadium, our results are in line with F10. 

\underline{Iron-peak elements.}
\underline{Chromium} abundance follows iron in S1, with [Cr/Fe] = $-0.05$, while S2 and S3 have higher [Cr/Fe]  of 0.2~dex.
For S1, our [Cr/Fe] is consistent within the error bars with those of F10, while for S2 and S3, F10 found lower [Cr/Fe] = $-0.54$ and $-0.34$~dex, respectively. This discrepancy can be partially explained with the NLTE effects.

\underline{Manganese} abundances are low in ComaBer stars, with [Mn/Fe] = $-0.37$ in  S3. We estimated upper limits [Mn/Fe] $< -0.61$ and $< -0.41$ for S1 and S2, respectively. Our estimations are consistent with those of F10.

ComaBer S1, S2, and S3  stars  have different [Ni/Fe] of $-0.50$, 0.12, and $-0.29$, respectively. At least three lines of Ni\ione\ were adopted for abundance determination. Similar [Ni/Fe] = $-0.56$, 0.08, $-0.19$ were found by F10 in S1 S2, and S3, respectively.

In ComaBer S2, S3, and S1 \underline{zinc} is underabundant with respect to iron, and [Zn/Fe] = $-0.11$, $-0.10$, and $< -0.23$~dex, respectively. Our values are close to those of F10.

\underline{Neutron capture elements.}
We found [Sr/Fe] = $-1.97$, $-1.82$ and $-1.63$ in ComaBer S1, S2, and S3, respectively. 
%In ComaBer S2, we  cannot determine  strontium abundance, since  the archival spectrum does not cover the strontium lines.
%For S1 and S3, we found 0.56~dex  and 0.24~dex higher [Sr/Fe] compared to F10.
These values are  higher compared to those of F10.

Barium abundances in ComaBer span a wide range, and  [Ba/Fe] = $-2.32$, $-1.64$, and $-1.02$ in S1, S2, and S3, respectively.
Our [Ba/Fe] in S1 is similar to that of F10, while for S2 and S3, we found 0.3~dex lower [Ba/Fe].

For ComaBer S3, we derived [Sr/Ba] = $-0.61$.
This value is close to the empirical estimates of pure r-process ratio: [Sr/Ba]$_r$ = $-0.44$ \citep{2005A&A...439..129B}  and $-0.38$ \citep{2017AA...608A..89M}. 
The higher [Sr/Ba] = 0.35 was found in ComaBer S1. 
See Section~\ref{ncapture} for a discussion of strontium and barium abundances in stars in different galaxies.

\subsubsection{Triangulum\ii\ UFD}

LTE abundance analysis of Tri\ii\ S40 star was presented by  \citet[][]{2017ApJ...838...83K}, \citet[][hereafter V17]{2017MNRAS.466.3741V}, and \citet[][hereafter J19]{2019ApJ...870...83J} using different high-resolution spectra.
We adopted spectral observations of K17, since they managed to measure the largest number of chemical species compared with the other two studies. When presenting our results, we focus mainly on the comparison with K17. We note that for the  chemical elements in  common (Na, Mg, K, Ca, Ti, Cr, Fe, and Ni) the abundance difference between different studies does not exceed 0.4~dex. Consistent within 0.1~dex abundances were found in K17 and J19, an exception is chromium, where the difference between the two studies amounts to 0.25~dex.   

\underline{$\alpha$-elements}. % in Tri\ii\ S40 are represented by Mg, Ca, and Ti.
Tri\ii\ S40 has moderate $\alpha$-enhancement with [$\alpha$/Fe] = 0.19, 0.47, and 0.09 for Mg, Ca, and Ti, respectively. Very similar [$\alpha$/Fe] ratios of 0.24, 0.40, and 0.14 were found by K17 for the same elements.  K17 determined an upper limit for [Si/Fe]$ < 0.76$ from weak Si\ione\ 6155 \AA\ line. We adopted Si\ione\ 4102 \AA\ line with EW = 84.3 m\AA\ for abundance determination and found a high [Si/Fe] = 0.80.

\underline{Sodium and Aluminium.}
Tri\ii\ S40 has very low [Na/Fe] = $-1.03$, while normal [Al/Fe] = $-0.56$. Low sodium was reported by K17, with [Na/Fe] = $-0.81$ in LTE. K17 found higher [Al/Fe] = 0. 

\underline{Potassium.}
We found a high [K/Fe] =  0.73. This value is lower compared to [K/Fe] = 0.88 found by K17 in LTE. 

\underline{Scandium and Vanadium}.
Tri\ii\ S40 shows nearly normal [V/Fe] = $0.14$, while enhanced [Sc/Fe] of 0.29. These ratios are consistent within the error bars with   those of K17.

\underline{Iron-peak elements.}
We found chromium abundance to be moderately enhanced with respect to iron, with [Cr/Fe] = 0.14, in contrast to low [Cr/Fe] = $-0.45$ found by K17 in LTE.

Tri\ii\ S40 has a high [Co/Fe] = 0.62. This value is higher than [Co/Fe] = 0.13 found by K17 in LTE.

%We found high [Ni/Fe] = 0.56. K17 also reported on nickel overabundance  with  [Ni/Fe]  = 0.53.
We found high [Ni/Fe] = 0.56, which is close to [Ni/Fe]  = 0.53 found in K17. 

A moderate enhancement was found for zinc, with [Zn/Fe] = 0.29. Slightly higher ratio [Zn/Fe] = 0.40 was obtained in K17.

\underline{Neutron-capture elements}
We found [Sr/Fe] =  $-1.94$, which differs from [Sr/Fe] = $-1.51$ determined by K17.
 The discrepancy in abundances is caused by observed spectra employed (see the Appendix). 
For S40, we adopted archival spectra, reduced in a slightly different manner, compared to K17. Each of Sr\ii\ lines is displaced at the edge of the echelle order, and the uncertainty in accounting for the scattered light can be the source of discrepancies in EWs (see the Appendix). 
For Sr\ii\ 4077\AA\ and 4215\AA\ lines, our best-fit synthetic spectra correspond to  EW = 59.6 and 45.1 m\AA, respectively, while K17 provides EW = 79.0 and 89.3 m\AA.
%	The discrepancy in abundances is caused by a difference in the EW. For Sr\ii\ 4077\AA\ and 4215\AA\ lines, we measured EW = 59.6 and 45.1 m\AA, respectively, while K17 provides EW= 79.0 and 89.3 m\AA.
Using the EWs from K17, we obtained a discrepancy of 0.4~dex between abundances from the two lines, while our measurements lead to similar abundances from both lines.
We obtained [Ba/Fe] = $-2.07$, while lower value [Ba/Fe] = $-2.37$ was found in K17. 
%For barium, the abundance discrepancy between the two studies can be partially explained with the NLTE effects.
See Section~\ref{ncapture} for a discussion of strontium and barium abundances in stars in different galaxies.

\begin{table}
	%\begin{longtable}{lccrrrr}
	\caption{The derived element abundances in the sample stars.}
	%	\centering
	\label{elements1col}
		\setlength{\tabcolsep}{0.7mm}
%	\begin{tabular}{|l|c|c|r|r|r|r|}
	\begin{tabular}{lcrrrrr}
		\hline
		Sp. &  $\eps_{\rm \odot}$ & N$_{\rm l}$ & $\eps_{\rm LTE}$ & $\eps_{\rm NLTE}$ & [El/H]$_{\rm }$ & [El/Fe]$_{\rm }$ \\
		\hline
		\multicolumn{7}{l}{Segue~1 S1 } \\
 CH    &   8.39 &    1 &   6.79 (0.10) &               &  --1.60 &   0.11 \\ 
 O  I  &   8.73 &    1 &   8.33 (0.10) &   8.17 (0.10) &  --0.56 &   1.15 \\ 
 Na I  &   6.33 &    2 &   4.39 (0.04) &   4.04 (0.04) &  --2.29 &  --0.58 \\ 
 Mg I  &   7.58 &    2 &   6.34 (0.01) &   6.33 (0.15) &  --1.25 &   0.47 \\ 
 Al I  &   6.47 &    1 &   4.12 (0.10) &   4.31 (0.10) &  --2.16 &  --0.45 \\ 
% Si I  &   7.52 &    2 &   6.29 (0.05) &   6.25 (0.07) &  --1.27 &   0.45 \\ 
 Si I  &   7.52 &    3 &   6.36 (0.12) &   6.32 (0.11) &  --1.20 &   0.51 \\ 
 K  I  &   5.12 &    1 &   4.62 (0.10) &   4.02 (0.10) &  --1.10 &   0.61 \\ 
 Ca I  &   6.36 &   16 &   5.26 (0.14) &   5.24 (0.15) &  --1.12 &   0.60 \\ 
 Sc II &   3.10 &    4 &   1.76 (0.24) &   1.76 (0.24) &  --1.34 &   0.37 \\ 
 Ti I  &   4.90 &    7 &   3.33 (0.23) &   3.36 (0.24) &  --1.54 &   0.17 \\ 
 Ti II &   4.90 &   23 &   3.68 (0.24) &   3.66 (0.25) &  --1.24 &   0.47 \\ 
 V  I  &   4.00 &    2 &   2.28 (0.05) &               &  --1.72 &  --0.03 \\ 
 Cr I  &   5.65 &    7 &   3.70 (0.15) &   4.33 (0.16) &  --1.32 &   0.39 \\ 
 Mn I  &   5.37 &    3 &   3.29 (0.04) &   3.59 (0.04) &  --1.78 &  --0.07 \\ 
 Fe I  &   7.50 &   58 &   5.81 (0.25) &   5.83 (0.25) &  --1.67 &   0.04 \\ 
 Fe II &   7.50 &   17 &   5.79 (0.18) &   5.79 (0.18) &  --1.71 &   0.00 \\ 
 Co I  &   4.92 &    2 &   3.00 (0.15) &               &  --1.92 &  --0.23 \\ 
 Ni I  &   6.23 &   12 &   4.61 (0.28) &               &  --1.62 &   0.07 \\ 
 Zn I  &   4.62 &    1 &   2.55 (0.10) &   2.60 (0.10) &  --2.02 &  --0.31 \\ 
		\multicolumn{7}{l}{Segue~1 S2 } \\
 CH    &   8.39 &    1 &   6.09 (0.10) &               &  --2.30 &   0.08 \\ 
 Na I  &   6.33 &    2 &   3.94 (0.26) &   3.42 (0.25) &  --2.91 &  --0.53 \\ 
 Mg I  &   7.58 &    3 &   5.82 (0.06) &   5.89 (0.16) &  --1.69 &   0.70 \\ 
 Al I  &   6.47 &    1 &   2.78 (0.10) &   3.13 (0.10) &  --3.34 &  --0.96 \\ 
 Si I  &   7.52 &    2 &   6.08 (0.12) &   6.07 (0.11) &  --1.45 &   0.93 \\ 
 K  I  &   5.12 &    1 &   3.73 (0.10) &   3.37 (0.10) &  --1.75 &   0.63 \\ 
 Ca I  &   6.36 &   12 &   4.43 (0.12) &   4.49 (0.13) &  --1.87 &   0.51 \\ 
 Sc II &   3.10 &    3 &   1.15 (0.08) &   1.15 (0.08) &  --1.95 &   0.43 \\ 
 Ti I  &   4.90 &    5 &   2.85 (0.11) &   2.99 (0.13) &  --1.91 &   0.48 \\ 
 Ti II &   4.90 &   20 &   3.00 (0.14) &   2.97 (0.16) &  --1.93 &   0.45 \\ 
 V  I  &   4.00 &    1 &   1.42 (0.10) &               &  --2.58 &  --0.26 \\ 
 Cr I  &   5.65 &    7 &   3.05 (0.20) &   3.73 (0.17) &  --1.92 &   0.46 \\ 
 Mn I  &   5.37 &    2 &   2.55 (0.10) &   2.85 (0.10) &  --2.52 &  --0.14 \\ 
 Fe I  &   7.50 &   74 &   5.18 (0.24) &   5.25 (0.24) &  --2.25 &   0.13 \\ 
 Fe II &   7.50 &   11 &   5.12 (0.13) &   5.12 (0.13) &  --2.38 &   0.00 \\ 
 Co I  &   4.92 &    3 &   2.36 (0.19) &               &  --2.56 &  --0.24 \\ 
 Ni I  &   6.23 &    2 &   3.68 (0.22) &               &  --2.55 &  --0.23 \\ 
		\multicolumn{7}{l}{Segue~1 S3 } \\ 
 CH    &   8.39 &    1 &   5.99 (0.10) &               &  --2.40 &  --0.11 \\ 
 Na I  &   6.33 &    2 &   3.66 (0.06) &   3.28 (0.02) &  --3.05 &  --0.76 \\ 
 Mg I  &   7.58 &    2 &   5.47 (0.02) &   5.52 (0.02) &  --2.06 &   0.23 \\ 
 Al I  &   6.47 &    1 &   3.49 (0.10) &   3.81 (0.10) &  --2.66 &  --0.37 \\ 
 Si I  &   7.52 &    1 &   5.78 (0.10) &   5.78 (0.10) &  --1.74 &   0.55 \\ 
 K  I  &   5.12 &    1 &   3.90 (0.10) &   3.63 (0.10) &  --1.49 &   0.80 \\ 
 Ca I  &   6.36 &   12 &   4.53 (0.13) &   4.61 (0.14) &  --1.75 &   0.54 \\ 
 Sc II &   3.10 &    3 &   0.98 (0.02) &   0.98 (0.02) &  --2.12 &   0.17 \\ 
 Ti I  &   4.90 &    6 &   3.01 (0.10) &   3.18 (0.12) &  --1.72 &   0.57 \\ 
 Ti II &   4.90 &   18 &   3.01 (0.23) &   2.98 (0.26) &  --1.92 &   0.37 \\ 
 Cr I  &   5.65 &    6 &   3.11 (0.15) &   3.82 (0.15) &  --1.83 &   0.46 \\ 
 Mn I  &   5.37 &    2 &   2.77 (0.12) &   3.07 (0.12) &  --2.30 &   0.00 \\ 
 Fe I  &   7.50 &   63 &   5.12 (0.18) &   5.20 (0.18) &  --2.30 &  --0.01 \\ 
 Fe II &   7.50 &   12 &   5.21 (0.10) &   5.21 (0.10) &  --2.29 &   0.00 \\ 
 Co I  &   4.92 &    1 &   2.64 (0.10) &               &  --2.28 &   0.10 \\ 
 Ni I  &   6.23 &    1 &   3.84 (0.10) &               &  --2.39 &  --0.01 \\ 
 Ba II &   2.13 &    1 &  --1.50 (0.10) &  --1.30 (0.10) &  --3.43 &  --1.14 \\ 
		\multicolumn{7}{l}{Segue~1 S4, CEMP-s star } \\ 
 CH    &   8.39 &    1 &   7.44 (0.10) &               &  --0.95 &   0.74 \\ 
 O  I  &   8.73 &    2 &   7.98 (0.09) &   7.83 (0.07) &  --0.90 &   0.79 \\ 
 Na I  &   6.33 &    2 &   5.01 (0.16) &   4.68 (0.10) &  --1.65 &   0.04 \\ 
 Mg I  &   7.58 &    4 &   6.26 (0.08) &   6.25 (0.01) &  --1.33 &   0.36 \\ 
 Si I  &   7.52 &    1 &   6.10 (0.10) &   6.05 (0.10) &  --1.47 &   0.22 \\ 
 K  I  &   5.12 &    1 &   4.00 (0.10) &   3.49 (0.10) &  --1.63 &   0.06 \\ 
 Ca I  &   6.36 &   15 &   5.10 (0.16) &   5.10 (0.22) &  --1.26 &   0.43 \\ 
 Sc II &   3.10 &    2 &   1.52 (0.37) &   1.52 (0.37) &  --1.58 &   0.11 \\ 
		\hline
	\end{tabular}
	%\end{longtable}
\end{table}

\begin{table}
	%\begin{longtable}{lccrrrr}
	\contcaption{}
	%	\centering
	\label{elements1col}
		\setlength{\tabcolsep}{0.7mm}	
%	\begin{tabular}{|l|c|c|r|r|r|r|}
	\begin{tabular}{lcrrrrr}
		\hline
		Sp. &  $\eps_{\rm \odot}$ & N$_{\rm l}$ & $\eps_{\rm LTE}$ & $\eps_{\rm NLTE}$ & [El/H]$_{\rm }$ & [El/Fe]$_{\rm }$ \\
		\hline
 Ti I  &   4.90 &    5 &   3.39 (0.18) &   3.55 (0.30) &  --1.35 &   0.34 \\ 
 Ti II &   4.90 &   19 &   3.61 (0.16) &   3.58 (0.17) &  --1.32 &   0.37 \\ 
 V  I  &   4.00 &    1 &   1.84 (0.10) &               &  --2.16 &  --0.44 \\ 
 Cr I  &   5.65 &    5 &   3.77 (0.09) &   4.40 (0.14) &  --1.25 &   0.44 \\ 
 Mn I  &   5.37 &    3 &   3.42 (0.12) &   3.72 (0.12) &  --1.65 &   0.04 \\ 
 Fe I  &   7.50 &   55 &   5.78 (0.19) &   5.80 (0.19) &  --1.70 &  --0.01 \\ 
 Fe II &   7.50 &   14 &   5.81 (0.16) &   5.81 (0.16) &  --1.69 &   0.00 \\ 
 % Co I  &   4.92 &    1 &   2.05 (0.10) &               &  --2.87 &  --1.15 \\ 
 Ni I  &   6.23 &    6 &   4.44 (0.16) &               &  --1.79 &  --0.07 \\ 		
		\multicolumn{7}{l}{Segue~1 S5, CEMP-no star } \\  
 CH    &   8.39 &    1 &   5.74 (0.10) &               &  --2.65 &   0.91 \\ 
 Na I  &   6.33 &    2 &   2.55 (0.11) &   2.43 (0.14) &  --3.90 &  --0.34 \\ 
 Mg I  &   7.58 &    5 &   4.68 (0.27) &   4.73 (0.24) &  --2.85 &   0.71 \\ 
 Al I  &   6.47 &    2 &   2.21 (0.16) &   2.78 (0.18) &  --3.69 &  --0.13 \\ 
 Si I  &   7.52 &    1 &   4.61 (0.10) &   4.54 (0.10) &  --2.98 &   0.58 \\ 
 Ca I  &   6.36 &    2 &   3.47 (0.06) &   3.60 (0.04) &  --2.76 &   0.80 \\ 
 Ti II &   4.90 &    9 &   1.76 (0.18) &   1.86 (0.17) &  --3.04 &   0.52 \\ 
 Cr I  &   5.65 &    2 &   2.11 (0.01) &   2.78 (0.01) &  --2.87 &   0.69 \\ 
 Fe I  &   7.50 &   15 &   3.78 (0.34) &   3.94 (0.35) &  --3.56 &   0.00 \\ 
% Fe II &   7.50 &    1 &   3.94 (0.10) &   3.94 (0.10) &  --3.56 &   0.00 \\ 
 Ni I  &   6.23 &    2 &   2.58 (0.36) &               &  --3.65 &   0.07 \\  
		\multicolumn{7}{l}{Segue~1 S6, CEMP-no star } \\ 
 CH    &   8.39 &    1 &   6.19 (0.10) &               &  $<$--2.20 &  $<$ 1.00 \\ 		
 Na I  &   6.33 &    2 &   3.04 (0.03) &   2.94 (0.00) &  --3.39 &  --0.20 \\ 
 Mg I  &   7.58 &    2 &   4.80 (0.17) &   4.92 (0.12) &  --2.66 &   0.53 \\ 
 Al I  &   6.47 &    1 &   2.69 (0.10) &   3.33 (0.10) &  --3.14 &   0.05 \\ 
 Si I  &   7.52 &    1 &   5.34 (0.10) &   5.29 (0.10) &  --2.23 &   0.96 \\ 
 Ca I  &   6.36 &    3 &   3.79 (0.26) &   3.90 (0.23) &  --2.46 &   0.73 \\ 
 Ti II &   4.90 &   10 &   1.95 (0.11) &   2.06 (0.10) &  --2.84 &   0.35 \\ 
 Cr I  &   5.65 &    4 &   2.26 (0.28) &   2.98 (0.23) &  --2.67 &   0.52 \\ 
 Fe I  &   7.50 &   12 &   4.12 (0.23) &   4.31 (0.24) &  --3.19 &  --0.00 \\ 
% Fe II &   7.50 &    1 &   4.31 (0.10) &   4.31 (0.10) &  --3.19 &   0.00 \\ 
		\multicolumn{7}{l}{Segue~1 S7, CEMP-no star } \\ 
 CH    &   8.39 &    2 &   7.21 (0.05) &               &  --1.18 &   2.30 \\ 
 Na I  &   6.33 &    2 &   3.13 (0.10) &   2.97 (0.15) &  --3.36 &   0.12 \\ 
 Mg I  &   7.58 &    3 &   4.85 (0.10) &   4.89 (0.16) &  --2.69 &   0.79 \\ 
 Al I  &   6.47 &    1 &   2.43 (0.10) &   2.75 (0.10) &  --3.72 &  --0.24 \\ 
 Si I  &   7.52 &    1 &   4.81 (0.10) &   4.80 (0.10) &  --2.72 &   0.76 \\ 
 Ca I  &   6.36 &    6 &   3.59 (0.04) &   3.70 (0.05) &  --2.66 &   0.81 \\ 
 Ti II &   4.90 &    7 &   2.28 (0.22) &   2.36 (0.22) &  --2.54 &   0.94 \\ 
 Cr I  &   5.65 &    3 &   1.89 (0.11) &   2.61 (0.18) &  --3.04 &   0.44 \\ 
 Mn I  &   5.37 &    1 &   0.83 (0.10) &   0.93 (0.10) &  --4.44 &  --0.96 \\ 
 Fe I  &   7.50 &   21 &   3.84 (0.14) &   3.99 (0.13) &  --3.51 &  --0.03 \\ 
 Fe II &   7.50 &    3 &   4.02 (0.04) &   4.02 (0.04) &  --3.48 &   0.00 \\ 
 Co I  &   4.92 &    3 &   1.54 (0.21) &               &  --3.38 &   0.28 \\ 
 Ni I  &   6.23 &    4 &   2.33 (0.18) &               &  --3.90 &  --0.24 \\ 
 Sr II &   2.90 &    2 &  --2.12 (0.09) &  --1.87 (0.05) &  --4.77 &  --1.29 \\ 
 Ba II &   2.13 &    1 &  --2.42 (0.10) &  --2.24 (0.10) &  --4.37 &  --0.89 \\ 
			\multicolumn{7}{l}{Tri II S40 } \\ 	
 Na I  &   6.33 &    2 &   2.66 (0.07) &   2.53 (0.15) &  --3.80 &  --1.03 \\ 
 Mg I  &   7.58 &    2 &   4.98 (0.06) &   5.08 (0.01) &  --2.50 &   0.27 \\ 
 Al I  &   6.47 &    2 &   2.88 (0.00) &   3.14 (0.04) &  --3.33 &  --0.56 \\ 
 Si I  &   7.52 &    1 &   5.55 (0.10) &   5.55 (0.10) &  --1.97 &   0.80 \\ 
 K  I  &   5.12 &    1 &   3.27 (0.10) &   3.08 (0.10) &  --2.04 &   0.73 \\  
 Ca I  &   6.36 &   11 &   3.88 (0.13) &   4.03 (0.12) &  --2.33 &   0.44 \\ 
 Sc II &   3.10 &    4 &   0.61 (0.12) &   0.61 (0.12) &  --2.49 &   0.29 \\ 
 Ti I  &   4.90 &    3 &   2.16 (0.02) &   2.44 (0.04) &  --2.46 &   0.32 \\ 
 Ti II &   4.90 &    8 &   2.21 (0.10) &   2.21 (0.11) &  --2.69 &   0.09 \\ 
   V  I  &   4.00 &    2 &   1.24 (0.01) &           &  --2.76 &   0.14 \\ 
 Cr I  &   5.65 &    7 &   2.49 (0.32) &   3.01 (0.28) &  --2.64 &   0.14 \\ 
 Mn I  &   5.37 &    2 &   2.47 (0.01) &   2.77 (0.01) &  -2.60 &   0.17 \\
 Fe I  &   7.50 &   40 &   4.60 (0.13) &   4.73 (0.14) &  --2.77 &   0.01 \\ 
 Fe II &   7.50 &    8 &   4.73 (0.11) &   4.73 (0.11) &  --2.77 &   0.00 \\ 
   Co I  &   4.92 &    1 &   2.64 (0.10) &            &  --2.28 &   0.62 \\ 
   Ni I  &   6.23 &    7 &   3.89 (0.09) &            &  --2.34 &   0.56 \\ 
 Zn I  &   4.62 &    1 &   2.09 (0.10) &   2.14 (0.10) &  --2.48 &   0.29 \\ 
 Sr II &   2.90 &    2 &  --1.87 (0.01) &  --1.82 (0.03) &  --4.72 &  --1.94 \\ 
 Ba II &   2.13 &    1 &  --2.94 (0.10) &  --2.71 (0.10) &  --4.84 &  --2.07 \\ 						 				
%		\hline
%		\multicolumn{7}{l}{Solar abundances from \citet{Lodders2009} } \\	  
%		\multicolumn{7}{l}{The standard deviations are indicated in parentheses. } \\       
		\hline
	\end{tabular}
	%\end{longtable}
\end{table}

\begin{table}
	%\begin{longtable}{lccrrrr}
	\contcaption{}
	%	\centering
	\label{elements1col}
		\setlength{\tabcolsep}{0.7mm}	
%	\begin{tabular}{|l|c|c|r|r|r|r|}
	\begin{tabular}{lcrrrrr}
		\hline
		Sp. &  $\eps_{\rm \odot}$ & N$_{\rm l}$ & $\eps_{\rm LTE}$ & $\eps_{\rm NLTE}$ & [El/H]$_{\rm }$ & [El/Fe]$_{\rm }$ \\
		\hline
		\multicolumn{7}{l}{Coma Berenices S1 } \\  
		CH    &   8.39 &   4 &   6.18 (0.07) &               &  --2.21 &  --0.09 \\ 
		Na I  &   6.33 &   2 &   3.54 (0.12) &   3.35 (0.18) &  --2.98 &  --0.86 \\ 
		Mg I  &   7.58 &   2 &   5.63 (0.08) &   5.69 (0.07) &  --1.89 &   0.24 \\ 
		K  I  &   5.12 &   1 &   3.10 (0.10) &   2.93 (0.10) &  --2.19 &  --0.07 \\  
		Ca I  &   6.36 &  16 &   4.29 (0.16) &   4.37 (0.20) &  --1.99 &   0.13 \\ 
		Sc II &   3.10 &   4 &   0.85 (0.15) &   0.85 (0.15) &  --2.25 &  --0.13 \\ 
		Ti I  &   4.90 &   5 &   2.51 (0.10) &   2.63 (0.09) &  --2.27 &  --0.15 \\ 
		Ti II &   4.90 &  12 &   2.66 (0.21) &   2.62 (0.25) &  --2.28 &  --0.16 \\ 
		% V  I  &   4.00 &   1 &   1.15 (0.10) &   1.15 (0.10) &  --2.85 &  --0.73 \\ 
		V  I  &   4.00 &   1 &   1.15 (0.10) &               &  --2.85 &  --0.64 \\ 
		Cr I  &   5.65 &   3 &   3.10 (0.19) &   3.47 (0.25) &  --2.18 &  --0.05 \\ 
		% Mn I  &   5.37 &   1 &$<$ 2.34       &$<$ 2.34       &$<$ --3.03 &$<$ --0.91 \\ 
		% Mn I  &   5.37 &   1 &$<$ 2.34       &$<$ 2.89       &$<$ --2.48 &$<$ --0.36 \\ 
		Mn I  &   5.37 &    1 &$<$   2.34 (0.10) &$<$   2.64 (0.10) & $<$ --2.73 &$<$  --0.61 \\
		Fe I  &   7.50 &  33 &   5.29 (0.18) &   5.35 (0.19) &  --2.15 &  --0.03 \\ 
		Fe II &   7.50 &  11 &   5.38 (0.14) &   5.38 (0.14) &  --2.12 &   0.00 \\ 
		% Ni I  &   6.23 &   3 &   3.52 (0.06) &   3.52 (0.06) &  --2.71 &  --0.59 \\
		Ni I  &   6.23 &   3 &   3.52 (0.06) &               &  --2.71 &  --0.50 \\ 
		Zn I  &   4.62 &   1 &$<$ 2.22 (0.10) &$<$ 2.27 (0.10) &$<$ --2.35 &$<$ --0.23 \\ 
		Sr II &   2.90 &   1 &  --1.44 (0.10) &  --1.19 (0.10) &  --4.09 &  --1.97 \\ 
		Ba II &   2.13 &   1 &  --2.41 (0.10) &  --2.31 (0.10) &  --4.44 &  --2.32 \\ 
		\multicolumn{7}{l}{Coma Berenices S2 } \\ 
		CH    &   8.39 &   4 &   5.89 (0.00) &               &  --2.50 &   0.02 \\ 
		Na I  &   6.33 &   2 &   3.59 (0.01) &   3.20 (0.05) &  --3.13 &  --0.61 \\ 
		Mg I  &   7.58 &   3 &   6.05 (0.02) &   6.13 (0.16) &  --1.45 &   1.07 \\ 
 Si I  &   7.52 &    1 &   6.22 (0.10) &   6.17 (0.10) &  --1.35 &   1.17 \\ 		
		% K  I  &   5.12 &   1 &   3.92 (0.10) &   3.60 (0.10) &  --1.52 &   1.00 \\
		K  I  &   5.12 &    1 &   3.92 (0.10) &   3.51 (0.10) &  --1.61 &   0.91 \\   
		Ca I  &   6.36 &  13 &   4.26 (0.18) &   4.31 (0.19) &  --2.05 &   0.48 \\ 
		% Ca II &   6.36 &   3 &   4.36 (0.01) &   4.31 (0.04) &  --2.05 &   0.47 \\ 
		Sc II &   3.10 &   3 &   1.14 (0.05) &   1.14 (0.05) &  --1.96 &   0.56 \\ 
		Ti I  &   4.90 &   7 &   2.67 (0.22) &   2.87 (0.21) &  --2.03 &   0.49 \\ 
		Ti II &   4.90 &  13 &   2.69 (0.19) &   2.70 (0.21) &  --2.20 &   0.32 \\ 
		% V  I  &   4.00 &   1 &   1.38 (0.10) &   1.64 (0.10) &  --2.36 &   0.16 \\ 
		V  I  &   4.00 &   1 &   1.38 (0.10) &               &  --2.36 &   0.31 \\
		Cr I  &   5.65 &   3 &   2.93 (0.27) &   3.31 (0.33) &  --2.34 &   0.18 \\ 
		% Mn I  &   5.37 &   1 & $<$ 2.14(0.10) & $<$ 2.73(0.10) & $<$ --2.64 & $<$ --0.12 \\
		Mn I  &   5.37 &   1 & $<$  2.14 (0.10) & $<$  2.44 (0.10) & $<$ --2.93 & $<$ --0.41 \\  
		Fe I  &   7.50 &  28 &   4.83 (0.12) &   4.91 (0.12) &  --2.59 &  --0.07 \\ 
		Fe II &   7.50 &   5 &   4.98 (0.15) &   4.98 (0.15) &  --2.52 &   0.00 \\ 
		% Ni I  &   6.23 &   5 &   3.68 (0.16) &   3.68 (0.16) &  --2.55 &  --0.03 \\ 
		Ni I  &   6.23 &   5 &   3.68 (0.16) &               &  --2.55 &   0.12 \\  
		Zn I  &   4.62 &   1 &   1.94 (0.10) &   1.99 (0.10) &  --2.63 &  --0.11 \\ 
		Sr II &   2.90 &    1 &  --1.55 (0.10) &  --1.44 (0.10) &  --4.34 &  --1.82 \\ 
		Ba II &   2.13 &   1 &  --2.15 (0.10) &  --2.03 (0.10) &  --4.16 &  --1.64 \\  		
		\multicolumn{7}{l}{Coma Berenices S3 } \\  
		CH    &   8.39 &   2 &   5.35 (0.14) &               &  --3.04 &  --0.65 \\ 
		Na I  &   6.33 &   2 &   4.17 (0.10) &   3.72 (0.14) &  --2.61 &  --0.22 \\ 
		Mg I  &   7.58 &   3 &   5.59 (0.03) &   5.69 (0.15) &  --1.89 &   0.49 \\ 
        K  I  &   5.12 &   1 &   3.16 (0.10) &   2.98 (0.10) &  --2.14 &   0.25 \\ 		
		Ca I  &   6.36 &  18 &   4.44 (0.09) &   4.47 (0.09) &  --1.89 &   0.49 \\ 
%		Ca II &   6.36 &   3 &   4.47 (0.04) &   4.46 (0.04) &  --1.90 &   0.49 \\ 
		Sc II &   3.10 &   4 &   0.29 (0.04) &   0.29 (0.04) &  --2.81 &  --0.42 \\ 
		Ti I  &   4.90 &   8 &   2.64 (0.10) &   2.84 (0.12) &  --2.06 &   0.32 \\ 
		Ti II &   4.90 &  17 &   2.94 (0.15) &   2.94 (0.15) &  --1.96 &   0.43 \\ 
%		V  I  &   4.00 &   1 &   1.22 (0.10) &   1.45 (0.10) &  --2.55 &  --0.16 \\ 
		V  I  &   4.00 &   1 &   1.22 (0.10) &               &  --2.55 &  --0.13 \\ 		
		Cr I  &   5.65 &   3 &   3.01 (0.12) &   3.45 (0.13) &  --2.20 &   0.19 \\
%        Mn I  &   5.37 &   1 &	 2.31        &   2.31        &  --3.06 &  --0.67 \\	
%        Mn I  &   5.37 &   1 &	 2.31        &   2.91        &  --2.46 &  --0.07 \\
        Mn I  &   5.37 &    1 &   2.31 (0.10) &   2.61 (0.10) &  --2.76 &  --0.37 \\ 	 
		Fe I  &   7.50 &  41 &   5.08 (0.12) &   5.16 (0.12) &  --2.34 &   0.05 \\ 
		Fe II &   7.50 &  17 &   5.12 (0.14) &   5.11 (0.13) &  --2.39 &   0.00 \\ 
%		Ni I  &   6.23 &   8 &   3.52 (0.04) &   3.52 (0.04) &  --2.71 &  --0.33 \\ 
		Ni I  &   6.23 &   8 &   3.52 (0.04) &               &  --2.71 &  --0.29 \\ 		
		Zn I  &   4.62 &   1 &   2.08 (0.10) &   2.13 (0.10) &  --2.49 &  --0.10 \\ 
		Sr II &   2.90 &   1 &  --1.20 (0.10) &  --1.12 (0.10) &  --4.02 &  --1.63 \\ 
		Ba II &   2.13 &   4 &  --1.32 (0.06) &  --1.28 (0.06) &  --3.41 &  --1.02 \\ 
		\multicolumn{7}{l}{Ursa Major II S1 } \\ 
		Na I  &   6.33 &   2 &   3.26 (0.06) &   3.08 (0.01) &  --3.25 &  --0.33 \\ 
		Mg I  &   7.58 &   5 &   4.94 (0.15) &   5.07 (0.27) &  --2.51 &   0.41 \\ 
		Ca I  &   6.36 &   6 &   3.56 (0.19) &   3.70 (0.19) &  --2.66 &   0.25 \\ 
		Ti I  &   4.90 &   3 &   2.20 (0.20) &   2.49 (0.21) &  --2.41 &   0.51 \\ 
		Ti II &   4.90 &   9 &   2.19 (0.22) &   2.20 (0.22) &  --2.70 &   0.22 \\ 
		Fe I  &   7.50 &  36 &   4.50 (0.15) &   4.66 (0.15) &  --2.84 &   0.08 \\ 
		Fe II &   7.50 &   7 &   4.58 (0.13) &   4.58 (0.13) &  --2.92 &   0.00 \\ 
		%		Ni I  &   6.23 &   2 &   3.67 (0.32) &   3.67 (0.32) &  --2.56 &   0.36 \\ 
		Ni I  &   6.23 &   2 &   3.67 (0.32) &               &  --2.56 &   0.44 \\ 
		Sr II &   2.90 &   1 &  --0.95 (0.10) &  --0.82 (0.10) &  --3.72 &  --0.80 \\ 
		Ba II &   2.13 &   1 &  --1.69 (0.10) &  --1.45 (0.10) &  --3.58 &  --0.66 \\ 		
%		\hline
%		\multicolumn{7}{l}{Solar abundances from \citet{Lodders2009} } \\	  
%		\multicolumn{7}{l}{The standard deviations are indicated in parentheses. } \\       
		\hline
		\multicolumn{7}{l}{Solar abundances from \citet{Lodders2009}. The standard} \\	  
		\multicolumn{7}{l}{ deviations are indicated in parentheses. For V\ione, Co\ione, and Ni\ione, } \\     
		\multicolumn{7}{l}{ [El/Fe] are calculated relative to Fe\ione\ in LTE.} \\  
		%		\multicolumn{7}{l}{ } \\  
		\hline		
	\end{tabular}
	%\end{longtable}
\end{table}

%The selected element abundance ratios, namely, [Mg/Fe], [Ca/Fe], [Ti/Fe], [Na/Mg], and [Ni/Fe] versus [Fe/H] are plotted in Fig.~\ref{abun_ratios}. For comparison, Fig.~\ref{abun_ratios} shows the abundance ratios for stars in different dwarf galaxies and the MW halo investigated by M17b and P19.

\section{Discussion}\label{discussion}

%\subsection{Chemical properties of Segue 1 UFD}\label{notes_segue}
\subsection{Incomplete mixing in Seg~1 and ComaBer UFDs}\label{mixing}

\underline{Segue~1.}

Seg~1 stars show a wide metallicity  range, from [Fe/H] = $-1.6$ to $-3.6$. 
%According to their chemical properties, stars in Segue~1 can be divided into two groups. 
They  could be subdivided into two groups according to their chemical properties.
The first group contains S1, S2, and S3 with [Fe/H] $> -2.4$, normal carbon abundances and [$\alpha$/Fe].  Star S4, classified as CEMP-s, gained its high carbon and n-capture element abundances from its binary companion, while its $\alpha$-elements and iron-peak elements are suitable for analysis of the chemical history in Segue~1. This star can also be added to the first group with its [Fe/H] = $-1.73$.

Another group contains carbon enhanced S5, S6, and S7 stars with  [Fe/H] $< -3.1$. 
%For Na, Al, $\alpha$-elements, these stars have higher [El/Fe] abundance ratios compared to stars in the first group
For Na, Al, $\alpha$-elements, these stars have higher [El/Fe] abundance ratios compared to the first group stars (see Fig.~\ref{elfe_patterns}).
Higher [Na/Fe] in more metal-poor stars is not surprising. 
%For [Na/Fe], 
M17b (see their Fig.~4) found nearly constant [Na/Fe] = $-0.4$ for stars with $-3 <$ [Fe/H] $< -2$, while more metal poor stars show an increasing trend with decreasing [Fe/H], up to solar [Na/Fe]. However, for $\alpha$-elements, constant  [$\alpha$/Fe] is expected.

In line with F14 and V13, we found all Seg~1 stars to be $\alpha$-enhanced. This argues for an absence of iron, produced in  SN\ione a and a short timescale of star formation.
We note that  Seg~1 S1 and S4 have fairly high [Fe/H] of $-1.6$ and high [$\alpha$/Fe]. 
%A relatively high [Fe/H] $\simeq -1.6$ should be reached  at timescale of less than 1~Gyr,  before the onset of iron production in SN\ione a. 
High star formation rate is required to achieve [Fe/H] $\simeq -1.6$ at timescale of less than 1~Gyr,  before the onset of iron production in SN\ione a.

Different stars in Seg~1 show different abundance ratios, which argues for an incomplete mixing of the interstellar medium during the formation of these stars.

\citet{2020arXiv200801259C} investigated how the structure of Pop\iii\ supernova ejecta affects the elemental abundance of extremely metal-poor stars and concluded that single ejecta can produce different abundance patterns of MP stars.
According to stellar evolution  models,  the  SN  ejecta  should  be  initially  stratified, where heavier elements, such as Fe, are in the inner layers and lighter elements, such as C, are in the outer layers. 
Such a scenario is one of the hypotheses that could explain
the observed splitting of Seg~1 stars into two groups (carbon enhanced, with [Fe/H] $<-3.1$ and carbon normal with [Fe/H] $> -2.4$). Accurate chemical enrichment modelling of Seg~1 is required to check this hypothesis.

From analyses of colour-magnitude diagram, coordinates, and proper motions, \citet{2020arXiv200705011M} found a low membership probability of p = 0.00142 and 0.00024 for Seg~1 S2 and S7, respectively. A membership of Seg~1  S3 and S6 with p = 0.30 and 0.37 is also in doubt. 
These probabilities were calculated for the Seg~1 structural parameters, d = 22.9~kpc, r$_{\rm 1/2}$ = 3.95 arcmins, and the eccentricity e = 0.34 \citep{2020arXiv200705011M}.  However, the structure of faint dwarf galaxies, especially at large radius, can be very complex, and it is known
that some of these systems have extended tidal features that are not well captured by the simple parametrisation \citep{2018ApJ...866...22L}. For example, \citet{2020arXiv201202309C} found members of Tucana\ii\ UFD in its outer region, up to 9 half-light radii. Thus S3 and S6 at distances of 9.54 and 4.5 r$_{\rm 1/2}$ could be actually members of Seg~1. Anyway, neglecting the above stars  would not change our results on the inhomogeneous gas mixing and splitting between carbon-rich and iron-rich stars.
%However,  neglecting the above stars would not change our results on an inhomogeneous mixing and splitting between carbon-rich and iron-rich stars.  

%Another chemical feature of Seg~1 stars is high [V/Fe] from 0.66 to 0.86, found in S1, S2, S3, and S4. In the remaining Seg~1 stars 

%\subsection{Chemical properties of Coma Berenices UFD}\label{notes_coma}

\underline{Coma Berenices.}

ComaBer stars show close [Fe/H] from $-2.6$ to $-2.1$, however, their [El/Fe] abundance patterns are different. 
S3 shows normal chemical composition similar to that of the VMP MW halo stars. An exception is low [Sc/Fe] = $-0.42$.
%\pm0.21$.

S2 is enhanced in Mg 
and depleted in sodium, which
results in extremely low [Na/Mg] = $-1.68$, which is the lowest value known to date. 
Due to the high magnesium abundance, this star has high hydrostatic to explosive $\alpha$-element abundance ratios, namely high [Mg/Ca] = 0.59. 
Stars with high  [Mg/Ca] are detected in
Hercules dSph \citep{2008ApJ...688L..13K}, Draco dSph \citep{2004ApJ...612..447F},  Grus\ii\ UFD \citep{2020ApJ...897..183H}, and the MW \citep{2008ApJ...684..588F,2013ApJ...762...28N,2014Natur.506..463K,Depagne2002,2008ApJ...672..320C}.
A high ratio of hydrostatic to explosive  $\alpha$-elements, as observed in these stars, can be produced in SN\ii\ with high progenitor mass \citep{hw2010}. 
In addition, there is a hint of inverted odd-even Z effect, namely, odd-Z elements Sc and V are enhanced with respect to their neighbouring even-Z elements Ti and Cr (see Fig.~\ref{elfe_patterns}). 
We note the large errors in abundance ratios: [Sc/Ti] = 0.24$\pm0.22$  and [V/Cr] = 0.37$\pm0.34$, which prevents us from making a solid conclusion. 

S1 shows very low [Na/Fe], it has
depleted titanium, vanadium, manganese, nickel, and zinc with respect to iron,
and is only moderately enhanced in $\alpha$-elements.
S1 is thus enhanced in iron, but not in the other chemical elements.
S1 also shows low [Na/Mg] encountered in the literature as a signature of explosions of Population\iii\ stars (see Sect.~\ref{footprint} for the discussion and references).

%Such an element abundance pattern can be interpreted as iron enhancement in this star.
%We doubt, that iron excess in S1 was produced by onset of iron production in SN Ia, as suggested by \citet{2013ApJ...767..134V}. 
%Thus, it seems doubtful, that . 

From analysis of medium resolution spectra, \citet{2013ApJ...767..134V} suggested that the iron excess in S1 and the lower [$\alpha$/Fe] abundance ratios compared to those of S2 and S3 are caused by an onset of iron production in SN\ione a.
To explain low [$\alpha$/Fe] in S1, there is no need to assume an onset of ubiquitous iron production in SNe\ione a throughout ComBer and extended star formation.  It can be explained by formation of S1 in a pocket locally enhanced by a single SN\ione a event \citep{2003ApJ...592..906I,2008AJ....135.1580K,2012ApJ...751..102V}.
Indeed, there are some prompt SN\ione a with a delay time of less than 500~Myr \citep{2006MNRAS.370..773M,2010AIPC.1314..233R}. 
Nucleosynthesis calculations of \citet{2020ApJ...895..138K} for SN\ione a explosions of sub-Chandrasekhar white dwarfs predict  [Mn,Ni,Zn/Fe] $< 0$, which is similar to what is found in S1.

Another possible explanation for the lower [$\alpha$/Fe] in S1 is its formation from a pocket, where the ejecta of the most massive Type II SNe are missing. This scenario was proposed by \citet{2015A&A...583A..67J} to explain the formation of a star in Sculptor dSph, ET0381 with [Fe/H] = $-2.5$, which is poor in  all elements with respect to iron. 
Our abundance analysis cannot solidly recognize the cause of low [$\alpha$/Fe] in S1.

Different chemical element abundance patterns  of ComaBer stars argue for an incomplete gas mixing during the formation of this galaxy.

\subsection{Footprints of Population\iii\ stars in UFDs.}\label{footprint}

Despite different [Mg/Fe] and [Na/Fe] found in  ComaBer~S1, Seg~1 S1, S2, S3, and S5, and Tri\ii\ S40, these stars have very similar low [Na/Mg] from $-1.3$ to $-1.0$. ComaBer S2 shows extremely low [Na/Mg] = $-1.68$, which is the lowest value found in the literature to date.  

Low sodium to magnesium abundance ratio of  [Na/Mg] $\simeq$ $-1.1$ dex was encountered in the literature and attributed to the odd-even effect, predicted for explosions of Population\iii\ stars \citep{2006NuPhA.777..424N,2006ApJ...653.1145K,tominaga2007,hw2010,2012ApJS..199...38L,2013ARA&A..51..457N,2018ApJ...857..111T,2019ApJ...870....2C}. According to the data from the Stellar Abundances for Galactic Archaeology Database \citep[SAGA\footnote{http://sagadatabase.jp}][]{2008PASJ...60.1159S,2017PASJ...69...76S}, there are eleven VMP stars with [Na/Mg] $< -1$.  Eight of them are in dwarf galaxies, while three are in the MW halo. In addition to the SAGA data, the recent study of \citet{2019ApJ...871..146F} reported on strong odd-even effect in J0023+0307, with [Na/Mg] = $-1.1$. While the statistics is poor, we note that stars with odd-even effect are more frequent in dwarf galaxies compared to the MW halo.

\begin{figure}
\includegraphics[trim=1.10cm 1.4cm 0cm 0cm,clip,width=90mm]{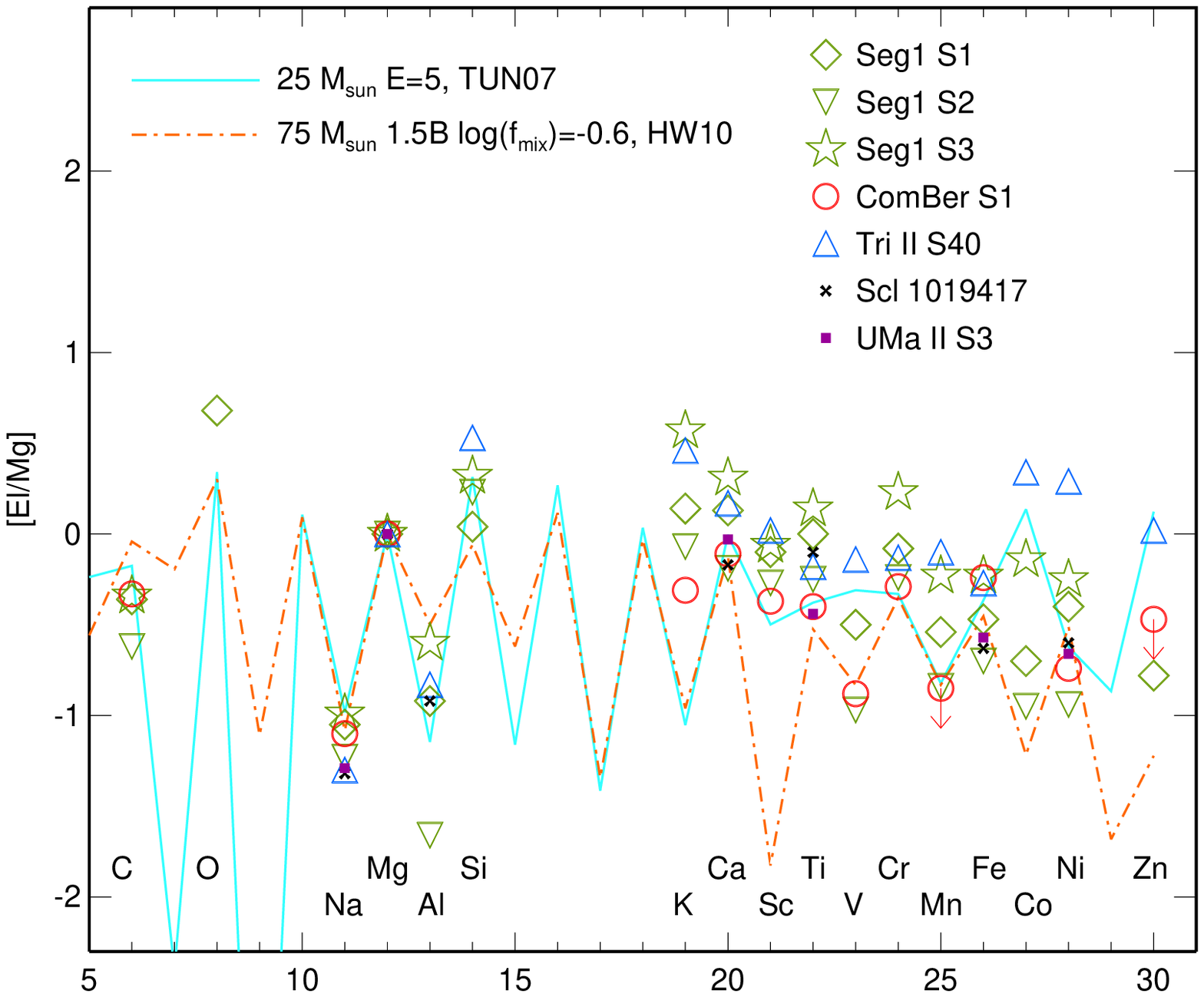}
\includegraphics[trim=1.10cm 1.4cm 0cm 0cm,clip,width=90mm]{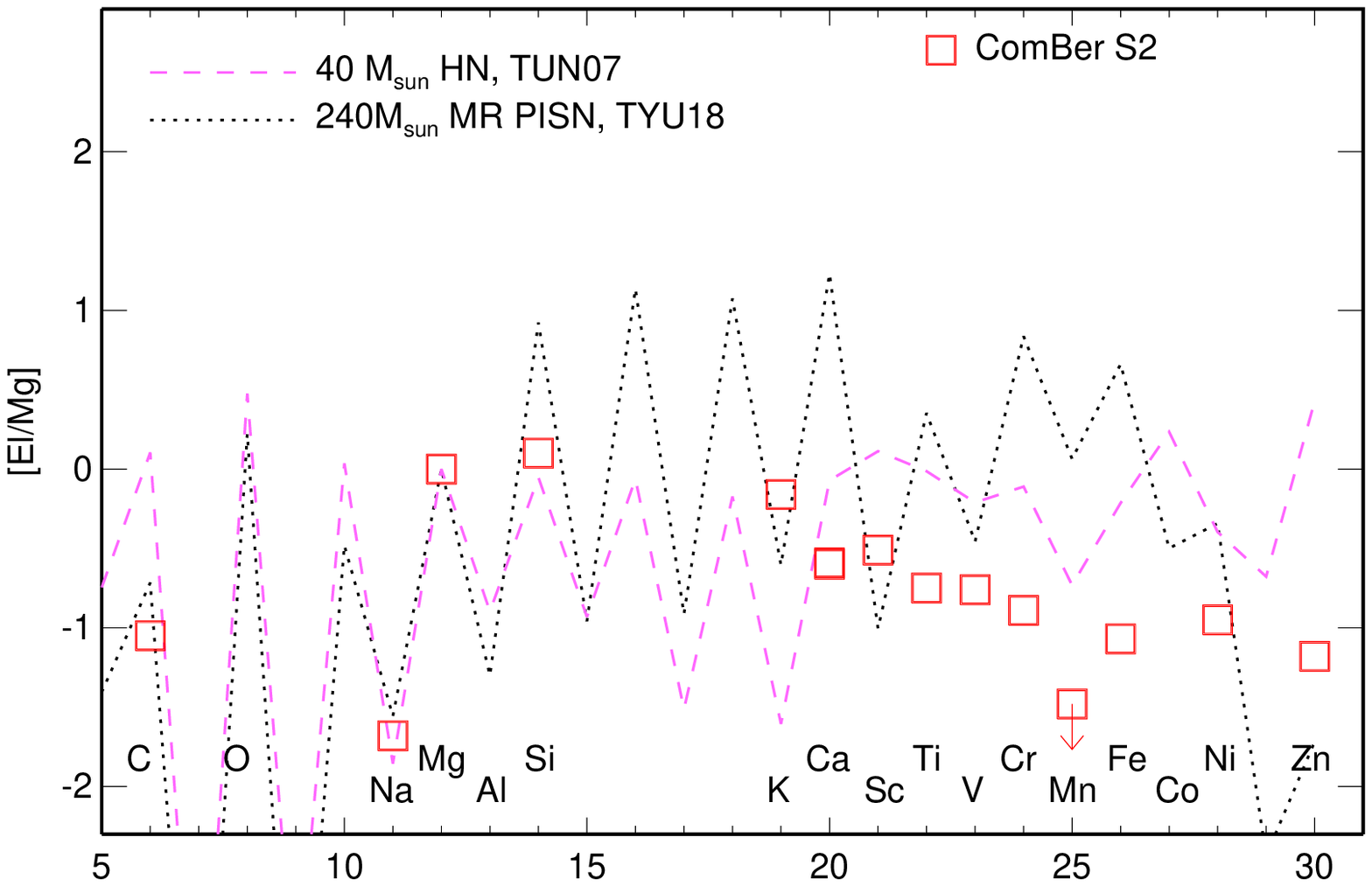}				
	\caption{Top panel: Element abundance patterns [El/Mg] of the sample stars with [Na/Mg] $\simeq -1.1$ and model predictions  for explosions of metal-free massive stars with  M$_{init}$ = 25M$_{\odot}$   \citep[][solid curve]{tominaga2007} and  M$_{init}$ = 75M$_{\odot}$ \citep[][dash-dotted curve]{hw2010}. 
		For comparison, we show  the NLTE [El/Mg] ratios of Scl1019417  and UMa\ii~S3 from M17b.
		See the legend for identification of the stars.
Bottom panel: Similar plot for ComaBer S2 star with [Na/Mg] = $-1.68$.
Dashed and dotted curves show [El/Mg] model predictions  for CCSN  with  M$_{init}$ = 40M$_{\odot}$ \citep{tominaga2007} and  PISN  with M$_{init}$ = 240M$_{\odot}$ \citep{2018ApJ...857..111T}.
}
	\label{pattern_tominaga}
\end{figure}

We selected our sample stars with [Na/Mg] $\le -1$, namely, ComaBer S1 and S2, Seg~1 S1, S2 and S3, and Tri\ii\ S40 to compare their abundances with predictions of nucleosynthesis in individual explosions of metal-free stars (Fig.~\ref{pattern_tominaga}). It is worth noting that the selected stars do not have any similar abundance ratios except the low [Na/Mg]. M17b found low [Na/Mg] = $-1.3$ in Scl1019417 from the Sculptor dSph and UMa\ii~S3.
% from the Ursa Major\ii\ UFD. 
In addition to the selected stars, we show in Figure~\ref{pattern_tominaga} the abundance patterns of Scl1019417 and UMa\ii~S3.

We tried to fit the derived abundance patterns for elements from C to Zn with the predictions of nucleosynthesis of core-collapse supernovae (CCSNe), hypernovae (HNe), and pair-instability supernova (PISN) of Population\iii\ stars from \citet[][hereafter TUN07]{tominaga2007}, \citet[][hereafter HW10]{hw2010}, and \citet[][hereafter TYU18]{2018ApJ...857..111T}, respectively.
Yields from various models of \citet{tominaga2007} provide [Na/Mg] of $-2.19$ to $-0.19$.
We selected the 25-5 and 40B models of TUN07,  with [Na/Mg] of $-1.07$ and $-1.86$, respectively, which are close to those in our sample stars. Here, the first number in the designation of models indicates an initial mass of a progenitor star (M$_{init}$) and the second symbol indicates a kinetic energy of an explosion at infinity, measured in Bethe (1 B = 10$^{51}$ erg), while B means a hypernova with E$_k$ = 30~Bethe. The model of \citet{hw2010}  was selected using the starfit.org on-line tool for fitting the stellar element abundance patterns with nucleosynthesis yields. It  has M$_{init}$ = 75M$_{\odot}$, E$_{k}$ = 1.5B, log(f$_{mix}$) = $-0.6$, where f$_{mix}$ is an ejection factor, and predicts [Na/Mg] = $-1.1$. The selected model of TYU18 has M$_{init}$ = 240M$_{\odot}$ and accounts for magnetic field and  rotation (MR PISN).
The yields of the selected models are plotted in Figure~\ref{pattern_tominaga}.

The only reasonable agreement was found between the predictions of the 25-5 model  of TUN07 and the abundance pattern of ComaBer~S1, where model predictions and determined abundance ratios are consistent for the ten chemical elements. For potassium,  model predictions are 0.4~dex  lower compared to the observations.
This element is known to be under-produced by models \citep[see, for example,][]{2006ApJ...653.1145K,2020ApJ...895..138K}. The [El/Mg] ratios of Scl1019417 and UMa\ii~S3 are consistent with the 25-5 model. An exception is the 0.3~dex higher [Ti/Mg] in Scl1019417 and about 0.2 dex lower [Fe/Mg] compared to the model predictions. We note that abundances of potassium  in Scl1019417 and UMa\ii~S3 are not available in M17b.

The 40B model of TUN07 predicts [Na/Mg] = $-1.86$. It was selected to fit the abundance pattern of ComaBer S2 with [Na/Mg] = $-1.68$. However, for the other elements, the model predicts  0.5 dex higher [El/Mg] compared to those observed is S2. There is nothing in common   between the model and the abundance pattern of ComaBer~S2 except the extremely low [Na/Mg]. This is because the model is Na-poor, while the star is Mg-enhanced.

Individual models predict various chemical element abundance ratios depending on M$_{init}$, explosion energy, and other properties of explosions. For example, there are TUN07 models predicting [Mg/Fe]  = 0.20 or 1.81~dex, while the majority of these models predict [Mg/Fe] from 0.3 to 0.5~dex. 
At the early epoch of massive galaxies formation, many SN\ii\ in each mass range enrich the interstellar medium, resulting in a settling of an average [Mg/Fe] of 0.3. This ratio is observed in the majority of MP stars in the MW and classical dSph galaxies. 
%In faint galaxies, only a subset of the full SNe mass range contributes, and the final [Mg/Fe] can differ from a standard average value, depending on stochastically sampled explosions, which have enriched a given galaxy. 
In faint galaxies, only a subset of the full SNe mass range contributes, and the final [Mg/Fe] can differ from a standard average value, depending on stochastically sampled explosions that have enriched a given galaxy. 
In addition, inhomogeneous mixing of gas results in different element abundance ratios in the next generation of stars. 
The same is true for other element abundance ratios, such as [Mg/Ca] or [Na/Mg].

Since there are model predictions of explosions of Pop\iii\ stars, producing low [Na/Mg], we suggest them to be responsible for the  low [Na/Mg] detected in MP stars. We failed to fit abundance patterns of any of our sample stars with a single nucleosynthesis event, however, we suggest that a small number of SN\ii\ explosions including those with low   [Na/Mg] contributed to the abundances found in the investigated stars.

A comparison of observed relative [El/Mg] abundance ratios with nucleosynthesis predictions should be performed with a caution. 
We note that our sample stars are too metal rich to be fitted with yields from only one explosion. For each selected model, its metallicity is about [Fe/H]$ \simeq -3$, which is lower compared to that of the sample stars ($-2.5 <$ [Fe/H] $< -2.1$). \citet{2020MNRAS.498.3703M} estimated analytically a limit for the dilution of metals produced by a single SN and recommended to account for not only element-to-element abundance ratios, but also the absolute abundance values when fitting stellar abundance patterns with nucleosynthesis yields.
In their calculations, the following assumptions were adopted: the SN being alone and isolated, the explosions being spherical and well-mixed, and the surrounding medium being homogeneous. The latter assumption is not always fulfilled  \citep[see, for example][]{2020arXiv200801259C}.
We also note that to date there is no a metal-poor star, whose extensive abundance pattern is reasonably well fitted with the yields from a single  nucleosynthesis episode, despite many attempts that were made \citep[see, for example,][]{tominaga2007,2015ApJ...809..136P,2020A&A...643A..49H}.

\subsection{Neutron-capture elements in classical and ultra-faint dwarf galaxies}\label{ncapture}

In the majority of EMP stars, n-capture elements are represented by Sr and Ba.
A number of the Galactic studies reported on a large scatter of [Sr/Ba], of about 2~dex, which decreases with increasing [Ba/H] \citep{2004ApJ...607..474H,2007A&A...476..935F,2011A&A...530A.105A}. 
In classical dSphs, abundances of Sr and Ba at the lowest metallicity are at or below the lowest values in the MW halo, however, the [Sr/Fe] and [Ba/Fe] ratios increase toward the solar ratio as [Fe/H] increases, similarly to that for the MW halo \citep{2009ApJ...701.1053C,2010ApJ...719..931C,2015A&A...583A..67J,2017AA...608A..89M}. 
Very low abundances of Sr and Ba were reported for the majority of UFDs \citep{2010ApJ...708..560F,2010ApJ...716..446S,Gilmore2013,2013A&A...554A...5K,2014ApJ...794...89K,2016ApJ...832L...3J,2016ApJ...817...41J,2018ApJ...852...99N}.
In the UFDs, the [Sr/Fe] and [Ba/Fe] ratios do not rise with metallicity, they scatter around a mean floor level of $-1$ and $-0.7$, respectively (M17b). At the same time, the r-process-rich galaxies were discovered, such as Reticulum\ii\ \citep{2016ApJ...830...93J,2016AJ....151...82R},
Tucana\iii\ \citep{2017ApJ...838...44H}
%(Hansen et al. 2017), 
and Grus\ii\ \citep{2020ApJ...897..183H},
% (Hansen et al. 2020), 
 providing evidence for a variety of evolution among UFDs.	
The origin of light neutron-capture  elements in VMP stars is not well-understood yet.

%The [Sr/Ba] abundance ratios split into two branches: a horizontal one, with a constant [Sr/Ba] abundance ratio, and a decreasing [Sr/Ba] with [Ba/H]. This behavior of [Sr/Ba] is explained by two nucleosynthesis processes occured in the early Galaxy, one  produces Sr and Ba  and another produces Sr, but not Ba. The first process is a rapid $\rm (r-)$ process of neutron captures. Its site is debated, see a review of \citet{2017ARNPS..67..253T}. For the second process, neither nuclear reaction, nor its site is known \citep[see, for example,][]{2007ApJ...671.1685M}. Observations show that these two processes operated in a different way: while the r-process produced a constant [Sr/Ba] abundance ratio at the timescale of [Ba/H] variation from $-5$ to $-2$, the second one had produced Sr within a single short period at the earliest epoch of galactic formation.

%-----------

In our earlier paper (M17b), it was found that  VMP stars in the four classical dSphs (Sculptor, Ursa Minor, Sextans, and Fornax) and in the MW halo form very similar trends on the [Sr/Ba] versus [Ba/H] plane, namely, the data split into two branches: a horizontal one, with a nearly constant [Sr/Ba] $\sim -0.5$, and a well-defined downward trend of [Sr/Ba] with [Ba/H].
Such a behavior of [Sr/Ba] can be explained by two nucleosynthesis processes that occurred  in the early Galaxy. The first process is a rapid (r-) process of neutron-capture nuclear reactions. Although astrophysical sites of the r-process are still debated \citep[for a review, see][]{2017ARNPS..67..253T}, a pure r-process [Sr/Ba]$_r$ ratio was estimated empirically as $-0.44$ \citep{2005A&A...439..129B}  and $-0.38$ (M17b).

Observations suggest that the second process operated in a different way: it had produced Sr within a single short period at the earliest epoch of galactic formation, with a negligible or absent contribution to Ba. 
 In the MW and classical dSph galaxies, this source operated at epoch with [Sr/H] ~$\simeq -3$ (see Fig.~\ref{srba_srh} in the Appendix, where [Sr/Ba] is plotted versus [Sr/H] in different galaxies using the data from M17b and this study). 
"The need for a second neutron-capture process for the synthesis of the first-peak elements" was concluded also by \citet{2007A&A...476..935F} from a tight anti-correlation of [Sr/Ba] with [Ba/H] that was obtained using  high-resolution, high-S/N spectroscopy for MW halo stars.
Various ideas have been proposed for the second neutron-capture process \citep[see, for example,][]{2017ApJ...836L..21N}. 
% \citep[see][for one of the most recent treatments]{2017ApJ...836L..21N}. 
However, the source(s) is (are) not
identified yet. We  note  studies of \citet{2013A&A...553A..51C} and \citet{rizzuti2021constraints} who use an inhomogeneous chemical evolution model for the MW halo and combine an r-process contribution with an s-process from fast-rotating massive (25~$M_\odot$) stars. They managed to reproduce the spread in observed [Sr/Ba] ratios taken from a collection of \citet{2010AN....331..474F}. However, this is probably not a final solution because the used observational data set is rather inhomogeneous and the theory is tested with [Sr/Ba] versus [Fe/H], but not [Ba/H].

M17b found also that the Sr/Ba ratios in the three UFDs (Boo\ione, UMa\ii, and Leo\iv) reveal different behavior compared with that of the MW halo and classical dSphs. 
%This result is supported by \citet{2017ApJ...835...23R,2019ApJ...870...83J} and \citet{2020A&A...641A.127R}, who found lower Sr/Ba ratio in the UFDs compared with that for the MW halo stars of close Ba abundance, by approximately one order of magnitude.  
This result is supported by \citet{2017ApJ...835...23R,2019ApJ...870...83J} and \citet{2020A&A...641A.127R}, who found nearly 1~dex lower [Sr/Ba] in the UFDs compared with that for the MW halo stars with close [Ba/H].  
%The offset of [Sr/Ba] is caused by lower Sr abundances in the UFDs compared to those in massive galaxies. Indeed, the observed range of [Ba/H] in the massive galaxies and UFDs under investigation is nearly the same, while the [Sr/H] range is about 1~dex lower in the UFDs compared to the massive galaxies. 
The offset of [Sr/Ba] is caused by nearly 1~dex lower Sr abundances in the UFDs compared to those in massive galaxies. 
%Indeed, the observed range of [Ba/H] in the massive galaxies and UFDs under investigation is nearly the same, while the [Sr/H] range is about 1~dex lower in the UFDs compared to the massive galaxies. 

\begin{figure}
\includegraphics[width=90mm]{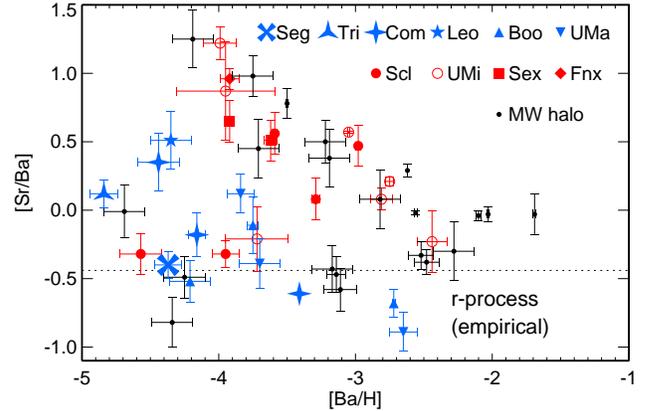}
	\caption{[Sr/Ba] NLTE abundance ratios as a function of [Ba/H]. Stars in Coma Berenices, S40  in Tri\ii, and S7 in Seg~1 are shown with four-pointed stars, a three-pointed star, and a cross, respectively. For comparison, we show NLTE abundance ratios from M17b for stars in classical dSphs (circles, squares, diamonds), UFDs (stars, triangles), and the MW halo (small circles). The dashed line indicates the empirical r-process ratio for [Sr/Ba].}
	\label{srba}
\end{figure}

In this study, we increased the statistics of homogeneous NLTE data on Sr and Ba abundances in the UFDs.  
Figure~\ref{srba} shows the [Sr/Ba] abundance ratios for ComaBer S1, S2, S3,  Seg~1 S7, and Tri\ii\ S40 stars together with the data from M17b for the UFDs, classical dSphs, and the MW halo.
Our results support the earlier findings: the [Sr/Ba] abundance ratios in the ComaBer stars, Seg~1 S7, and the Tri\ii~star are lower than that in the stars of close Ba abundance in the classical dSphs and MW halo. 
However,  [Sr/Ba] in UFDS is not constant and it can be higher than the pure r-process ratio [Sr/Ba] = $-0.4$, but not as high as in massive galaxies.
%These data allow to better constrain the behavior of [Sr/Ba] in UFDs.

%-----------

%From analyses of VMP stars in dwarf galaxies and the MW halo, \citet{2017AA...608A..89M} and \citet{2019ApJ...870...83J} found that the decreasing [Sr/Ba] branch in UFD is located 1~dex below that of classical dwarfs and MW halo. This offset is caused by lower Sr abundances in UFD compared to those in massive galaxies. The range of [Ba/H] and [Ba/Fe] in massive galaxies and UFD is nearly the same, while [Sr/H] in UFD is about 1~dex lower compared to those in massive galaxies.

%Figure~\ref{srba} shows the  [Sr/Ba] NLTE abundance ratio plotted versus [Ba/H] in Coma Berenices S1, S3, and Tri\ii\ S40  stars together with the data from \citet{2017AA...608A..89M} for classical dSphs (Sculptor, Ursa Minor, Sextans, and Fornax), UFDs (BooI, UMaII, LeoIV), and the MW halo stars. In line with earlier studies \citep{2017AA...608A..89M,2019ApJ...870...83J}, our results for Coma Berenices stars confirm lower [Sr/Ba] abundance ratio in UFD compared to those in classical dwarfs and MW halo stars. We increased a statistics of the UFDs with homogeneous data on NLTE Sr and Ba abundances. These data allow better constrain the behavior of [Sr/Ba] in UFDs.

While the sites of n-capture elements  production are not understood, observations raise one more question.  
Why do smaller galaxies have lower [Sr/Ba] compared to their more massive counterparts?
To explain the difference in Sr abundance between different galaxies, some studies  appeal to the idea of the initial mass function (IMF) sampling \citep{2013ApJ...774..103L}, and the lack of the most massive stars in UFDs.
Regardless of this idea, we note a correlation between [Sr/Ba] ratio and a fraction of binary stars in different dwarf galaxies.
%These value corresponds to the present fraction of binary stars	and spans from 0.14$^{+0.28}_{-0.05}$ in Carina dSph \citep{2017AJ....153..254S} to 0.78$^{+0.09}_{-0.08}$ in Ursa Minor dSph \citep{2018AJ....156..257S}. 

To date, the binary fraction has been determined in a dozen dSphs \citep{Martinez_2011,2013ApJ...779..116M,2013ApJ...771...29G,2017AJ....153..254S,2018AJ....156..257S,2019MNRAS.487.2961M} and it spans from 0.14$^{+0.28}_{-0.05}$ in Carina dSph
to 0.78$^{+0.09}_{-0.08}$ in Ursa Minor (UMi) dSph. 
The [Sr/Ba] ratio in Ursa Minor spans in wide range from close to the r-process ratio [Sr/Ba] = $-0.23 \pm 0.23$, up to [Sr/Ba] = $1.22 \pm 0.12$.
In contrast to UMi, Carina stars show values  near the r-process, [Sr/Ba] = $-0.33$ and $-0.59$ \citep{2012ApJ...751..102V}.
\citet{Martinez_2011} found a low binary fraction of less that 0.2 in Segue~1, and Seg~1 S7 has [Sr/Ba] = $-0.40$, while, in the other Seg~1 stars, Sr and Ba lines are too weak to be measured ([Sr/H] $< -4.9$ and [Ba/H] $< -4.2$, according to estimates of F14).
The Reticulum\ii\ UFD, with a boundary value of 0.5 for the binary fraction \citep{2019MNRAS.487.2961M}, misses a second source of Sr: the two most metal-poor stars ([Fe/H] $< -3$) have very low abundances of Sr and Ba, such that they cannot be measured, while the higher metallicity stars have strongly enhanced n-capture elements originated from the r-process \citep{2016ApJ...830...93J,2016AJ....151...82R}.
The MW and five dSphs, namely, Ursa Minor, Sextans, Sculptor, Fornax, and Draco show a binary fraction of more than 0.5 together with a wide range of [Sr/Ba], up to 1.3~dex. 

\begin{figure}
	\includegraphics[width=90mm]{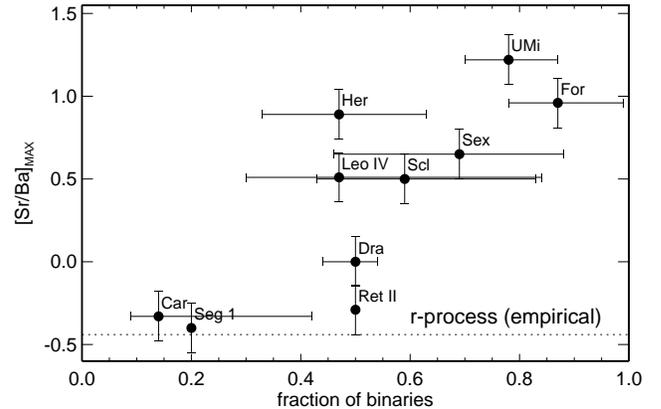}
	\caption{
		%For different dwarf galaxies, their maximal [Sr/Ba] found to the date  is plotted versus the fraction of binary stars.  
		The maximal [Sr/Ba] ratios found to date in different dwarf galaxies as a function of fraction of binary stars in the galaxy. See Sect.~\ref{ncapture} for details.}
	\label{nbf}
\end{figure}

%Fig.~\ref{nbf} shows the maximal [Sr/Ba] ratios found to date in different dwarf galaxies as a function of a fraction of binary stars in the galaxy. 
Fig.~\ref{nbf} shows the maximal [Sr/Ba] ratios found to date in different dwarf galaxies as a function of their fraction of binaries. 
A linear trend is found: the larger spread in [Sr/Ba] is observed in the galaxies with the larger fraction of binaries.  The data on [Sr/Ba] were taken from this study (Segue~1), M17b  (Fornax, Leo\iv, Sculptor, Sextans, Ursa~Minor),   \citet[][Hercules]{2016A&A...588A...7F},  \citet[][Reticulum\ii]{2016ApJ...830...93J}, \citet[][Draco]{2009ApJ...701.1053C}, and \citet[][Carina]{2012ApJ...751..102V}. 
% For Draco, strontium and barium  abundances were determined in \citet{2009ApJ...701.1053C}. 
%For, Leo\ii\ and Reticulum\ii, with binary fraction of 0.3 \citep{2017AJ....153..254S} and 0.5 \citep{2019MNRAS.487.2961M}, respectively, we cannot make a solid conclusion on the correlation between binary fraction and [Sr/Ba]. 
%Reticulum\ii\ is an outstanding UFD, known with its extremely r-process enhanced  stars \citep{Ji_ret2}. 
%For Leo\ii, with binary fraction of 0.3 \citep{2017AJ....153..254S}, spectral observation in the region covering Sr and Ba lines had never been performed yet. This hypothesis can be checked in future  with high resolution observations of stars in Leo\ii.
%If the additional source of strontium (besides the r-process) existed in Reticulum~II, we would not be able to trace it against the background of high level or r-process with [Sr/Fe]$\simeq0.3$ and [Ba/Fe]$\simeq1.2$ \citep{Ji_ret2}.

%Although these value corresponds to the present fraction of binary stars, the binary fraction of the Population\iii\ stars in different galaxies may differ as well.
%Although the  above fractions of binary stars  correspond to the present time, the binary fraction of  Population\iii\ stars in different galaxies likely differed.
Although the  above fractions of binary stars  correspond to the present time, the binary fraction  in different galaxies  likely differed at the earliest epoch of their formation.
We suggest a hypothesis that the second source (besides the r-process) of Sr is attributed to  massive metal-poor binary stars.	
%This hypothesis can be checked in the future  with high resolution observations of stars in Leo\ii\ UFD, a galaxy with a  low binary fraction of 0.3   \citep{2017AJ....153..254S}.
This hypothesis can be checked in the future  with high resolution observations of stars in Leo\ii\ UFD, which has a  low binary fraction of 0.3   \citep{2017AJ....153..254S}.
It is worth noting that Sr has recently been directly detected 	in a neutron-star merger \citep{2019Natur.574..497W}, which does 	require a binary system.

While the cause of a difference in [Sr/Ba] between UFDs and massive galaxies is not clear, this property can be used to shed light on the nature of stellar stream progenitors. 
Measurements of Sr and Ba abundances in a given stream could allow to  figure out whether a destroyed dwarf galaxy was an UFD or a classical dSph.
%, the nature of its progenitor could be uncovered, and the 
%One can figure out whether a destroyed dwarf galaxy was an UFD or a classical dSph by measuring Sr and Ba abundances in a stellar stream.
%The abundances of Sr and Ba in a stellar stream can shed light on the nature of its progenitor and to deduce, whether a destroyed dwarf galaxy was an UFD or a classical dSph.	

%It is worth noting that dwarf spheroidal galaxies (dSps) and UFDs have a different fraction of binary stars \citep{2013ApJ...779..116M,2017AJ....153..254S,2018AJ....156..257S}. A fraction of binaries spans from 20 to 87 percent  in case of Carina and Fornax dSphs, respectively. These data correspond to the present fraction of binary stars. However, the binary fraction of the Population\iii\ stars in different galaxies may differ as well. We noted that Fornax, Ursa Minor, and Sextans have the largest binary fraction of 87, 78, and 71 \%, respectively, and these dSphs show the highest [Sr/Ba] abundance ratio up to 1.0~dex, 1.2~dex, and 0.7~dex, respectively, while Carina galaxy shows low [Sr/Ba] = $-0.33$ and $-0.59$ \citep{2012ApJ...751..102V} and a very low binary fraction of 20 \%. Unfortunately, for other dwarf galaxies, either binary fraction or [Sr/Ba] are not available in the literature. We suggest a hypothesis that the second source (besides the r-process) of Sr can be attributed to the massive Pop\iii\ binary stars. 

\section{Conclusions}
\label{con}

%We revisited S1, S2, and S3 metal-poor stars ([Fe/H] $<$ -2) in Coma Berenices UFD investigated by \citet{2010ApJ...708..560F}.
%We revisited S1, S2, and S3 in Coma Berenices UFD and S1, SB, and SC stars in Segue~1 system investigated earlier by \citet{2010ApJ...708..560F} and \citet{2014ApJ...786...74F}, respectively.
We revisited chemical element abundances of three stars in Coma Berenices UFD, seven stars in Segue~1 UFD, and one star in Tri\ii\ UFD, using high resolution spectral observations from the original papers of \citet{2010ApJ...708..560F}, \citet{2014ApJ...786...74F}, \citet{2010ApJ...722L.104N}, and \citet{2017ApJ...838...83K}.
Our study took advantage of new photometric observations in the visible and infra-red bands obtained with the 2.5~m telescope of the SAI MSU Caucasian observatory and  chemical abundance determinations taking into account  NLTE effects. 
The procedure of photometric  observations and  data reduction  
was tested, and high data accuracy for faint objects was achieved. The errors in the derived magnitudes are within 0.03~mag.

We determined atmospheric parameters and abundances for up to eighteen chemical elements from carbon to barium in the sample stars. 
For the first time, we present  abundance determination of  oxygen in Seg~1 S1 and S4, silicon in ComaBer S2 and Tri\ii\ S40, potassium in Seg~1 S1$-$S6 and ComaBer S1$-$S3, and barium in Seg~1 S7.
Our conclusions  can be summarized as follows.

\begin{itemize}

\item 
In Segue~1 and Coma Berenices, different stars show different chemical element abundance patterns, which argues for stochastic star formation and incomplete gas mixing in these UFDs. This is in line with the earlier findings of \citet{2010ApJ...708..560F} and \citet{2014ApJ...786...74F}.

\item
In each galaxy, we found stars with low [Na/Mg] of $-1.1$ dex.
Coma Berenices S2 shows extremely low [Na/Mg] = $-1.68$, which is the lowest value known to date.
We interpret this chemical property as a footprint of nucleosynthesis in Pop\iii\ stars.  This odd-even effect was not blurred due to a small number of nucleosynthesis events that contributed to chemical abundances of the sample stars.

	\item 
We increased the statistics of homogeneous data on Sr and Ba NLTE abundances in UFDs. 
Our results for Coma Berenices, Segue~1, and Triangulum\ii\ stars confirm that
ultra-faint dwarf galaxies are distinct from massive galaxies in Sr production: their [Sr/H] is nearly 1~dex lower, while [Ba/H] is similar \citep{2017AA...608A..89M,2019ApJ...870...83J}. 
However, in UFDs, just as in massive galaxies, [Sr/Ba] is not constant and it can be higher than the pure r-process ratio, but not as high as in massive galaxies.

	\item 
We propose that the additional Sr production at the earliest epoch of galactic formation could be attributed to evolution of massive metal-poor binaries.

\end{itemize}

%\section*{Acknowledgments}
{\it Acknowledgments}.
%This work was supported in part by M.V.Lomonosov Moscow State University Program of Development.
The authors acknowledge the support from the Program of development of M.V. Lomonosov Moscow State University (Leading Scientific School ''Physics of stars, relativistic objects and galaxies'').
T.S., L.M., and P.J. are indebted to the International Space Science Institute (ISSI), Bern, Switzerland, for supporting and funding the international team ''The Pristine'' and to the ''ChETEC'' COST Action (CA16117), supported by COST (European Cooperation in Science and Technology).
T.S. and M.N. are grateful to Foundation for the Advancement of Theoretical Physics and Mathematics 'BASIS' for financial support (20-1-3-10-1).
T.S. is grateful to K.~Plotnikov for language editing support.
A.F. acknowledge support from NSF grant AST-1716251 and thanks the Wissenschaftskolleg zu Berlin for their wonderful Fellow's program and generous hospitality.
This study is based on spectral observations retrieved from the Keck Observatory Archive (C168Hr), which is operated by the W. M. Keck Observatory and the NASA Exoplanet Science Institute (NExScI), under contract with the National Aeronautics and Space Administration.
We made use of the StarFit database http://starfit.org/, SAGA database http://sagadatabase.jp/, the SIMBAD database, operated at CDS, Strasbourg, France; NASA/IPAC Infrared Science Archive, which is operated by the Jet Propulsion Laboratory, California Institute of Technology, under contract with the National Aeronautics and Space Administration; MARCS; and VALD databases.
We also thank the referee for useful comments and suggestions.

%T.S., L.M., and Yu.P. acknowledge financial support from the Presidium RAS Programme P-12. 

\section{Data availability}

The data used in this article will be shared on request to the corresponding author.

\bibliography{mp,dwarf_gal,stellar_parameter1,photo,nlte1,mashonkina,ref,gfvalues,mp_stars}
\bibliographystyle{mn2e}

%\bsp

\section{Appendix}

%\begin{figure}
%	\includegraphics[trim=0.0cm 1.45cm 0.0cm 0.0cm,clip,width=90mm,height=22mm]{/home/sitamih/jablonka/copy_plot/plots/idl_data/crfe_feh.ps}	
%	\includegraphics[trim=0.0cm 1.45cm 0.0cm 0.65cm,clip,width=90mm,height=22mm]{/home/sitamih/jablonka/copy_plot/plots/idl_data/mnfe_feh.ps}
%	\includegraphics[trim=0.0cm 1.45cm 0.0cm 0.65cm,clip,width=90mm,height=22mm]{/home/sitamih/jablonka/copy_plot/plots/idl_data/cofe_feh.ps}
%	\includegraphics[trim=0.0cm 1.45cm 0.0cm 0.65cm,clip,width=90mm,height=22mm]{/home/sitamih/jablonka/copy_plot/plots/idl_data/nife_feh.ps}
%	\includegraphics[trim=0.0cm 0.30cm 0.0cm 0.65cm,clip,width=90mm,height=22mm]{/home/sitamih/jablonka/copy_plot/plots/idl_data/znfe_feh.ps}
	%	\includegraphics[trim=0.0cm 1.42cm 0.0cm 0.6cm,clip,width=90mm]{/home/sitamih/jablonka/copy_plot/plots/idl_data/vmg_mgh.ps}
	%	\includegraphics[trim=0.0cm 0.00cm 0.0cm 0.6cm,clip,width=90mm]{/home/sitamih/jablonka/copy_plot/plots/idl_data/kv_mgh.ps}
	%	\includegraphics[trim=0.0cm 1.42cm 0.0cm 0.6cm,clip,width=90mm]{.ps}
	%	\includegraphics[trim=0.0cm 1.42cm 0.0cm 0.6cm,clip,width=90mm]{.ps}
	%	\includegraphics[trim=0.0cm 0.00cm 0.0cm 0.6cm,clip,width=90mm]{.ps}
%	\caption{The NLTE chemical abundance ratios for iron-peak elements in different galaxies.  For Co and Ni, LTE abundances from Co\ione\ and Ni\ione\ are plotted relative to LTE abundance from Fe\ione.}
%	\label{iron_ratios}
%\end{figure}

\begin{figure}
	\includegraphics[trim=0.0cm 1.42cm 0.0cm 0.0cm,clip,width=90mm,height=29mm]{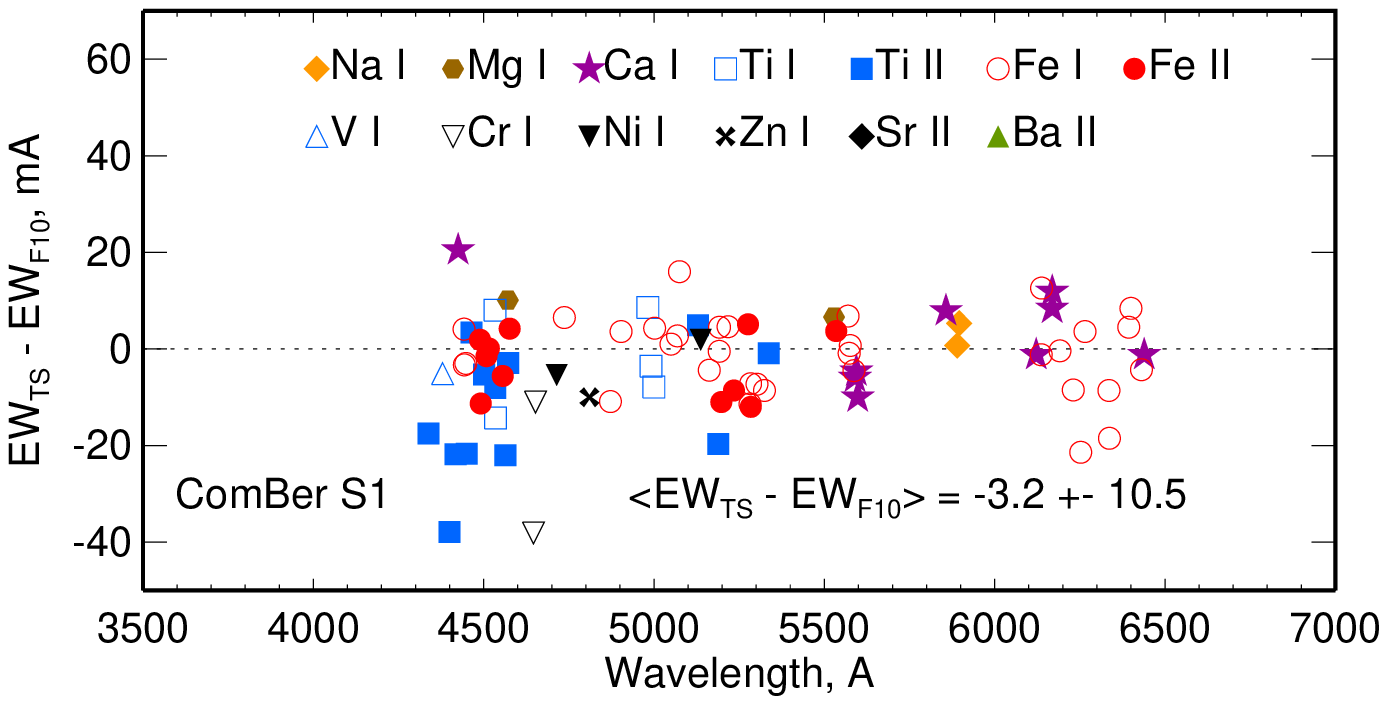}	
	\includegraphics[trim=0.0cm 1.42cm 0.0cm 0.6cm,clip,width=90mm,height=19mm]{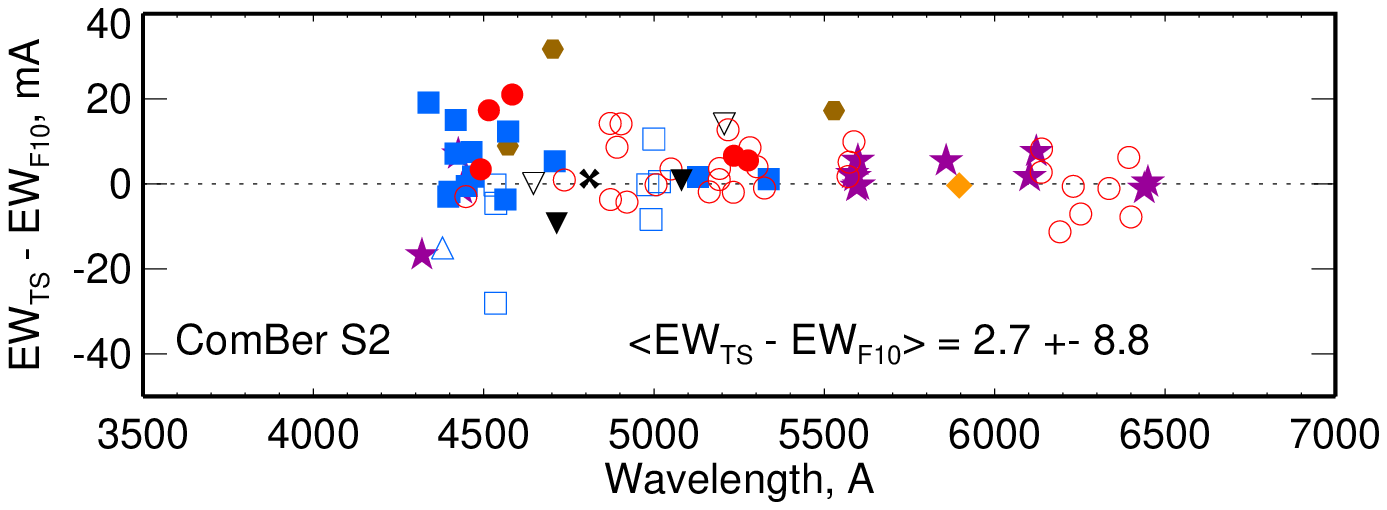}
	\includegraphics[trim=0.0cm 1.42cm 0.0cm 0.6cm,clip,width=90mm,height=19mm]{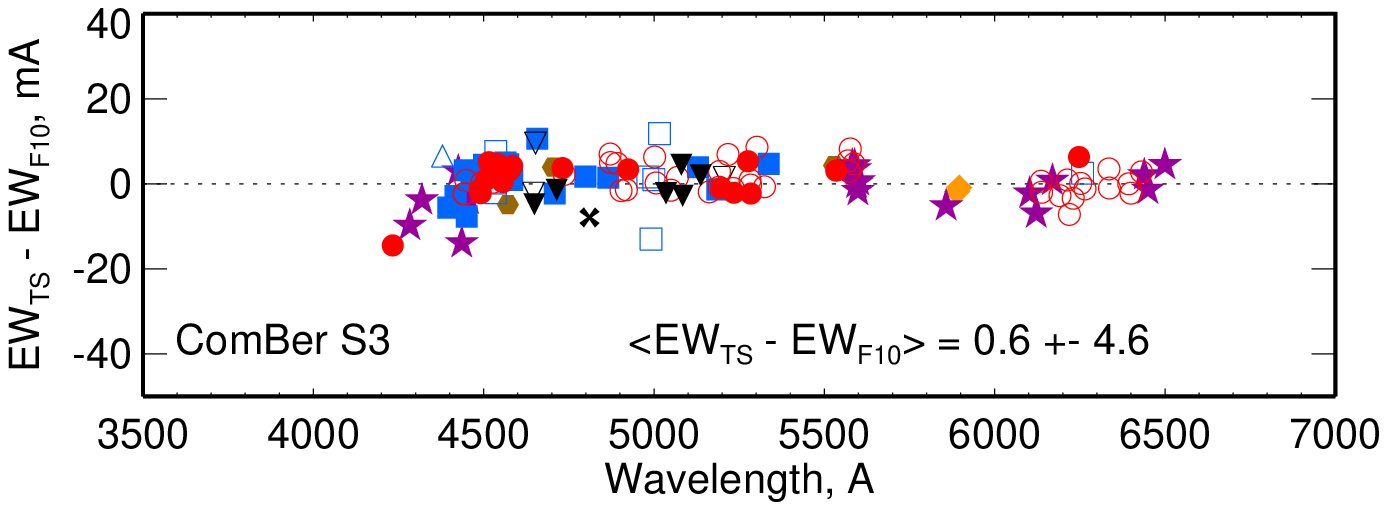}
	\includegraphics[trim=0.0cm 1.42cm 0.0cm 0.6cm,clip,width=90mm,height=19mm]{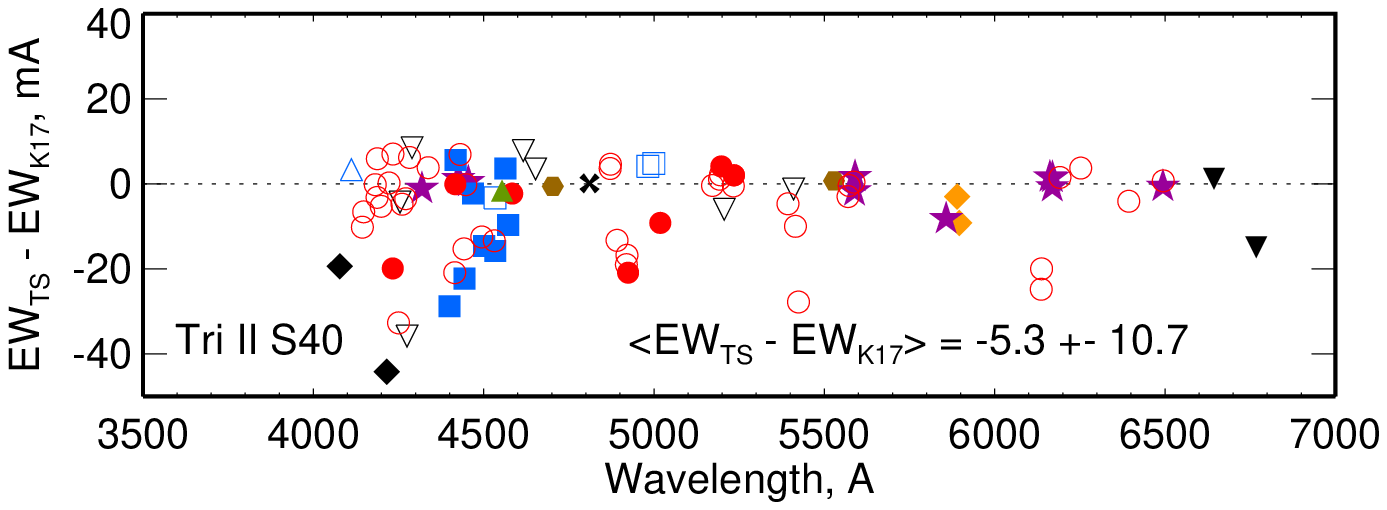}
	\includegraphics[trim=0.0cm 1.42cm 0.0cm 0.6cm,clip,width=90mm,height=19mm]{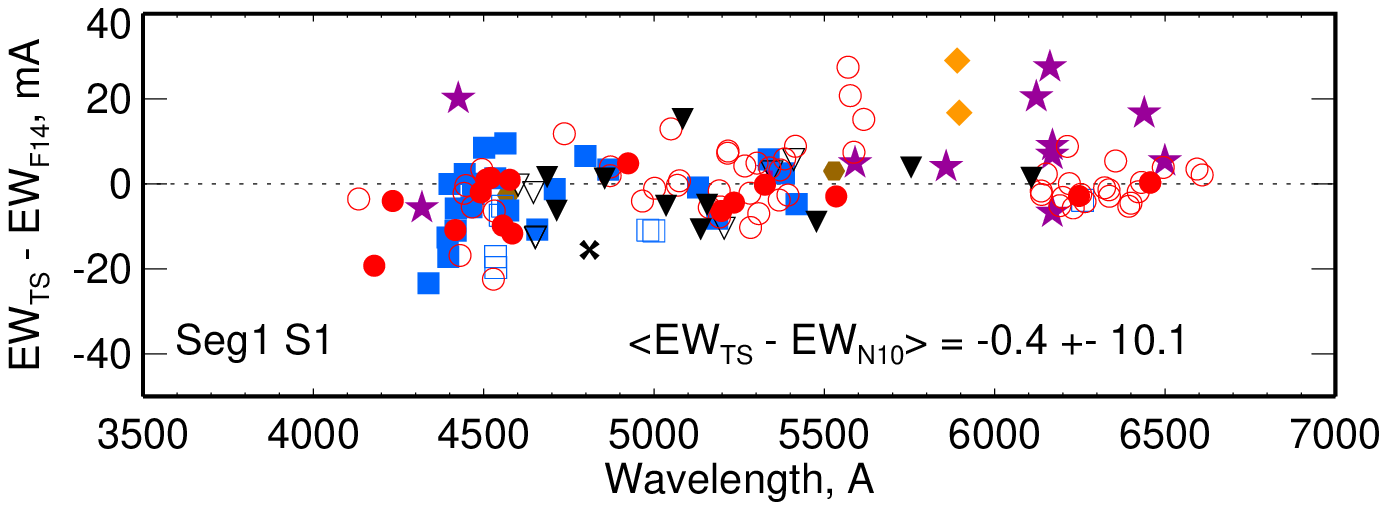}
	\includegraphics[trim=0.0cm 1.42cm 0.0cm 0.6cm,clip,width=90mm,height=19mm]{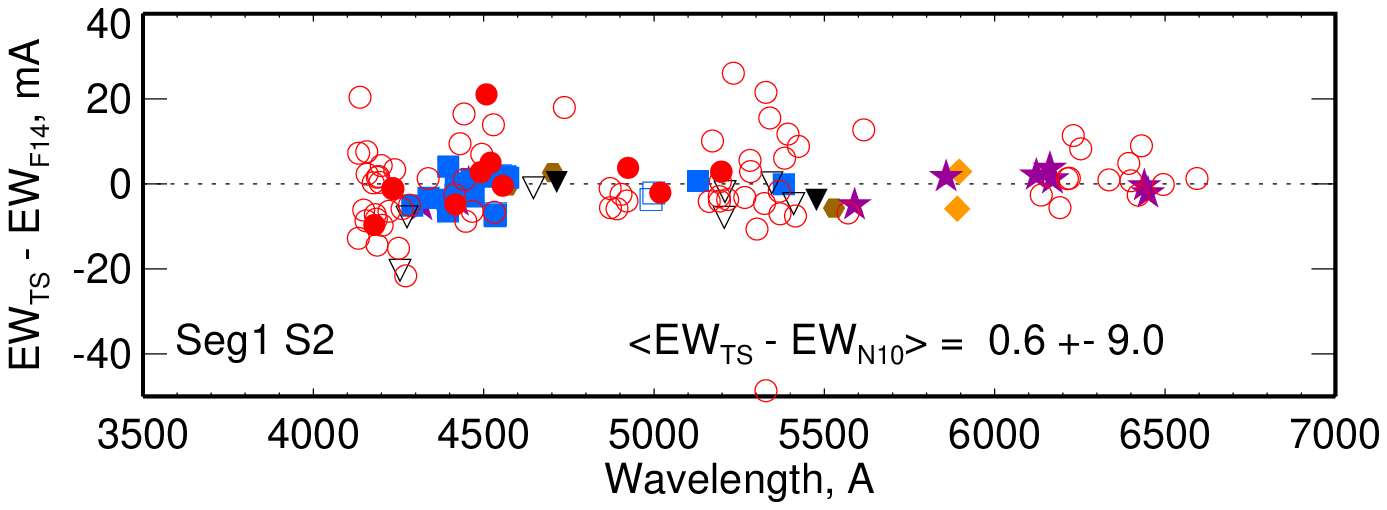}
	\includegraphics[trim=0.0cm 1.42cm 0.0cm 0.6cm,clip,width=90mm,height=19mm]{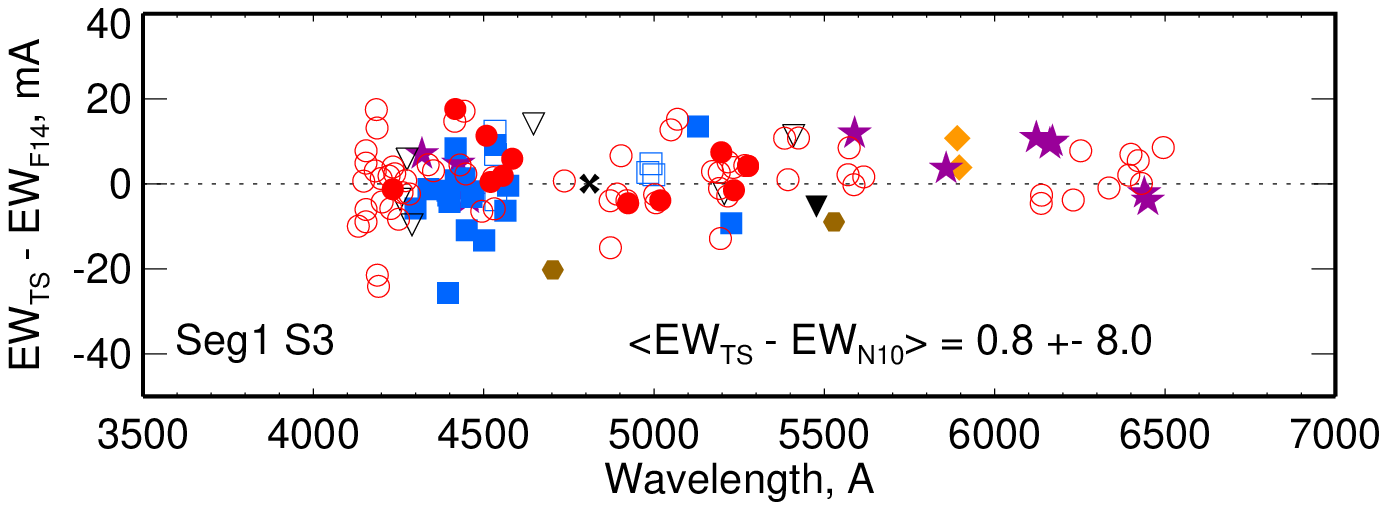}
	\includegraphics[trim=0.0cm 1.42cm 0.0cm 0.6cm,clip,width=90mm,height=19mm]{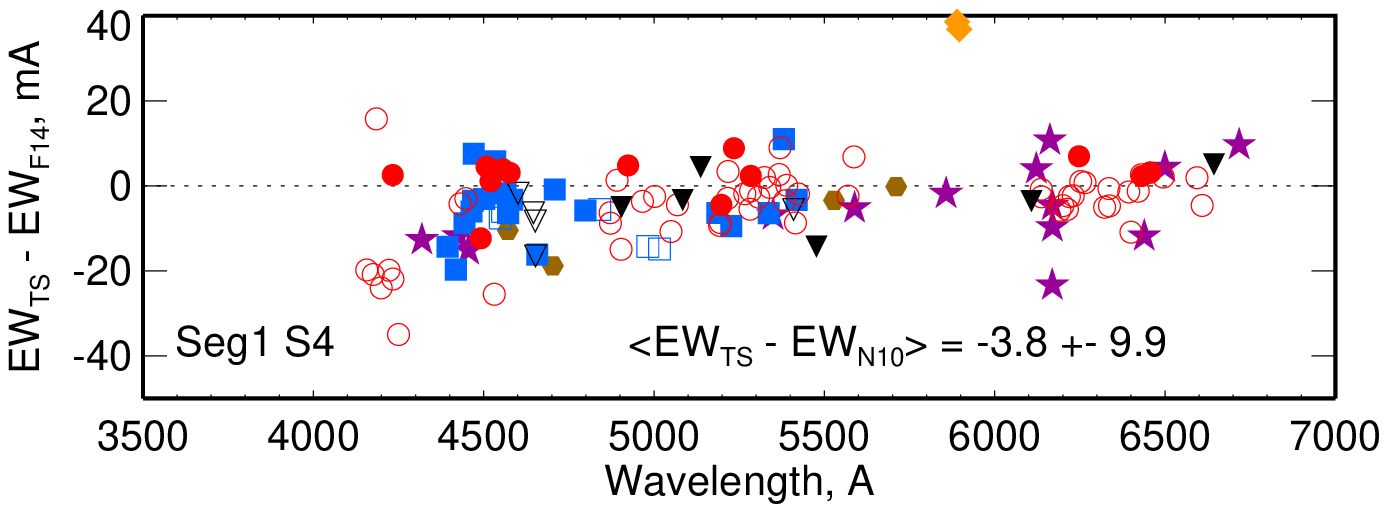}
	\includegraphics[trim=0.0cm 1.42cm 0.0cm 0.6cm,clip,width=90mm,height=19mm]{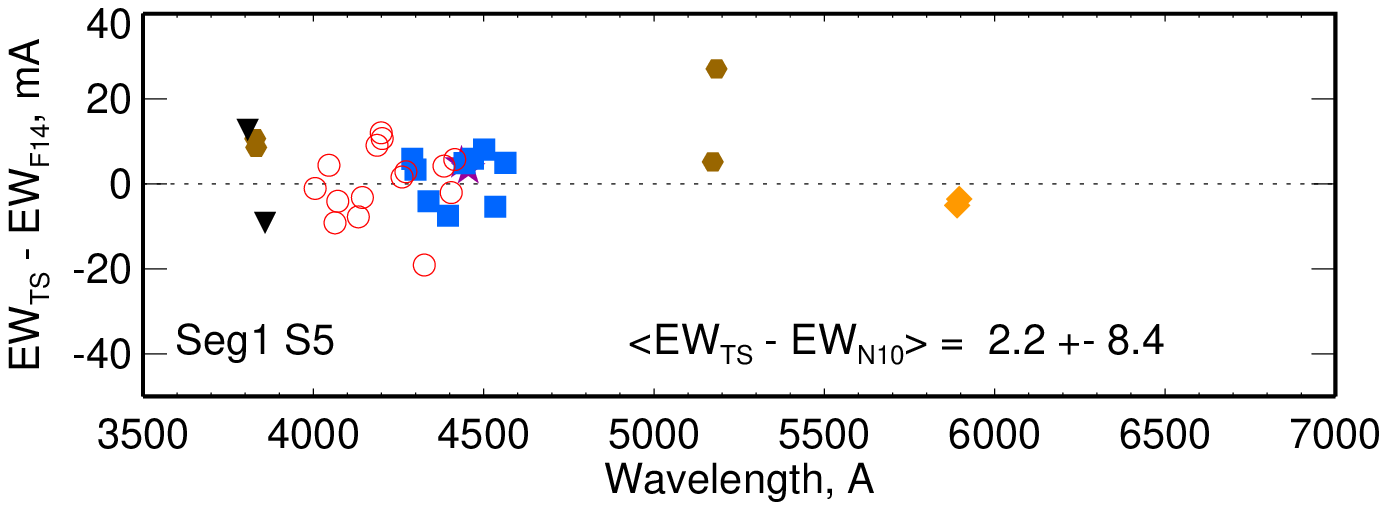}
	\includegraphics[trim=0.0cm 1.42cm 0.0cm 0.6cm,clip,width=90mm,height=19mm]{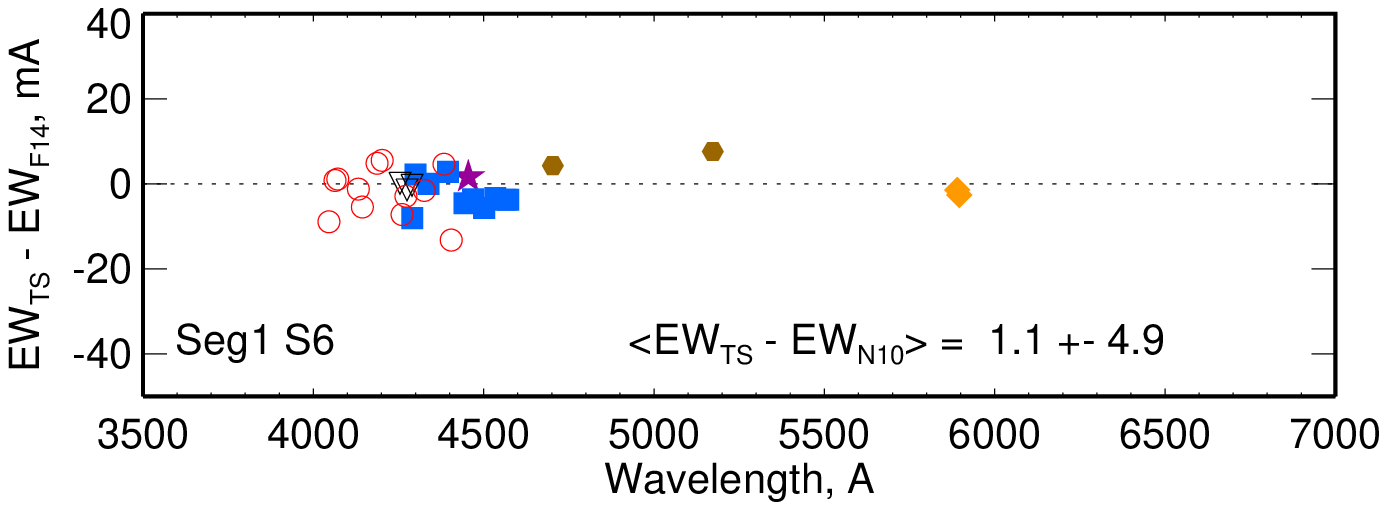}
	\includegraphics[trim=0.0cm 0.00cm 0.0cm 0.6cm,clip,width=90mm,height=19mm]{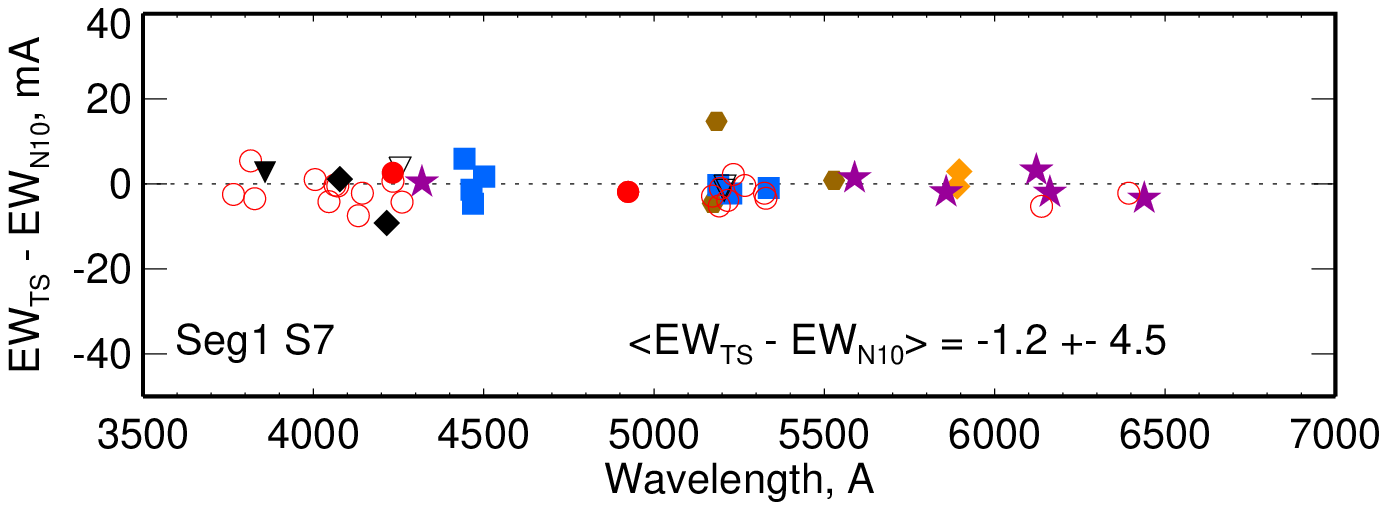}
	\caption{Differences in equivalent width (EW) between this study (TS) and the literature data. %\citet[F10]{2010ApJ...708..560F}
		% for lines of different species in Coma Berenices S1 (top panel), S2 (middle panel), and S3 (bottom panel). 
		For each star, the average difference in EW between TS and the literature data is indicated.}
	\label{ewcomp}
\end{figure}

	%Поскольку для некоторых звёзд мы используем архивные спектры, обработанные иным способом нежели в первоначальных статьях, то мы измерили ЭШ и сравнили их с тем, что получено в литературе.
	%Since, for some stars, we adopted archival spectra, reduced in a different manner, compared to the original papers, we present  a comparison of the equivalent widths (EWs) measured in our study and in the original papers. %(Figure~\ref{ewcomp}, in the Appendix).
	For some stars, we adopted archival spectra, reduced in a different manner, compared to the original papers. Thus, for comparison, we measured  the equivalent widths (EWs) and presented  a comparison with the original papers.
	For EW measurements, we apply the following procedure. We fit the synthetic spectrum to the observed spectrum and measure the equivalent width of the synthetic line profile. 
	Figure~\ref{ewcomp}  shows the differences in the equivalent widths between this study and the literature data \citep{2010ApJ...708..560F,2014ApJ...786...74F,2017ApJ...838...83K,2010ApJ...722L.104N}.
	We found a reasonable agreement with the literature data, and, for each star, the average difference in equivalent widths  does not exceed the standard deviation.
	The brighter star, the smaller the scatter in the equivalent width measurements. A large discrepancy of 38 m\AA, was found for the Ti\ii\ 4399~\AA\ and Cr\ione\ 4646~\AA\ lines in ComaBer S1 star. This difference is caused by a lower signal to noise ratio in the blue spectrum region. We also note a large difference between this study and K17 found for Sr\ii\ 4077 and 4215\AA\ lines in Tri\ii\ S40. 
	Each of these lines is displaced at the edge of the echelle order, and the uncertainty in accounting for the scattered light can be the source of discrepancies in EWs.
	Our EWs are smaller and lead to consistent strontium abundance, while, using EWs from K17, leads to a discrepancy of 0.4~dex between the two lines.

\begin{figure}
	\includegraphics[trim=0.0cm 0.0cm 0.0cm 0.0cm,clip,width=90mm]{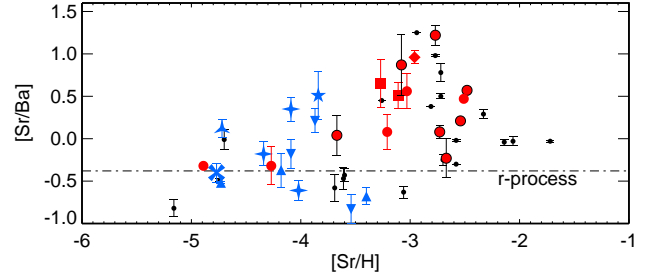}	
	\caption{The NLTE [Sr/Ba] abundance ratios for stars in different galaxies. Note a steep [Sr/Ba] increase at [Sr/H] $\simeq -3$ in the MW (black symbols) and classical dSph galaxies (red symbols) and lower [Sr/H] in UFDs (blue symbols) compared to those in massive galaxies.  The designations are the same as in Fig.~\ref{srba}.  }
	\label{srba_srh}
\end{figure}

\begin{figure}
	\includegraphics[trim=0.0cm 1.45cm 0.0cm 0.0cm,clip,width=90mm]{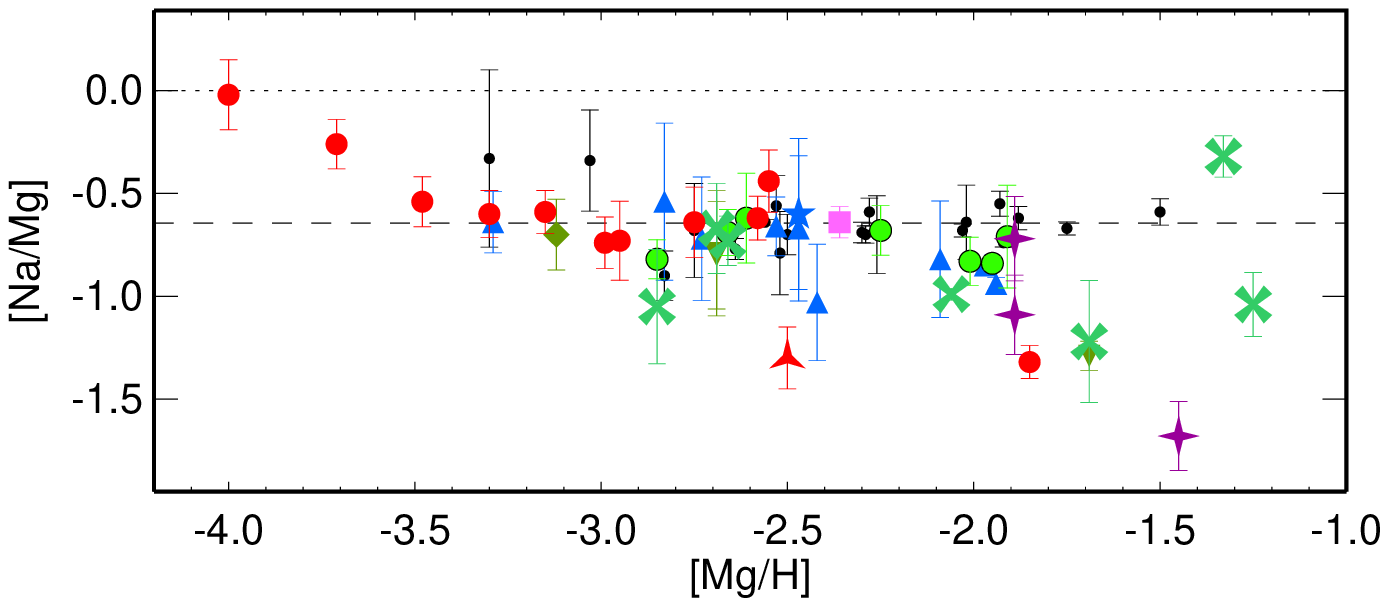}	
	\includegraphics[trim=0.0cm 0.30cm 0.0cm 0.65cm,clip,width=90mm]{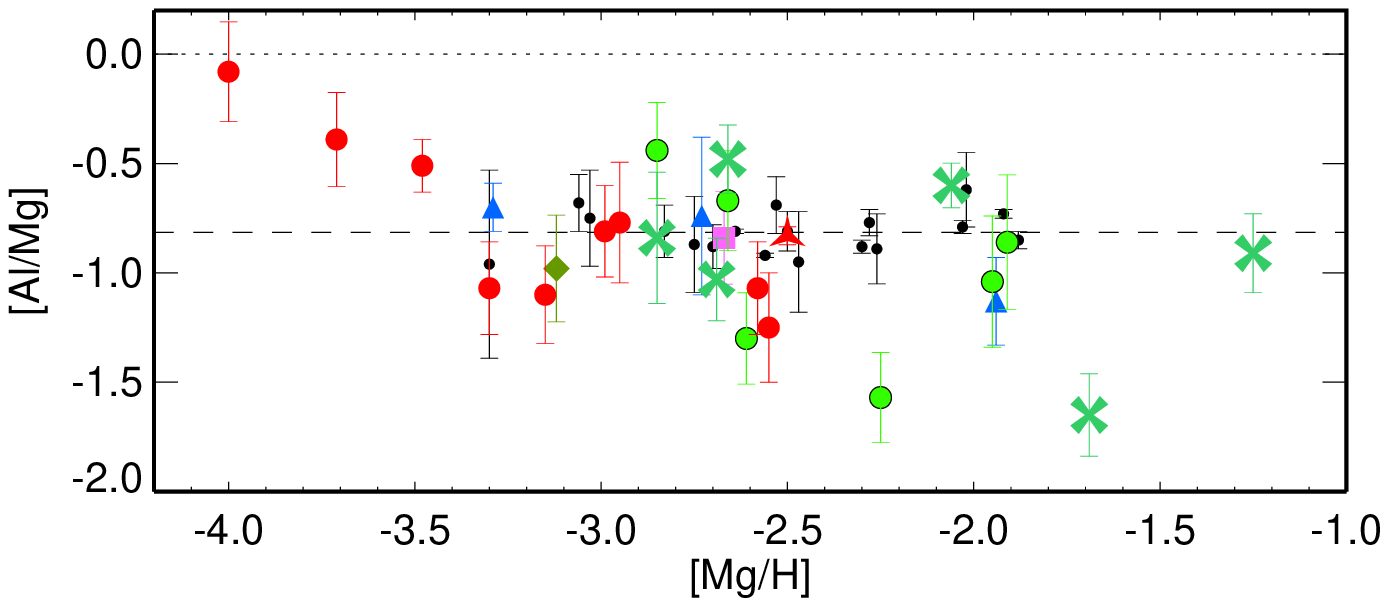}
	\caption{The NLTE chemical abundance ratios for odd-Z elements in different galaxies. Some stars have significantly lower [El/Mg] compared to the mean value.}
	\label{odd_ratios}
\end{figure}

\begin{figure}
	\includegraphics[trim=0.0cm 1.45cm 0.0cm 0.0cm,clip,width=90mm]{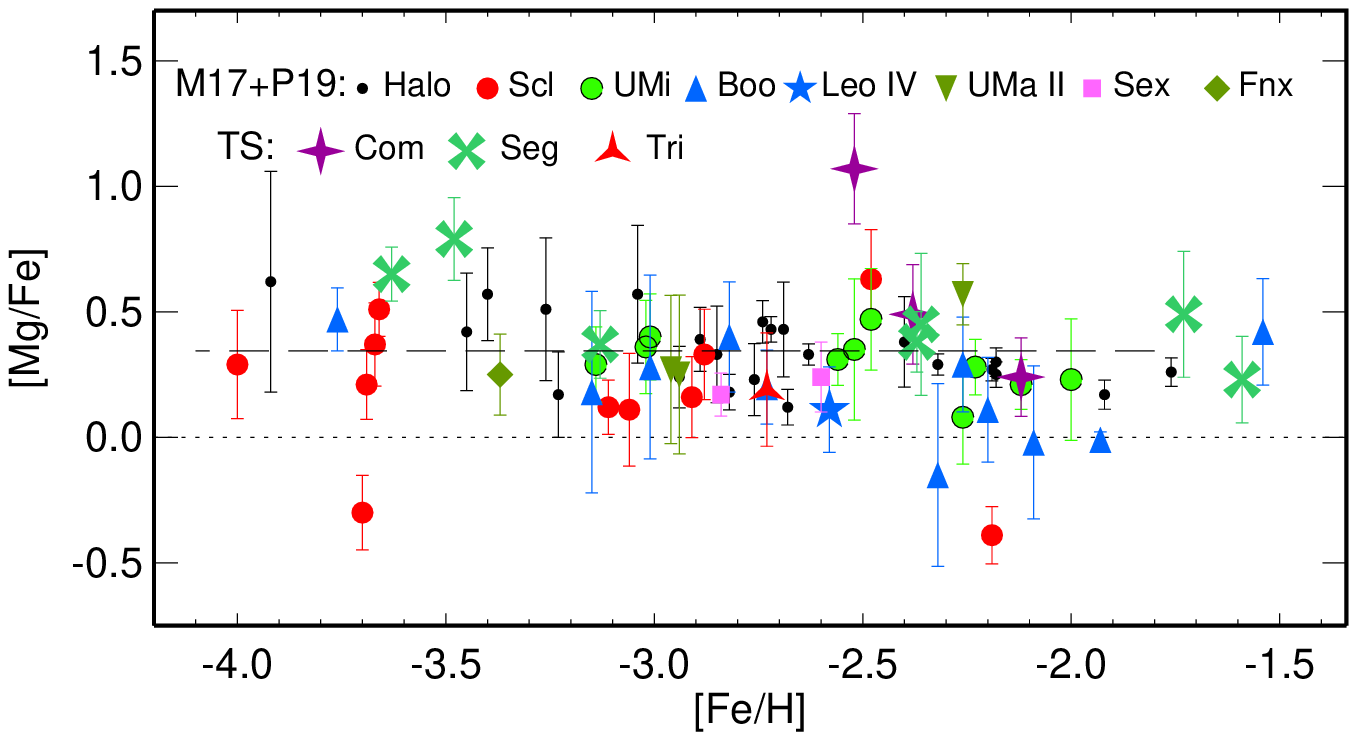}	
	\includegraphics[trim=0.0cm 1.45cm 0.0cm 0.65cm,clip,width=90mm]{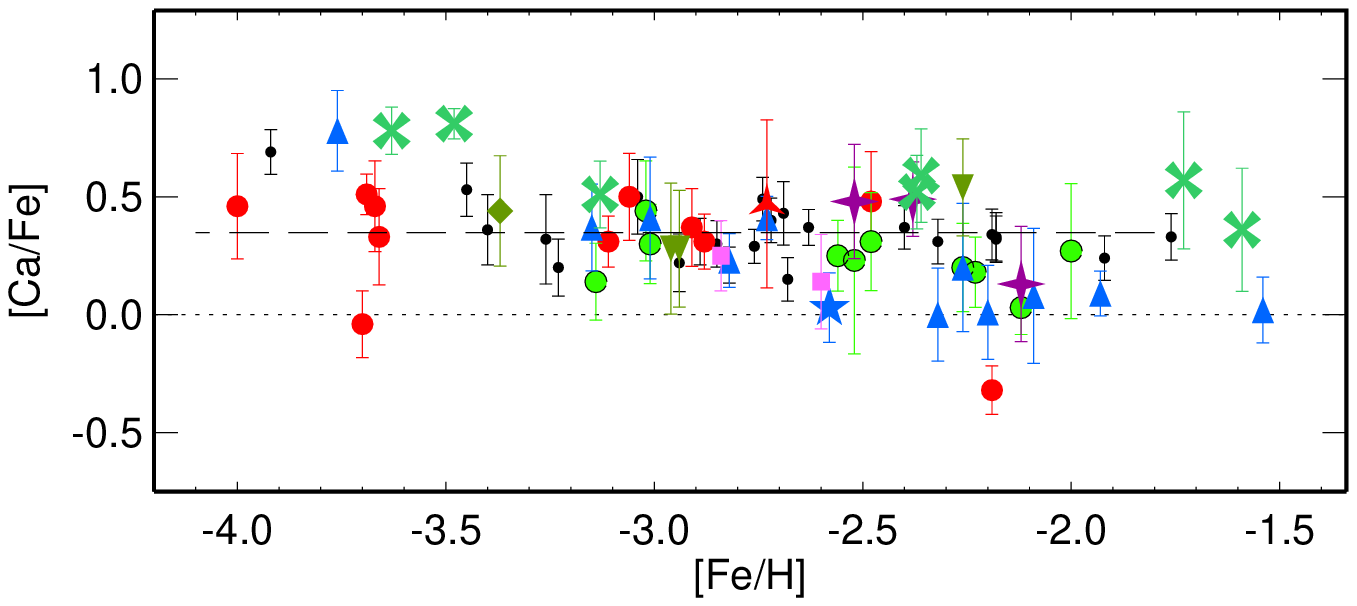}
	\includegraphics[trim=0.0cm 1.45cm 0.0cm 0.65cm,clip,width=90mm]{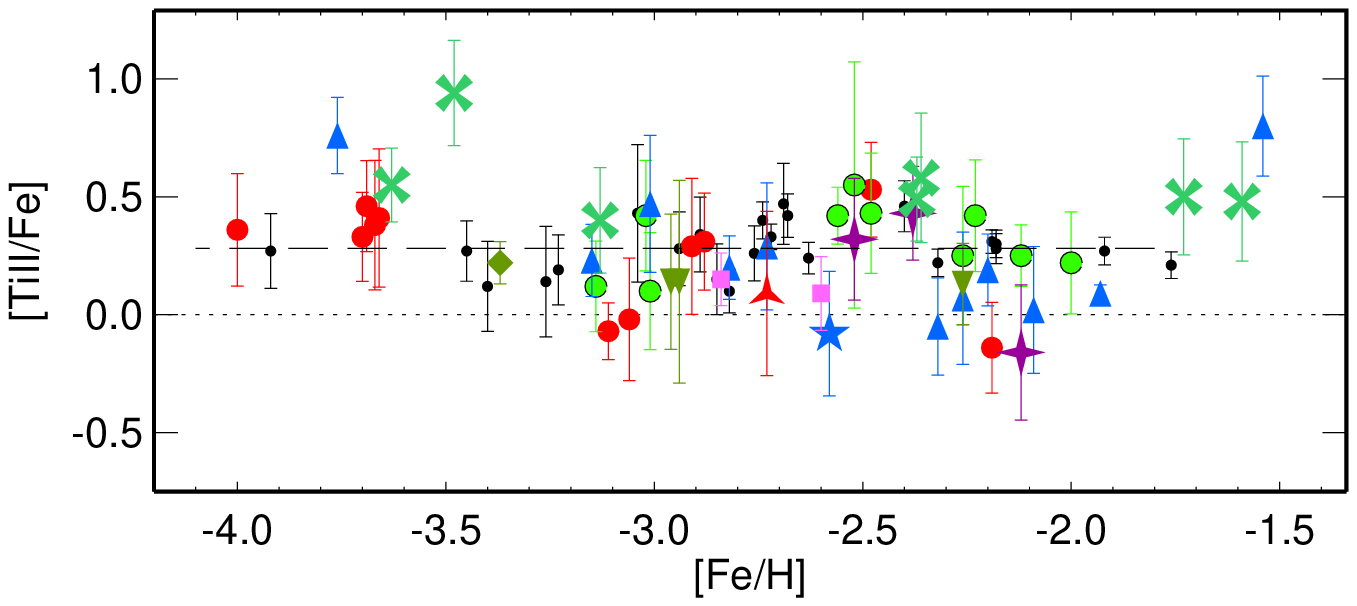}
	\includegraphics[trim=0.0cm 1.45cm 0.0cm 0.65cm,clip,width=90mm]{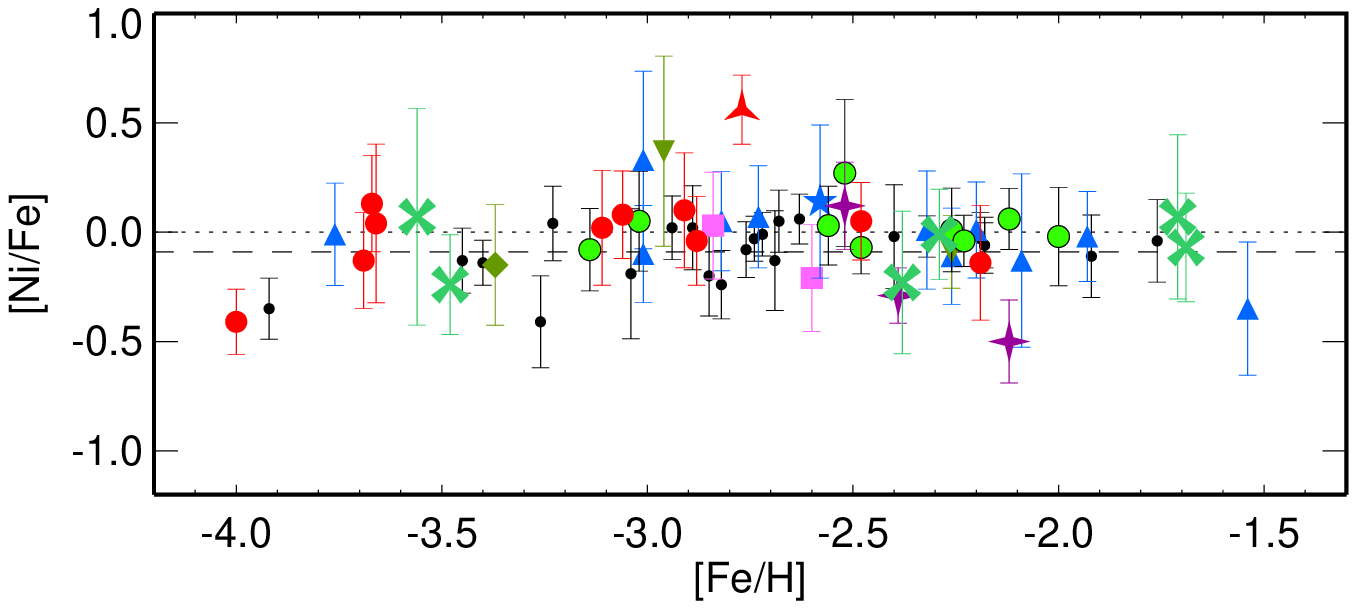}
	\includegraphics[trim=0.0cm 0.30cm 0.0cm 0.65cm,clip,width=90mm]{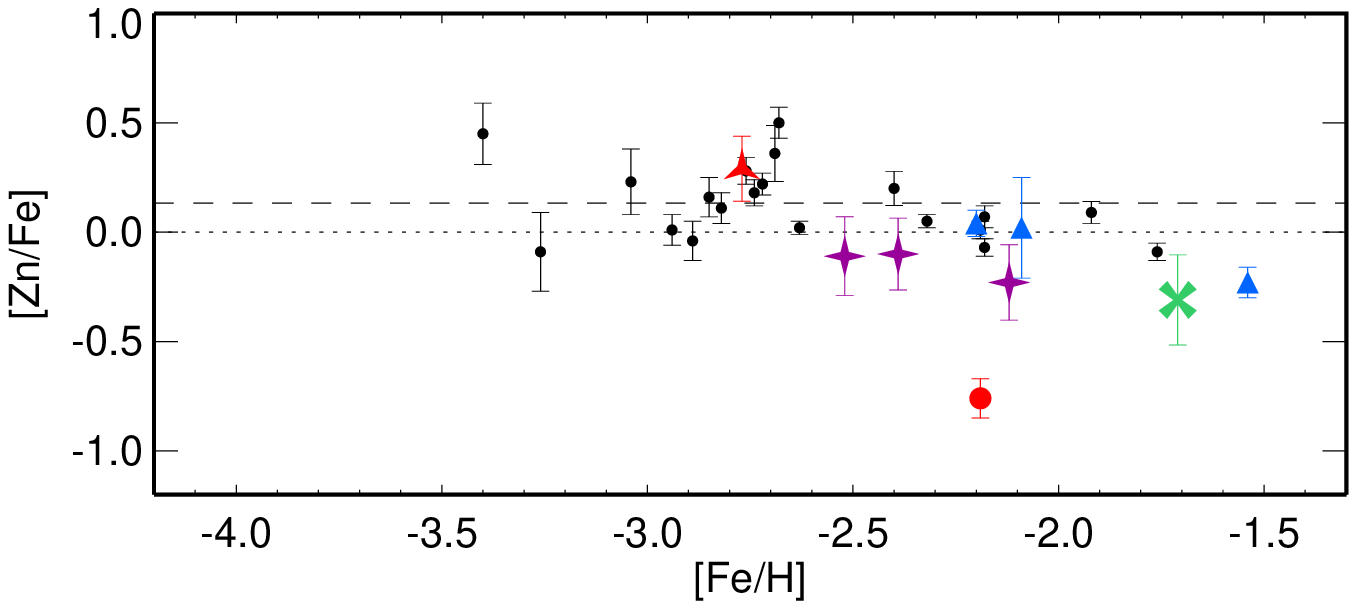}
	\caption{The NLTE chemical abundance ratios in different galaxies. Mean abundance ratios calculated for the MW halo stars are shown with dashed lines. 
			For [Ni/Fe],  LTE abundance from Ni\ione\ relative to LTE abundance from Fe\ione\ is plotted.
		[$\alpha$/Fe] ratios are high in all Seg~1 stars. %Mean abundance ratios calculated for the MW halo stars are shown with dashed lines.
	}
	\label{tot_ratios}
\end{figure}

\end{document}